\documentclass[useAMS]{mn2e}
\usepackage{mnras_cite}
\usepackage{epsfig}

\def\gsim{\ga}
\def\lsim{\la}

\def\HI{{\rm H}{\sc i}}
\def\HII{{\rm H}{\sc ii}}
\def\H2{{\rm H}$_2$}
\def\HeI{{\rm He}{\sc i}}
\def\HeII{{\rm He}{\sc ii}}
\def\d{{\rm d}}

\begin{document}

\title[The Epoch of Reionization]{The Epoch of Reionization}
\author[Andrew~J.~Benson, Naoshi~Sugiyama, Adi~Nusser, Cedric~G.~Lacey]{Andrew~J.~Benson$^1$, Naoshi~Sugiyama$^2$, Adi~Nusser$^{2,3}$, Cedric~G.~Lacey$^4$\\
1. Department of Physics, University of Oxford, Keble Road, Oxford OX1 3RH, United Kingdom (e-mail: abenson@astro.ox.ac.uk)\\
2. NAOJ, 2-21-1 Osawa, Mitaka, Tokyo, Japan (e-mail: naoshi@yso.mtk.nao.ac.jp)\\
3. Technion - Israel Institute of Technology, Technion City, Haifa 32000, Israel (e-mail: adi@physics.technion.ac.il)\\
4. Institute for Computational Cosmology, University of Durham, Durham, DH1 3LE, United Kingdom (e-mail: Cedric.Lacey@durham.ac.uk)\\
}

\maketitle

\begin{abstract}
We have modelled the process of reionization of the intergalactic medium (IGM) by photoionization by galaxies, in order to learn
what galaxy formation in the framework of the CDM model predicts for the epoch of reionization.  We use a sophisticated
semi-analytic model of galaxy formation to track the formation of these galaxies, their influence on the IGM, and the
back-reaction of the state of the IGM on further galaxy formation. Our study represents a much more complete and physically
consistent modelling of reionization than has been conducted in the past. In particular, compared to previous work by ourselves
and others, our new calculations contain significant improvements in the modelling of the effects of reionization of the IGM on
the collapse of baryons into dark matter halos (this is now computed self-consistently from the properties of model galaxies), and
in the model for the cooling and condensation of gas within halos (our new model includes photoheating from a self-consistently
computed ionizing background and also includes cooling due to molecular hydrogen).  We find that reionization can be achieved by
$z\sim 10$--$20$ in a $\Lambda$CDM cosmological model with $\sigma_8\approx 0.9$.  However, a cosmological model with a running
spectral index is only able to achieve reionization before $z\approx 9$, and thus be consistent with an optical depth of $0.1$, if
very extreme assumptions are made about the physics of feedback at high redshifts.  We also consider the specific galaxy formation
model recently discussed by Baugh et al., which includes a top-heavy IMF in starbursts, and find that it is able to reionize the
Universe by $z\approx 12$. The previous results assume that all of the ionizing photons produced by stars in galaxies are able to
escape and ionize the IGM. If this is not the case, then the redshift of reionization could be substantially reduced. We find that
extended periods of partial reionization and double reionizations can occur in models in which the first stars formed via cooling
by \H2 molecules, are very massive, and in which the escape fraction of ionizing photons $\sim $10--30\%.  Such models do not
fully reionize until $z\approx$ 6--7, but predict an electron scattering optical depth as large as $0.15$. Models with lower
$\sigma_8=0.7--0.8$ as suggested by the recent Wilkinson Microwave Anisotropy Probe (WMAP) three year data have reduced redshifts
of reionization, but can be consistent with the lower optical depth also suggested by the WMAP three year data.
\end{abstract}

\section{Introduction}

While gas in the intergalactic medium (IGM) is highly ionized in the
present day Universe, it is known to have been mostly neutral in the
past. Evidence for this comes from observations of the Gunn-Peterson
effect in the spectra of high redshift quasars \cite{becker01,djorg01,fan06}
and from the value of the optical depth to scattering measured from
the cosmic microwave background (CMB; \pcite{spergel03}). Thus, there
was a period in the history of the Universe when it transitioned from
being neutral to ionized---the epoch of reionization.

Measurements of the CMB polarization made during the first year of operation of the Wilkinson Microwave Anisotropy Probe (WMAP)
satellite\footnote{While this paper was being completed, results based upon the first three years of data from WMAP became
available (Spergel et al. 2006). These show a much lower optical depth to reionization. We will comment upon these results in \S\protect\ref{sec:disc}.}
have been used to estimate a value $\tau=0.17 \pm 0.04$ for the electron scattering optical depth to reionization \cite{kogut03},
which, in the simple case that the IGM reionized instantaneously, implies a reionization redshift $z_{\rm reion}=17\pm 5$
\cite{spergel03}.  This reionization redshift is significantly higher than was previously thought on the basis of the observed
Gunn-Peterson effect and on the basis of theoretical modelling (see, for example,
\pcite{haiman97,go97,baugh98,vs99,bruscoli00,chiu00,ciardi00,benson01}).  Although the WMAP measurement could be explained in
scenarios in which reionization happened twice, or happened only partially for some time \cite{cen03,wylo03,nasel04},\footnote{The
physical plausibility of these scenarios has been questioned by \protect\scite{furl05}. We present evidence, in
\S\protect\ref{sec:enhance} that such double or partial reionization may in fact occur under plausible physical conditions.} the
simple interpretation (that the Universe was fully reionized at $z=17$ and remained so ever since) implies the presence of a
significant population of ionizing sources at high redshift. Coupled with the hints that the effective spectral index of the
matter power spectrum is changing with scale (i.e. a ``running spectral index''), this scenario becomes hard to explain in the
standard cold dark matter (CDM) scenario of structure and galaxy formation.\footnote{The running spectral index reduces the
amplitude of fluctuations on small, galactic scales relative to that on large scales. As a result there are many fewer haloes
present at high redshift with mass sufficient to form galaxies.}

This tension has been explored recently by several authors
\cite{hh03,rs03,yoshi03b,ciardi03,onken04}, who generally reach the conclusion
that $z_{\rm reion}=17$ is incompatible with the running spectral
index model, and hard to achieve even in a power-law model. One
possible escape from this apparent problem is to reionize the Universe
in some other way, i.e. without relying on ionizing photons from stars
in high redshift galaxies. One such mechanism which has received
attention recently is a contribution to reionization from particle
decays during the cosmic dark ages \cite{chen04,hansen04,kasuya04,kasuya04b},
another is a contribution from ``mini-QSOs'' (see, for example,
\pcite{rico04}).

In this paper, we undertake detailed calculations of the galaxy
formation process at high redshift to investigate whether galaxies
alone can ionize the Universe sufficiently early to explain the WMAP
measurement of the optical depth. For this purpose we employ the {\sc
galform} semi-analytic model of galaxy formation
\cite{cole00,benson02}. The advantage of our model over previous
attempts to address this question is that it incorporates much more of
the actual physics of galaxy formation (e.g. it follows the cooling of
gas within haloes and incorporates a model of star formation and
feedback), even if this physics is treated in a simplified way.
The model used in this paper is significantly improved over that
employed in our earlier study of reionization in \scite{benson01}. The
effects of reionization on subsequent galaxy formation (due to the
heating of the IGM and the subsequent rise in the Jeans mass, and due
to photoheating of gas in haloes) are now self-consistently computed
within our model. Furthermore, we now include a detailed treatment of
cooling due to molecular hydrogen in our calculations. Thus, we are
able to provide a more reliable answer as to what the CDM model of
structure formation predicts for the epoch of reionization.

The remainder of this paper is arranged as follows. In
\S\ref{sec:model} we outline the model used to follow the reionization
of the Universe. In \S\ref{sec:results} we examine the reionization
histories predicted by our models and assess the importance of the
various physical ingredients of the models.  In \S\ref{sec:2res} we
present results for the reionization epoch and the imprint of
reionization on the cosmic microwave background. Finally, in
\S\ref{sec:disc} we assess the implications of our results. Several
Appendices describe further details of the modelling, and present
additional results.

\section{Modelling}
\label{sec:model}

\subsection{Galaxy Formation Modelling}

We employ the {\sc galform} semi-analytic model of galaxy formation
throughout this work to predict the properties of the galaxy
population as a function of redshift. Most importantly for the current
work is the ability of this model to compute the star formation rates
(and associated luminosities in ionizing photons) needed to predict
the reionization history of the Universe. A full description of the
{\sc galform} model is beyond the scope of this paper. We instead
refer the reader to previous papers which document the model in
more detail \cite{cole00,benson02}. Here, we give a very brief
description of key components of the model and provide somewhat more
detail about two ingredients which are critical to the current work:
the cooling of hot gas in haloes (see \S\ref{sec:GFcool}) and star
formation (see \S\ref{sec:GFSF}). It should be noted that the major
innovation with respect to Paper~I is our improved model of gas
cooling, which  now includes cooling due to \H2\, and which is
affected by feedback from the reionization process. The implications
of this new modelling are discussed in \S\ref{sec:GFcool}.

The basic structure of the {\sc galform} calculation can be summarized
as follows:
\begin{itemize}
\item The hierarchical formation of dark matter haloes is followed
using the extended Press-Schechter theory \cite{bond91,bower91}. This
allows the construction of merger trees which form the backbone of the
galaxy formation calculation.
\item A galaxy can form in each branch of the merger tree if the hot
 gas in the halo is able to cool and condense to the halo centre (see
 \S\ref{sec:GFcool}) where it forms a rotationally supported disk.
\item This condensed gas is assumed to begin to form stars at a rate
governed by simple, phenomenological rules (see \S\ref{sec:GFSF}). As
a consequence of this star formation, supernovae inject energy into
the galaxy causing some of the condensed gas to be re-ejected from the
galaxy. This feedback loop is crucial to producing a population of
galaxies which matches the observed low-redshift Universe
\cite{benson03}. Star formation also produces metals, enriching the
gas in the galaxy. When this enriched gas is ejected from the galaxy
by supernova explosions, the hot atmosphere of gas pervading the halo
also becomes enriched in metals.
\item When two branches of the merger tree join, the most massive
galaxy from the branches becomes the central galaxy of the new
branch. Any other galaxies become satellite galaxies, free to orbit
within the gravitational potential of the new dark matter
halo. Satellite galaxies are no longer able to gain gas from the
surrounding hot atmosphere.
\item Satellite galaxies experience dynamical friction as they orbit,
which eventually forces them to merge into the galaxy residing at the
halo centre. These mergers are assumed to transform stellar disks into
stellar spheroids if the masses of the two galaxies are
comparable. Thus, in this model, mergers are the drivers of
morphological evolution. Mergers of this type may also trigger bursts
of star formation (see \S\ref{sec:GFSF}).
\end{itemize}

\subsection{Cooling of Gas in Haloes}
\label{sec:GFcool}

When a dark matter halo reaches virial equilibrium, it is assumed to
have accreted into itself some mass of gas from the surrounding
IGM. For high mass haloes, the mass of gas accreted is equal to
$\Omega_{\rm b}/(\Omega_0-\Omega_{\rm b})$ times the dark matter mass
of the halo (where $\Omega_0$ and $\Omega_{\rm b}$ are the mean
densities of total and baryonic mass in the Universe in units of the
critical density). For lower mass haloes, the mass of gas accreted is
reduced, due to the finite gas pressure in the IGM, in accordance with
the filtering mass formalism of \scite{gnedin00}. The filtering mass,
$M_{\rm F}$, is discussed in detail in Appendix~\ref{app:filt}. The
mass of gas accreted by a halo of mass $M_{\rm tot}$ is then given by
\begin{equation}
M_{\rm gas} = {(\Omega_{\rm b}/\Omega_0)M_{\rm tot} \over [1 + (2^{1/3}-1)M_{\rm F}(z)/M_{\rm tot}]^3}.
\end{equation}

The gas is assumed to be heated by shocks as it accretes into the
halo, reaching a uniform temperature equal to the halo virial
temperature. The gas is then assumed to be distributed through the halo
with a density profile
\begin{equation}
\rho_{\rm gas}(r) \propto {1 \over r^2 + c^2},
\end{equation}
where $c$ is a scale radius which we set equal to $r_{\rm s}/3$, where
$r_{\rm s}$ is the scale radius of the dark matter density profile
(assumed to have the NFW form).

To compute the rate of cooling of the gas, we construct a table of net
radiative cooling rates as a function of temperature, density,
metallicity and redshift. (The redshift dependence arises through the
inclusion of photoheating from a time-varying ionizing background
which is computed self-consistently from the emission from the model
galaxies.) This net cooling/heating rate includes contributions from
atomic processes (e.g. recombinations, collisional excitations etc.),
Compton cooling \cite{peebles68}, cooling due to molecular hydrogen
(see \S\ref{sec:H2cool}) and photoheating by the ionizing background
(see \S\ref{sec:IGMevol}). Given these cooling rates and the assumed
gas density profile and temperature for each halo, we then calculate,
at each time step of our calculation, the cooling radius, defined as
the radius in the halo at which the local cooling time equals the time
since the halo formed. (A halo is defined to form anew whenever its
mass is doubled by mergers.) Any gas within this cooling radius is
assumed to accrete onto the galaxy forming at the centre of the halo,
provided that the time for the gas to free-fall to the halo centre is
less than the age of the halo. Occasionally, for example when the
ionizing background is rapidly rising, the cooling radius at a given
timestep may be smaller than that at the previous timestep. In such
cases no gas is allowed to cool until the cooling radius once again
begins to increase.

The inclusion of cooling due to \H2\ is crucial to the present work
as, at high redshifts, this is the only cooling channel available to
gas in the majority of halos. Without \H2\ cooling included, we risk
seriously underestimating the epoch of reionization. The IGM pressure
and the ionizing background act as sources of negative feedback on
star formation (i.e. star formation leads to a rise in the IGM
pressure and ionizing background, which in turn suppress later star
formation). By including these effects self-consistently in our model
we are therefore able to check whether this negative feedback is
sufficient to significantly quench star formation. It should be noted
that if we neglect \H2\ cooling and set the filtering mass and
ionizing background to zero our revised cooling model produces results
identical to those of Paper~I (i.e. it simply follows the cooling
prescription of Cole et al. (2000) with the addition of Compton
cooling). We will examine the effects of including \H2\ cooling in our
calculations in \S\ref{sec:h2help}.

The effects of including the filtering mass and ionizing background in
our cooling model have been explored in detail by
\scite{benson02}. \scite{benson02} find that it is the filtering mass
which has the greatest effect on galaxy formation at low redshifts,
resulting in a suppression of star formation after the epochs of \HI\
and \HeI\ reionization. For the \scite{cole00} model, \scite{benson02}
found this suppression to be quite small. However, for a model with
weak feedback from supernovae the suppression effect of the filtering
mass was significant (reducing the global star formation rate by a
factor of two). This occurs because, in models with weak SNe feedback,
star formation can otherwise progress efficiently in low mass halos,
and therefore such haloes make a significant contribution to the
global star formation rate. Once the filtering mass rises star
formation is shut off in these low mass halos, thereby greatly
suppressing the global star formation rate. The filtering mass also
acts to reduce the number of faint galaxies which can form, as
discussed in detail by \scite{benson02b}. Note that the filtering mass
begins to rise as the process of reionization begins, but does not
reach its maximum value until significantly after the epoch of
reionization. As such, we may expect it to have only minimal impact on
the epoch of reionization, a point which we will explore in
\S\ref{sec:feedmechs}.

\subsection{Star Formation and Supernova Feedback}
\label{sec:GFSF}

Once gas has cooled inside a dark matter halo it is assumed to
collapse, conserving its angular momentum, until it reaches a radius
where it becomes rotationally supported against further collapse. At
this point, the gas is assumed to form an exponential disk. The gas in
this disk then begins to form stars at a rate $\psi$ given by
\begin{equation}
\psi = M_{\rm cold}/\tau_\star,
\end{equation}
where $M_{\rm cold}$ is the total mass of gas available in the galaxy
and $\tau_\star$ is a characteristic timescale for star
formation. Following \scite{cole00} we adopt the parameterization
\begin{equation}
\tau_\star =\epsilon_\star^{-1}\tau_{\rm disk} (V_{\rm disc}/200\hbox{ km/s})^{\alpha_\star},
\end{equation}
where $\epsilon_\star$ and $\alpha_\star$ are dimensionless
parameters, $V_{\rm disc}$ is the circular velocity of the disk at its
half-mass radius and $\tau_{\rm disc}$ is the dynamical time of the
disk at that radius.

We assume that supernova explosions in the galaxy cause gas to be
ejected from the galaxy at a rate $\dot{M}_{\rm eject}=\beta\psi$,
where
\begin{equation}
\beta = (V_{\rm disc}/V_{\rm hot})^{-\alpha_{\rm hot}},
\label{eq:SNfeedback}
\end{equation}
where $\alpha_{\rm hot}$ is a dimensionless parameter and $V_{\rm
hot}$ is a parameter with dimensions of speed.

Star formation can also occur in a burst mode as the result of a
violent merger between two galaxies. When such a merger occurs, all of
the gas in the two merging galaxies undergoes star formation on a very
short timescale. We use the same rules as for quiescent star formation
in disks, but using the dynamical properties of the newly formed
spheroid instead.

We compute the chemical enrichment of each galaxy using the
instantaneous recycling approximation, with yield of heavy elements
$p$ and recycled fraction $R$ (as defined by \scite{cole00}---we use
the same values for these parameters as did \scite{cole00} unless
otherwise noted), including the exchange of metals between the galaxy
and the halo gas. Once the star formation history and associated
chemical enrichment history of a galaxy have been determined, we
compute the SED of the galaxy at each timestep using a stellar
population synthesis model based on the Padova stellar evolution
tracks \cite{granato00}.  This allows us to compute the ionizing
photon luminosity for each galaxy.

\subsection{Evolution of the IGM and Ionizing Background}
\label{sec:IGMevol}

We model the process of reionization and its feedback on the IGM and
galaxy formation using the methods of Benson et al. (2001; hereafter,
Paper~I) and \scite{benson02}. Briefly, we compute the volume averaged
emissivity in ionizing photons at each redshift by summing the
ionizing luminosity of all galaxies in our calculation. This
emissivity is used to evolve a distribution of gas elements drawn from
an appropriate PDF \cite{benson01}. We track the ionization (H and He)
and thermal state of these gas elements as the Universe evolves. From
the resulting thermal history of the IGM we are able to compute the
filtering mass describing the effects of the IGM pressure on the
collapse of baryons into haloes. This allows us to estimate the mass
of baryons accreted into each dark matter halo forming at any
redshift. We also have, at each redshift, the spectrum of the ionizing
background, which allows us to compute the rate of photoionization
heating experienced by gas in dark matter haloes. These two processes
(reduced accretion due to the IGM pressure and photoheating of halo
gas), which act to suppress galaxy formation, are incorporated into a
second iteration of our galaxy formation calculation. This iterative
process is repeated until convergence is reached (i.e. the input
ionizing background and filtering mass are consistent with those
resulting from the galaxy formation calculation) as in
\scite{benson01}. Schematically, the iterative procedure works as
follows:
\begin{enumerate}
\item Initially assume that there is no ionizing background (and
associated photoheating) and that the filtering mass is always zero.
\item Run the {\sc galform} galaxy formation calculation.
\item Compute the evolution of the ionizing background based on the
{\sc galform} calculation.
\item Compute associated photoheating rates for gas of varying
densities, temperatures and metallicities.
\item Compute the evolution of the filtering mass based on the {\sc
galform} calculation.
\item Using these photoheating rates and filtering mass, go back to
step (ii) and repeat.
\end{enumerate}
Iteration stops once the star formation history produced in {\sc
galform} calculations on successive iterations is sufficiently well
converged.

\scite{benson01} and \scite{benson02} considered cooling due to both
atomic processes and Compton cooling due to the scattering of cosmic
microwave background photons from hot electrons. Here we employ an
improved cooling model which is described in detail in
Appendix~\ref{sec:newcool}. To explore the high redshifts and low mass
scales of interest here we must also consider the effects of cooling
due to molecular hydrogen (see \S\ref{sec:H2cool}). We must also
reconsider Compton cooling in the non-equilibrium regime (i.e. before
free electrons in the gas have had sufficient time to recombine to
reach their abundance in ionization equilibrium at the density of the
halo gas) which occurs in some of the lower mass haloes that we
consider (see \S\ref{sec:Compt}).

\subsection{Escape Fractions}

A very important parameter in our models is the fraction, $f_{\rm
esc}$, of ionizing photons produced in a galaxy which escape into the
IGM. For most of our models (and unless stated otherwise), we assume
$f_{\rm esc}=1$. The only exceptions are the models H$_2$WeakFB and
H$_2$WeakFB-VMS (to be defined in \S\ref{sec:GFpar}), for which we also present some results for lower
escape fractions (for the purpose of producing models with ``double
reionization''---see \S\ref{sec:enhance}). The assumption that $f_{\rm
esc}=1$ is obviously an extreme one (although such an assumption has
been used in other works studying reionization, e.g. \pcite{hh03}),
since it ignores the absorption of ionizing photons by gas and dust in
the galaxy, but it should result in the earliest possible reionization
if other parameters are held fixed. We note, however, that
observational evidence (and simple physical models for $f_{\rm esc}$)
imply a much lower escape fraction. Therefore, in
Appendix~\ref{app:fesc} we present a simple, physically motivated
calculation of $f_{\rm esc}$ which is in reasonable agreement with
observational data, and show the resulting reionization histories.

\subsection{Calculation of Filling Factor}

To compute the filling factor of \HII\ we follow the growth of ionized
regions around each halo in the galaxy formation model using the
methodology of Paper~I.  Specifically, we compute the total emission
rate of ionizing photons from stars in each dark matter halo, $S(t)$,
and compute the number of hydrogen atoms which will be ionized by this
radiation by solving
\begin{equation}
{{\rm d}N_{\rm H} \over {\rm d}t} = S(t) - \alpha_{\rm H}^{(2)} a^{-3} f_{\rm clump}n_{\rm H} N_{\rm H},
\end{equation}
where $N_{\rm H}$ is the total number of hydrogen atoms ionized,
$\alpha_{\rm H}^{(2)}$ is the case B recombination coefficient for
hydrogen (for a temperature of $10^4$K), $a$ is the cosmic expansion
factor, $f_{\rm clump}$ is the clumping factor for ionized gas, and
$n_{\rm H}$ is the mean number density of hydrogen atoms in the
Universe. We compute the clumping factor following the method
described in Paper~I for $f_{\rm clump}^{\rm (haloes)}$, but with the
cosmological parameters appropriate to the current work. In this
model, the main contribution to the clumping factor comes from gas
residing in haloes with virial temperatures around $10^4$K, which have
deep enough potential wells to retain gas, but which are cool enough
to be not completely collisionally ionized.  The specific values for
the clumping factors used in this work will be discussed in
\S\ref{sec:cospar}.

Note that we do not assume any particular geometry for the ionized
region around each source. Ionized regions from two or more sources
may overlap (particularly just before the epoch of reionization or if
the sources are strongly clustered). This does not affect our results,
as we are simply counting the total number of hydrogen atoms which
have been ionized.

The total number of hydrogen atoms, $N_{\rm H,tot}$, which will have
been ionized in some representative volume of the Universe, $V$, is
then found by simply summing over the contributions from galaxies in
haloes of all masses
\begin{equation}
N_{\rm H,tot} = \sum_i N_{\rm H,i} n_i V,
\end{equation}
where $N_{\rm H,i}$ is the number of hydrogen atoms which have been
ionized by galaxies in halo $i$, of which there are $n_i$ such haloes
per unit volume. The sum is taken over all haloes. The mean ionized
fraction in volume $V$ is found by simply dividing by the total number
of hydrogen atoms in the volume, $n_{\rm H}V$ where $n_{\rm H}$ is the
mean density of hydrogen atoms in the Universe. If we assume a uniform
IGM, we can identify this mean ionized fraction with the filling
factor of ionized regions,
\begin{equation}
F_{\rm fill} =  {1 \over n_{\rm H}} \sum_i N_{\rm H,i} n_i.
\end{equation}
We consider the Universe to have been fully reionized once $F_{\rm
fill}=1$. Note that with these definitions $F_{\rm fill}$ can exceed
unity. In such cases, the Universe is fully ionized and ionizing
photons will begin to build up an ionizing background rather than be
absorbed by neutral hydrogen atoms.

The IGM evolution model of \scite{benson02} gives an alternative means
of following the reionization of the Universe. In that model, the mean
ionization state of the IGM is determined by computing the ionization
and recombination rates in the presence of an evolving ionizing
background, allowing for a distribution of IGM densities. While these
two methods for following reionization are quite different, we find
that they predict very similar redshifts of reionization, with the
method of \scite{benson02} typically predicting $z_{\rm reion}$ to
occur roughly $\Delta z\approx 1$ earlier. Throughout this work we
will show results based on the approach described in this section.

\subsection{Cosmological and Galaxy Formation Models}

We will explore three different cosmological models and several
different models of galaxy formation in this work. The details of
these models are outlined below.

\subsubsection{Cosmological Parameters}
\label{sec:cospar}

For the cosmological parameters we adopt the values determined from
the first-year WMAP data set \cite{spergel03}. In particular, we
consider cosmological models with the best fit parameters for a
power-law $\Lambda$CDM model with constraints from WMAP alone (Table~7
of \pcite{spergel03}) and a running spectral index $\Lambda$CDM model
with constraints from WMAP, the 2dF Galaxy Redshift Survey and
Lyman-$\alpha$ forest observations (Table~8 of \pcite{spergel03}). We
refer to these cosmological models as WMAP PL and WMAP RSI
respectively, and their parameters are given in
Table~\ref{tb:cosmos}. We also compute reionization histories in the
cosmological model corresponding to the specific galaxy formation
model of \scite{baugh05}, discussed further below.

We include these three cosmologies because they have complementary
strengths and weaknesses.  The cosmologies based on WMAP results give
good fits to the available CMB data (by definition). However, we do
not have a set of galaxy formation model parameters in these
cosmologies which result in galaxy populations with properties which
match local observations. The \scite{baugh05} model, on the other
hand, was not designed to match the CMB data (although its
cosmological parameters are not too far from those measured by WMAP),
but does reproduce the properties of galaxies in the local and
high-redshift Universe.

\begin{figure}
\psfig{file=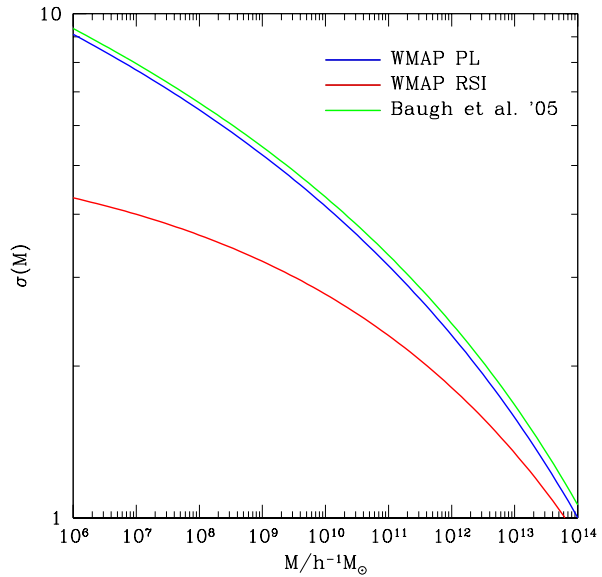,width=80mm}
\caption{The fractional root mean square fluctuation of the density averaged in spheres, $\sigma(M)$, as a function of mass $M$,
  for each of our cosmological models: WMAP RSI (red line), WMAP PL (blue line) and Baugh et al. (2005; green line).}
\label{fig:sigmaM}
\end{figure}

We show in Figure~\ref{fig:sigmaM}, for all three cosmological models,
the fractional fluctuation of the density measured in spheres,
$\sigma(M)$, for the linear density field extrapolated to $z=0$. The
quantity $\sigma(M)$ determines the abundances of dark matter haloes
at different redshifts. It is clearly seen that the WMAP RSI model has
much smaller fluctuations on small scales.

\begin{table*}
\caption{Cosmological parameter sets used in this work. Values are taken from (\protect\pcite{spergel03}; WMAP) and
(\protect\pcite{boom03}; Boomerang). $n_{\rm s}$ is the scalar spectral index at $k=0.05$Mpc$^{-1}$, while ${\rm d}n_{\rm s}/{\rm
d}\ln k$ is the running of the spectral index as defined by eqn.~(5) of \protect\scite{spergel03}.}
\label{tb:cosmos}
\begin{tabular}{lccccccr@.l}
Name & $\Omega_0$ & $\Lambda_0$ & $\Omega_{\rm b}$ & $H_0$/km s$^{-1}$ Mpc$^{-1}$ & $\sigma_8$ & $n_{\rm s}$ (at $k=0.05$Mpc$^{-1}$) & \multicolumn{2}{c}{${\rm d}n_{\rm s}/{\rm d}\ln k$} \\
\hline
WMAP PL          & 0.270 & 0.730 & 0.0463 & 72.0 & 0.90 & 0.99 &  0 & 000 \\
WMAP RSI         & 0.268 & 0.732 & 0.0444 & 71.0 & 0.83 & 0.93 & -0 & 031 \\
Baugh et al. '05 & 0.300 & 0.700 & 0.0400 & 70.0 & 0.93 & 1.00 & 0 & 000\\
\hline
Boomerang '03    & 0.300 & 0.700 & 0.0469 & 69.6 & 0.85 & 0.95 & 0 & 000\\
\hline
\end{tabular}
\end{table*}

\begin{figure}
\psfig{file=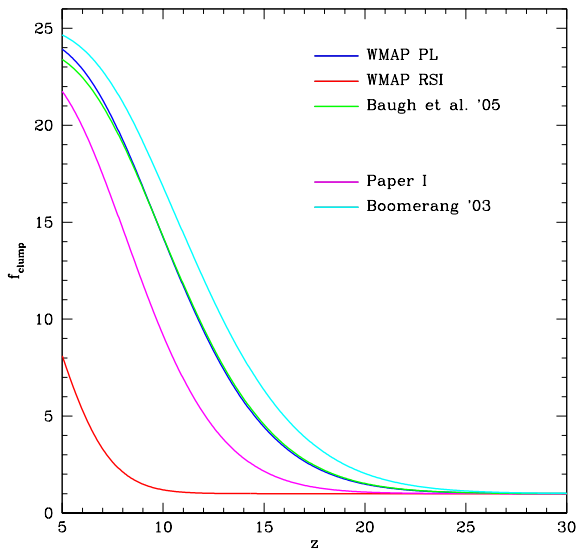,width=80mm}
\caption{The clumping factor, $f_{\rm clump}$, as a function of
redshift for the three cosmological models considered in this
work. Also shown, for reference, are the clumping factors
corresponding to the cosmological parameters of Paper~I (see
Appendix~\protect\ref{app:p1comp}) and the ``Boomerang '03'' cosmology
discussed in \S\protect\ref{sec:disc}.}
\label{fig:clump}
\end{figure}

The IGM clumping factor which we use depends only on the cosmological
model, and is shown as a function of redshift in
Fig.~\ref{fig:clump}. For the WMAP PL cosmology it is very similar to
that found in Paper~I ($f_{\rm clump}\approx 18$ at $z=10$, dropping
to approximately $2.5$ at $z=20$ for example), while for the WMAP RSI
cosmology it is much lower ($f_{\rm clump}\approx 1.5$ at $z=10$,
dropping to close to $1$ at $z=20$) due to the reduction of
small-scale fluctuations in the latter cosmology. Inclusion of the
clumping factor makes almost no difference to our results for the WMAP
RSI cosmology, as the clumping factor is still small by the time
reionization occurs. However, in the WMAP PL cosmology, clumping can
delay reionization by $\Delta z=2$--4 (the earlier reionization
occurs, the less effect clumping has). In the Baugh et al. '05
cosmology, the clumping factor is comparable to that of the WMAP PL
cosmology.

\subsubsection{Galaxy Formation Parameters}
\label{sec:GFpar}

To model the formation of galaxies in the Universe we employ the
semi-analytic model of galaxy formation {\sc galform}. A full
description of this model can be found in \scite{cole00} and
\scite{benson02}. For our ``Standard model'' in the WMAP RSI and WMAP
PL cosmologies, we adopt parameters which are very similar to those of
\scite{cole00}. (We ignore the cosmological parameters specified by
\scite{cole00} and instead use those described in \S\ref{sec:cospar}.)
However, we use the improved merging calculation of \scite{benson02},
which results in somewhat fewer mergers on average than the simpler
model used by \scite{cole00}. Due to the change in cosmological
parameters and in the merger rates, we adjust the values of a few of
the parameters in the galaxy formation model, increasing the star
formation efficiency $\epsilon_\star$ from $0.005$ to $0.0067$ and
reducing the mass-to-light ratio parameter $\Upsilon$ from $1.38$ to
$1.00$.

We regard the resulting model as being a plausible model for galaxy
formation in the high-redshift Universe, since it has been shown to
reproduce well the properties of galaxies in the low-redshift
universe, and we simply assume that the same physical processes apply
to galaxy formation at high redshift as at low redshift. Our model is
therefore motivated by what we currently know about the local
Universe---the extrapolation to high redshifts is a simple assumption
as our current understanding of the physics that governs the high
redshift Universe does not allow us to devise more accurate physical
prescriptions for early times.  Note, however, that the \scite{cole00}
model assumed $\Omega_{\rm b}=0.02$. The higher baryon densities in
the cosmologies considered here will result in this model producing
too many galaxies of all luminosities at $z=0$. We do not attempt to
correct for this deficiency for two reasons: 1) we are interested here
in determining how early reionization can occur---overproduction of
galaxies will push reionization to higher redshifts and so give us a
conservative answer to the question of whether a particular model can
reionize sufficiently early to match the WMAP results; and 2) while
the \scite{cole00} model works well at $z=0$ it does not match
observational data for high-redshift galaxies and, in any case, we are
here extrapolating our model to $z>20$ where observational constraints
on galaxy formation are practically non-existent. It is for these
reasons that we consider our model to be plausible but not definitive.

While there are numerous variations of the model parameters which it
would be interesting to explore, we limit ourselves here to examining
a handful which might be expected to have a particularly large impact
on the epoch of reionization. These various models are summarized in
Table~\ref{tb:GFmods}.

\begin{figure}
\psfig{file=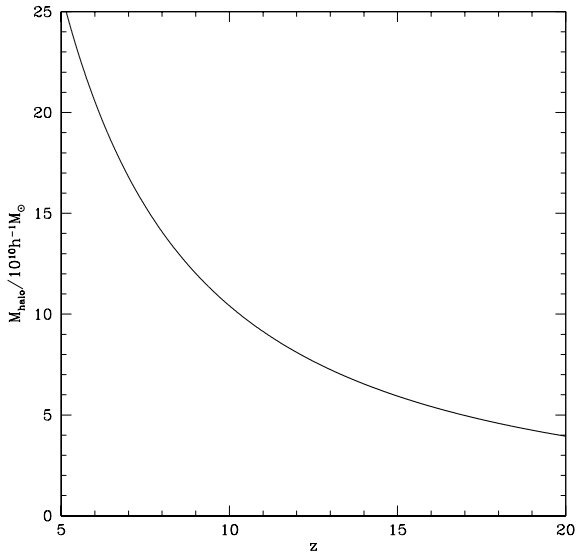,width=80mm}
\caption{The mass of a halo with virial velocity corresponding to the
feedback parameter, $V_{\rm hot}=200$km/s, is shown as a function of
redshift for the WMAP PL cosmology (results for the WMAP RSI and Baugh
et al. '05 cosmologies are practically identical).}
\label{fig:Mvhot}
\end{figure}

\begin{itemize}
\item {\bf Standard:} For the WMAP RSI and WMAP PL cosmologies, this
model is very similar to that described in \scite{cole00}---the
differences from \scite{cole00} are the parameter changes described in
\S\ref{sec:GFpar} and the inclusion of Compton cooling, which is
necessary for studying the high redshifts of interest here. It assumes
that all stars form with a \scite{kennicutt83} initial mass function
(IMF), with slope $x=0.4$ for $0.1 < m < 1 M_{\odot}$ and $x=1.5$ for
$1 < m < 120 M_{\odot}$. (A Salpeter IMF has $x=1.35$.) For this IMF,
the total number of ionizing photons (with $h\nu >$13.6 eV) produced
per $M_{\odot}$ of stars formed is $N_{\rm Lyc}/M_*= 1.0\times
10^{53}$, for a metallicity $Z= 0.1 Z_{\odot}$. Note that, for this
model (and all others unless stated otherwise) we adopt a value of
$V_{\rm hot}=200$km/s for the feedback parameter. This value, chosen
to reproduce the low abundance of faint galaxies in the $z=0$
Universe, results in quite strong feedback. For the \scite{baugh05}
cosmology, the Standard model corresponds to the galaxy formation
parameters given by \scite{baugh05}. Note that this includes a higher
value for $V_{\rm hot}=300$km/s. Figure~\ref{fig:Mvhot} shows the halo
mass with virial velocity equal to $V_{\rm hot}=200$km/s as a function
of redshift. At high redshifts, the vast majority of haloes able to
form will be strongly affected by feedback for $V_{\rm hot}\ge
200$km/s.

\item {\bf MetFree:} In this model, we compute stellar luminosities
and spectra assuming that all stars are be metal free (while retaining
the chemical enrichment of the interstellar and intergalactic gas, as
this affects gas cooling rates), thereby enhancing the emission of
ionizing photons from galaxies. In this case, we have $N_{\rm
Lyc}/M_*= 2.0\times 10^{53}$.

\item {\bf TopHeavy:} We assume an extreme IMF which has the Salpeter
slope $x=1.35$, but only covers the mass range $10 < m < 100
M_{\odot}$, i.e. it is truncated for masses below $10M_\odot$. As
massive stars produce the majority of ionizing photons, this model is
expected to produce earlier reionization. For this IMF, we have
$N_{\rm Lyc}/M_*= 7.1\times 10^{53}$.

\item {\bf WeakFB:} This model has very weak feedback from supernova
explosions, such that even the lowest mass galaxies which form in the
model will not experience significant mass loss due to feedback. In
{\sc galform}, the strength of feedback is controlled by two
parameters $V_{\rm hot}$ and $\alpha_{\rm hot}$, as defined in
eqn.(\ref{eq:SNfeedback}). We fix $\alpha_{\rm hot}=2$ throughout this
work, and use $V_{\rm hot}=200$(300)km/s for the WMAP (Baugh et
al. '05) cosmologies except for models with weak feedback, for which
we assume $V_{\rm hot}=1$km/s (see Table~\ref{tb:GFmods}). The lower
value of $V_{\rm hot}=1$km/s corresponds to an efficiency of supernova
energy injection and of mass ejection which is lower by a factor
$4\times 10^4$ than for the standard case $V_{\rm hot}=200$km/s, so it
is fairly extreme. This weak feedback will greatly enhance the
efficiency of star formation in low-mass haloes, and hence will also
increase the production of ionizing photons at high redshift.

\item {\bf H$_2$:} This model is identical to the Standard model
except that it incorporates cooling due to molecular hydrogen---this
should allow galaxies to form earlier, thereby shifting the epoch of
reionization to higher redshifts.

\item {\bf H$_2$WeakFB:} This model combines a weak feedback with
molecular hydrogen cooling.

\item {\bf H$_2$WeakFB-VMS:} This model investigates the possibility
that stars forming via \H2\ cooling (``Population~III'') may be very
massive stars (VMS), with $m \sim 100-1000 M_{\odot}$
\cite{abel00,abel02,bromm02}. If all of the stars are very massive,
then the total ionizing luminosity will be greatly enhanced relative
to stars formed with a normal IMF. Therefore, in this model, we
increase the UV luminosity of any stars forming in a halo with virial
temperature below $10^4$K by a factor of 20 relative to the Kennicutt
IMF at all metallicities.\footnote{Our choice of a UV enhancement
factor of 20 is motivated by the results of \protect\scite{bromm01},
who find an enhancement factor of 10--20. (Their enhancement is
relative to a Salpeter IMF. We use the \scite{kennicutt83} IMF here,
but the total ionizing photon production per unit mass of star
formation for this IMF differs from that of the Salpeter IMF by
approximately 10\% for sub-Solar metallicities.) We have taken the
value at the upper end of their range to maximize the effects. We also
note that theoretical considerations suggest that very massive stars
should form when the metallicity of the ISM is below a critical value,
and not merely when gas cooling occurs through the \H2\ channel. If we
model VMS formation in this way (i.e. we enhance the UV emissivity
only for stars with a metallicity below, for example, $10^{-4}Z_\odot$
\protect\cite{schneider02,bromm03} we find little enhancement in the
net UV emission. This occurs because the duration of the sub-critical
metallicity phase of star formation is very short. However, our
current model does not account for the different heavy element yield
of these Population III stars---a low yield could dramatically extend
the duration of the sub-critical metallicity phase of star
formation. We defer a detailed examination of this possibility to a
future work.}  We therefore have $N_{\rm Lyc}/M_*= 2.0\times
10^{54}$.

\end{itemize}

\scite{baugh05} recently presented a {\sc galform} model employing a
significantly different set of galaxy formation parameters from those
used by \scite{cole00}. We will use the \scite{baugh05} parameters to
describe our ``Standard model'' in the \scite{baugh05} cosmology. The
cosmological parameters of this model are listed in
Table~\ref{tb:cosmos}.  The \scite{baugh05} model was constrained to
produce a good match to observations of sub-mm galaxies and
Lyman-break galaxies at high redshifts, as well as the properties of
galaxies at low redshifts. In order to match the numbers of sub-mm
galaxies, \ncite{baugh05} assumed that stars which form in
merger-induced bursts have a top-heavy IMF (with slope $x=0$ covering
the mass range $0.1 < m < 120 M_{\odot}$), while stars which form
quiescently in disks have a \scite{kennicutt83} IMF (as in our
Standard model). In the \ncite{baugh05} model, the fraction of stars
formed in the burst mode increases with redshift, with the burst mode
dominating the total star formation density at $z\gsim 3$ (see Fig.~1
in \ncite{baugh05}). In subsequent papers, Le Delliou et al. (2005a,b)
have investigated the properties of Ly$\alpha$-emitting galaxies at
high redshift, and Nagashima et al. (2005a,b) have investigated the
chemical enrichment of elliptical galaxies and intracluster gas in the
same model. Given the success of the \ncite{baugh05} model in
reproducing observations of various types of star-forming galaxies at
high redshift, as well as observations of chemical abundances at low
redshift, it is of interest to see what it predicts for the
reionization of the IGM. We have therefore used the galaxy formation
and cosmological parameters of the \scite{baugh05} model to compute
the reionization history of the Universe\footnote{Note that
\protect\scite{baugh05} included a simple model of photoionization
feedback in their calculation, in which they suppressed the cooling of
gas in dark matter haloes with virial velocities below 60km/s for
$z<6$. As we are here modelling reionization and photoionization
feedback in much greater detail, we remove this simple suppression
from our implementation of the \protect\scite{baugh05} model.}.

For stars formed in bursts in the \ncite{baugh05} model, the number of ionizing photons produced per $M_{\odot}$ of stars formed
$N_{\rm Lyc}/M_*= 1.7\times 10^{54}$ for a metallicity $Z= 0.1 Z_{\odot}$, i.e. nearly 20 times larger than for a Kennicutt
IMF. The only model variants we consider for the \ncite{baugh05} cosmology at Standard, H$_2$ and H$_2$WeakFB\footnote{Model
H$_2$WeakFB-VMS is not considered, as the Baugh et al. (2005) model already includes very massive stars which are assumed to form
in bursts.}.

\begin{table*}
\caption{Names and details of the galaxy formation models used in this
work. Column~1 describes the model while column~2 gives the name by
which we refer to the model throughout this work. Column~3 lists the
changes in model parameters made relative to the standard model.}
\label{tb:GFmods}
\begin{center}
\begin{tabular}{lll}
\hline
{\bf Model} & {\bf Name} & {\bf Variation from standard model} \\
\hline
Standard model & Standard & --- \\
Metal free stellar spectra & MetFree & $Z_{\star}=0$ \\
Top-heavy IMF & TopHeavy & IMF Salpeter above $10M_\odot$, zero below; $R=0.91$; $p=0.14$ \\
Weak feedback & WeakFB & $V_{\rm hot}=1$km/s \\
Compton + H$_2$ cooling and dynamical heating & H$_2$ & H$_2$ cooling on \\
Compton + H$_2$ cooling + weak feedback & H$_2$WeakFB & H$_2$ cooling on; $V_{\rm hot}=1$km/s \\
Compton + H$_2$ cooling + weak feedback & H$_2$WeakFB-VMS & H$_2$ cooling on; $V_{\rm hot}=1$km/s; UV flux $\times 20$ for stars formed \\
\hspace{6mm} + VMS for $T_{\rm vir}<10^4$K & & \hspace{6mm} via \H2\ cooling  \\
\hline
\end{tabular}
\end{center}
\end{table*}

\section{Results}
\label{sec:results}

Below, we present results from our calculations of reionization. We
begin with a summary of the key results, before concentrating on the
importance of H$_2$ cooling and photoionization feedback in setting
the epoch of reionization. A comparison of the predicted epochs of
reionization to the observational determinations and an exploration of
the signal imprinted in the cosmic microwave background (CMB) by the
patchy reionization process are given in \S\ref{sec:2res}.

Some further results, including comparisons to Paper~I and tests of
the convergence of our models are given in
Appendix~\ref{app:tests}. Appendix~\ref{app:fesc} briefly examines the
effects on the reionization history of using physically motivated
calculations of the escape fraction of ionizing photons.

Figure~\ref{fig:summary} shows key results from all of our
calculations. Each row corresponds to a different model (as indicated
along the right-hand edge of the figure) and within each panel
different colours correspond to different cosmologies: WMAP RSI (red),
WMAP PL (blue) and Baugh et al. '05 (green). The three columns
respectively show the average star formation rate per unit volume, the
ionizing luminosity density and the \HII\ filling factor (as indicated
by labels at the top of the figure) all as functions of redshift. It
is clear from this figure that the suppression of small mass haloes in
the WMAP RSI cosmology leads to a severe reduction in the star
formation rate, ionizing luminosity and \HII\ filling factor compared
to our other cosmological models. This effect is larger than that of
any of the other variations we consider in this work. As such, it may
be possible to use a measured reionization epoch to place strong
constraints on the cosmological model. It is also clear that
reionization redshifts high enough to explain the WMAP measurement
of the optical depth \emph{can} be attained in the WMAP PL and Baugh
et al. '05 cosmologies, albeit only in rather extreme models. These
points will be discussed further in the remainder of this section and
in \S\ref{sec:2res}.

\begin{figure*}
\psfig{file=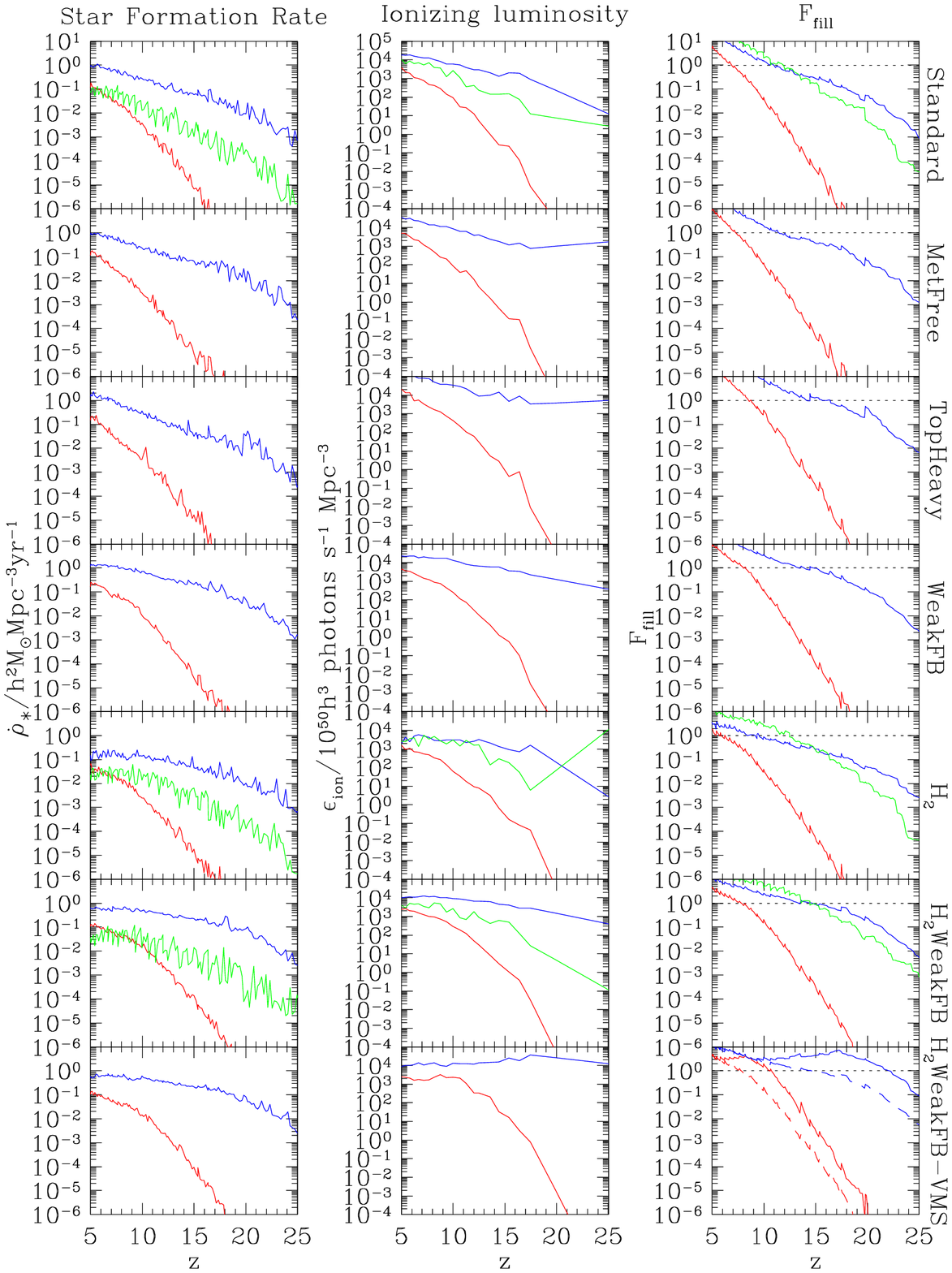,width=190mm}
\vspace{-20mm}
\caption{}
\label{fig:summary}
\end{figure*}

\begin{figure*}
\addtocounter{figure}{-1}
\caption{\emph{(cont.)} The evolution of key physical parameters in
our models. The three columns show mean star formation rate density,
ionizing luminosity and \HII\ filling factor as indicated at the top
of each panel. Each row corresponds to a different galaxy formation
model, as indicated on the right-hand side of the figure. The three
lines in each panel correspond to three different cosmological models:
WMAP RSI (red), WMAP PL (blue) and Baugh et al. (2005; green---not
shown for models MetFree, TopHeavy, WeakFB and H$_2$WeakFB-VMS). In the \HII\ filling
factor columns, the horizontal dashed line indicates the point of
complete reionization, while dashed coloured lines for the
H$_2$WeakFB-VMS model show the contribution of Pop~I and II stars to
the total filling factor.}
\end{figure*}

\subsection{Does H{\boldmath $_2$} Cooling Help?}
\label{sec:h2help}

\begin{figure}
\psfig{file=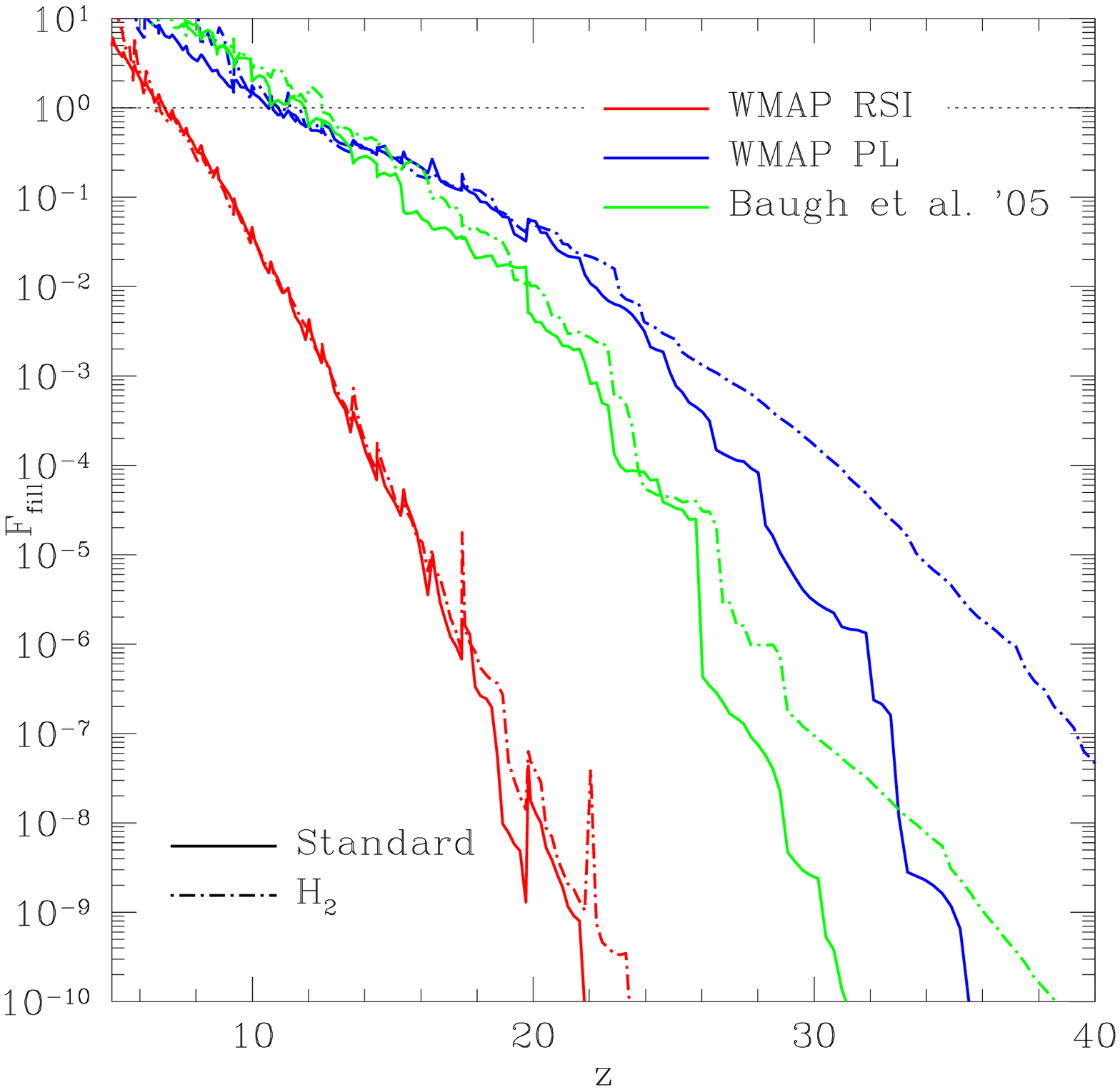,width=80mm}
\caption{The \HII\ filling factor as a function of redshift for models
with (dot-short dashed lines) and without (solid lines) \H2\
cooling. Colours indicate the cosmological model as indicated in the
figure. Results are shown to higher redshifts and lower filling
factors than in Fig.~\protect\ref{fig:summary}.}
\label{fig:FfillH2}
\end{figure}

We can examine whether the inclusion of \H2\ cooling channels in our
calculations actually aids in pushing the epoch of reionization to
higher redshifts. Figure~\ref{fig:FfillH2} shows the \HII\ filling
factor as a function of redshift for models with and without cooling
due to \H2. It can clearly be seen that the addition of \H2\ cooling
channels has almost no effect on the epoch of reionization. At earlier
redshifts, \H2\ cooling \emph{does} result in an enhancement in the
filling factor (e.g. by a factor of $10^4$ at $z=35$ in the WMAP PL
cosmology). However, by the time $F_{\rm fill}$ is approaching unity,
the ionizing emissivity is dominated by emission from haloes which are
able to cool through atomic processes, and so the addition of \H2\
cooling channels does not help to increase $z_{\rm reion}$
significantly.

\begin{figure*}
\psfig{file=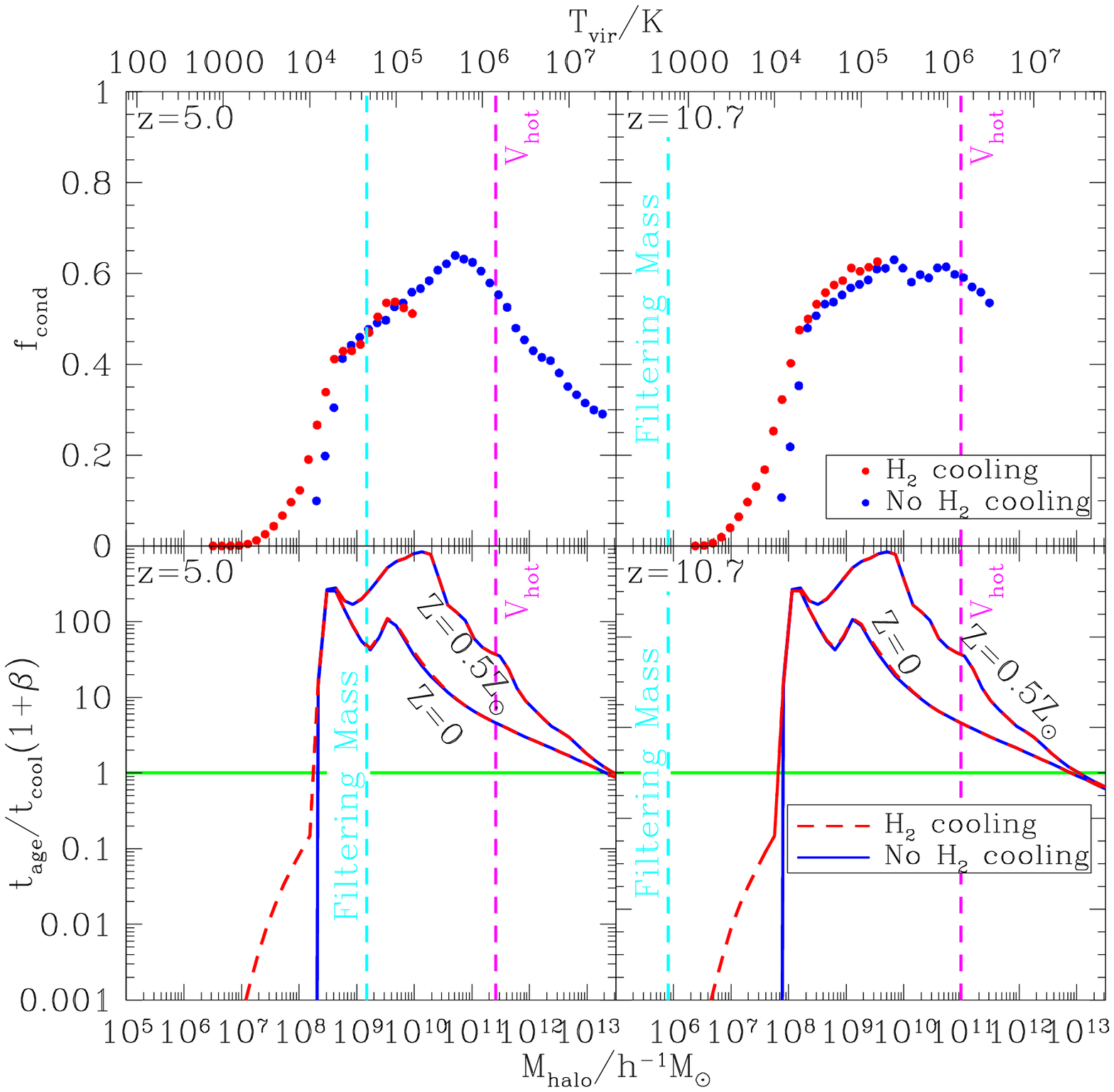,width=160mm,bbllx=0mm,bblly=75mm,bburx=190mm,bbury=270mm,clip=} 
\vspace{-15mm}
\caption{Measures of the ability of baryonic material to cool into the
galactic phase are shown for $z=5$ (left-hand panels) and $z=10$
(right-hand panels) for models \emph{without} the effects of
photo-ionization and photo-dissociation feedback. Vertical dashed magenta
lines indicate the halo mass for which the virial temperature
corresponds to the feedback parameter $V_{\rm hot}$. Vertical, dashed
cyan lines indicate the filtering masses for the two
models. \emph{Upper panels:} The average fraction of baryonic material
in haloes which has been able to cool and condense into galaxies, as a
function of halo mass in the Standard model in the WMAP RSI
cosmology. Blue circles indicate the cooled fraction for a model with
no molecular hydrogen cooling, while red circles show results for a
model with molecular hydrogen cooling. \emph{Lower panels:} Cooling
measures for metallicities $Z=0$ and $Z=0.5 Z_{\odot}$ (the upper
curves at large halo masses): solid blue lines are for the model
without molecular hydrogen cooling, while dashed red lines include
molecular hydrogen cooling (and Compton cooling). The cooling measure
is the ratio $t_{\rm age}/(1+\beta)t_{\rm cool}$. Here $\beta$ is the
ratio of rate of mass ejection due to supernova feedback to the star
formation rate in model galaxies. The horizontal green lines indicate
the point where the cooling measure equals 1 (i.e. the point above
which cooling is efficient in the sense described in the text.}
\label{fig:fcond_nophoto}
\end{figure*}

To understand the effects of cooling through molecular hydrogen, we
can explore the cooled baryonic fraction in haloes as a function of
their mass. The upper panels of Fig.~\ref{fig:fcond_nophoto} show the
condensed baryonic fractions of haloes at redshifts 5 (left-hand
panel) and 10 (right-hand panel) for models without the effects of
photo-ionization and photo-dissociation included. Blue circles
indicate the cooled fraction, $f_{\rm cond}$ (i.e. the mass of
material which has been able to cool and condense into the galactic
phase), for a model with no molecular hydrogen cooling, while red
circles show results for a model with molecular hydrogen cooling. We
indicate, with a vertical dashed magenta line, the halo mass for which
the virial temperature corresponds to the supernova feedback parameter
$V_{\rm hot}$ in our semi-analytic model. Vertical, dashed cyan lines
indicate the filtering masses for the two models. In the lower panels
of Fig.~\ref{fig:fcond_nophoto} are curves which show the cooling
measure $t_{\rm age}/(1+\beta)t_{\rm cool}$, the ratio of the age of
the Universe at the specified redshift $t_{\rm age}$ to the cooling
time $t_{\rm cool}$ (calculated at 200 times the mean density of the
Universe, which is approximately the mean density of haloes), and
divided by $1+\beta$, where $\beta$ (given by
eqn.~\ref{eq:SNfeedback}) is the ratio of the mass ejection rate due
to supernova feedback to the star formation rate in model
galaxies. Solid blue lines are for the model without molecular
hydrogen cooling and without Compton cooling, while dashed red lines
include molecular hydrogen cooling (and Compton cooling). We show
curves for primordial gas and for gas of half Solar metallicity (the
upper curves at large halo masses).

Our cooling measure, the quantity $t_{\rm age}/(1+\beta)t_{\rm cool}$,
is interesting since it roughly determines whether a halo has had time
to turn most of its supply of gas into stars. Roughly speaking, a
galaxy will accrete (from its surrounding hot atmosphere) all
available gas in a time $t_{\rm cool}$. Assuming haloes to have
survived for a period roughly equal to $t_{\rm age}$, the ratio of
these two timescales then determines whether or not a halo has been
able to accrete all available gas. However, much of this gas will be
ejected once again due to feedback. This gas may be subsequently
re-accreted. Consequently, a galaxy must accrete its hot halo
$\sim(1+\beta)$ times before it can turn the majority of the available
gas into stars.\footnote{Assuming the cooling/reheating loop to occur
$n$ times (and assuming cooling and reheating to happen
instantaneously), the fraction of available gas eventually bound into
stars will be $1-(\beta/(1+\beta))^n$. For $n=1+\beta$ this converges,
for large $\beta$ to $1-1/{\rm e}\approx 0.63$.}

When $t_{\rm age}/(1+\beta)t_{\rm cool}$ is greater than one (the
point $t_{\rm age}/(1+\beta)t_{\rm cool}=1$ is indicate by green
horizontal lines in Fig.~\ref{fig:fcond_nophoto}), the effective
cooling time is short compared to the age of the Universe and we do
not expect the cooling function to have strongly determined the shape
of the $f_{\rm cond}$ curve. When it is less than unity, the effective
cooling time is long and so the shape of the $f_{\rm cond}$ curve
should have been determined strongly by the shape of the cooling
curve.

From Fig.~\ref{fig:fcond_nophoto} we can see that the inclusion of
\H2\ cooling makes no difference to the fraction of condensed mass in
haloes with virial temperatures above about $2 \times 10^4$K. \H2\
cooling does however, result in a tail of low mass haloes (with virial
temperatures below $10^4$K) which are able to condense some of their
gas, unlike in the case of cooling through purely atomic
processes. However, the amount of gas condensed is minimal, with
$\lsim 10$\% of the total available gas being condensed (a number
which falls rapidly to zero in lower mass haloes). This inefficient
condensation occurs because $t_{\rm age}/t_{\rm cool}(1+\beta) \ll 1$
(typically $10^{-3}$--0.1) for these haloes due mostly to the fact
that $\beta \gg 1$ (typically, $\beta \gsim 100$ for these
haloes)---this makes star formation a very inefficient process in
these galaxies, with the majority of the gas which was able to
condense being re-ejected due to feedback.

Further plots of the condensed fraction for all cosmologies and models
can be found in Appendix~\ref{app:fcond}.

\subsection{Effects of including photoionization and photodissociation feedback}
\label{sec:feedmechs}

\begin{figure*}
\psfig{file=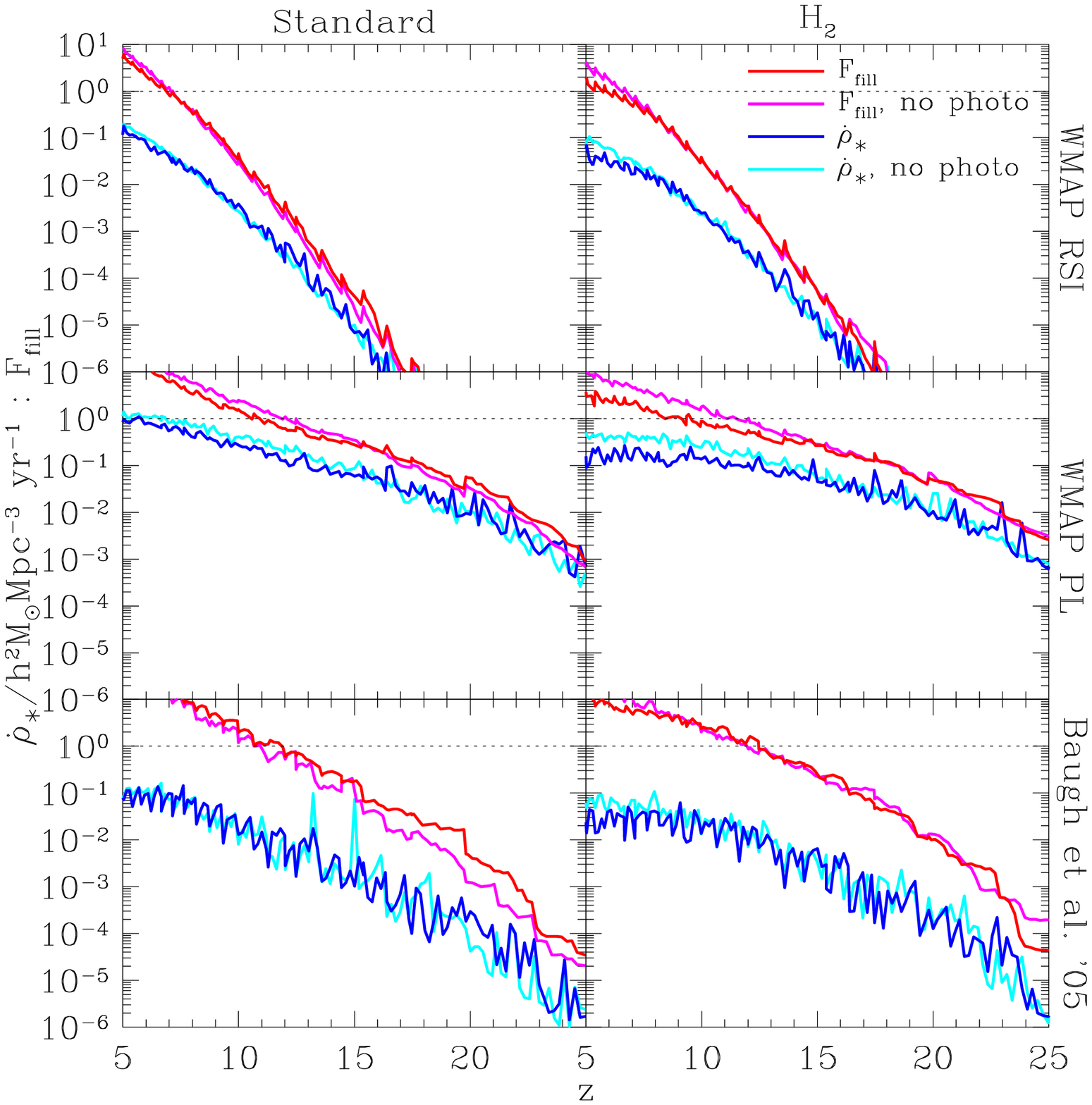,width=170mm,bbllx=0mm,bblly=75mm,bburx=195mm,bbury=270mm,clip=}
\caption{Star formation histories (blue and cyan lines) and \HII\
filling factors calculated using ionized spheres (red and magenta
lines). Cyan and magenta lines show results for models where
photoionization and photodissociation feedback are neglected, while
blue and red lines show results for models including these feedback
mechanisms. Results are shown for Standard and \H2\ models in WMAP PL,
WMAP RSI and Baugh et al. '05 cosmologies as indicated for each column
and row.}
\label{fig:phot}
\end{figure*}

We have run all models both with and without the effects of
photoionization feedback (i.e. the suppression of accretion due to the
IGM pressure and photoheating of gas in haloes by an ionizing
background) and photodissociation feedback (i.e. the destruction of
H$_2$ molecules by the dissociating background). In
Fig.~\ref{fig:phot} we show the resulting star formation rates and
\HII\ filling factors as a function of redshift. Photoionization and
photodissociation feedback reduce the redshift of reionization in
general due to the suppression of galaxy and star
formation. Fig.~\ref{fig:phot} also illustrates once more that the
inclusion of \H2\ cooling does little to increase the redshift of
reionization. In fact, it appears that including \H2\ cooling actually
reduces the redshift of reionization in the curves plotted here. This
is a numerical artifact, arising from the fact that the most massive
haloes simulated in our calculations are of lower mass in models
which include \H2\ cooling (see \S\ref{sec:ulim}). Once this is corrected
for, reionization redshifts are almost identical whether or not \H2\
cooling is included (see \S\ref{sec:epoch}).

\begin{figure*}
\psfig{file=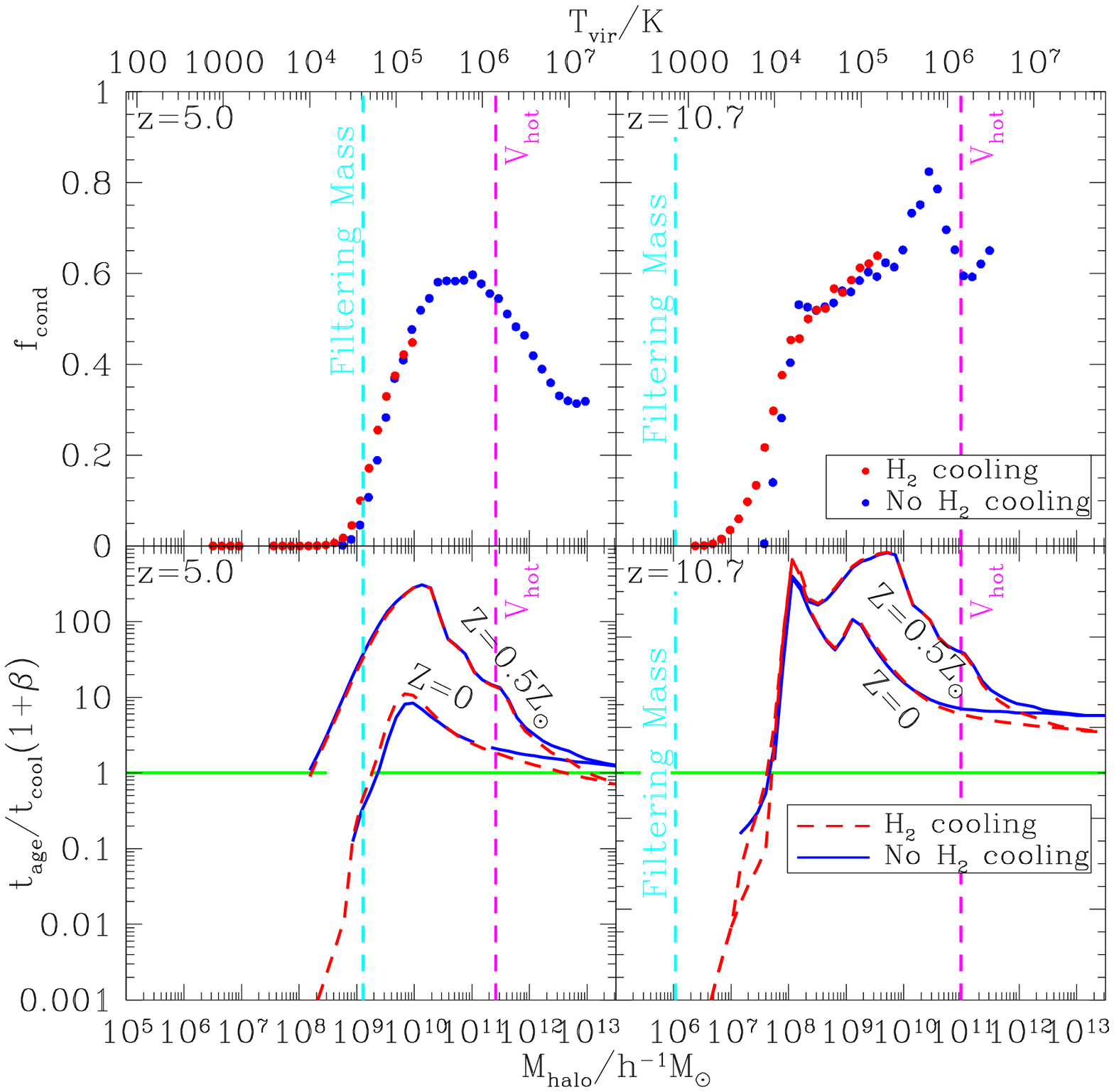,width=160mm,bbllx=0mm,bblly=75mm,bburx=190mm,bbury=270mm,clip=} 
\vspace{-15mm}
\caption{As Fig.~\protect\ref{fig:fcond_nophoto} but for models
\emph{including} the effects of photo-ionization and photo-dissociation
feedback.}
\label{fig:fcond_photo}
\end{figure*}

Figure~\ref{fig:fcond_photo} shows the same information as
Fig.~\ref{fig:fcond_nophoto} but for models which now include the
effects of photo-ionization and photo-dissociation feedback. By
comparing this figure with Fig.~\ref{fig:fcond_nophoto}, it can be
seen from this figure that the main causes of the suppression of
galaxy formation are photoheating of hot gas in haloes and
photodissociation of H$_2$. The filtering mass is approximately the
same in models with and without photoionization and photodissociation
feedback included\footnote{Note that, even when photoionization
feedback is not included the filtering mass still rises due to the
heating of the IGM by ionizing photons emitted by galaxies. All that
is left out of the calculation is the effect of the filtering mass on
subsequent galaxy formation.} (it is not until after reionization that
the filtering mass has the chance to respond to the increase in the
temperature of the gas). At $z=5$ we can clearly see that the cooling
times in haloes with virial temperatures below around $10^5$K are
greatly increased due to this photoheating and photodissociation. At
$z=10$ the effect is much weaker, due both to the lower background at
that redshift and the fact that gas in haloes is denser at higher
redshifts and so less affected by photoheating.

\begin{figure}
\psfig{file=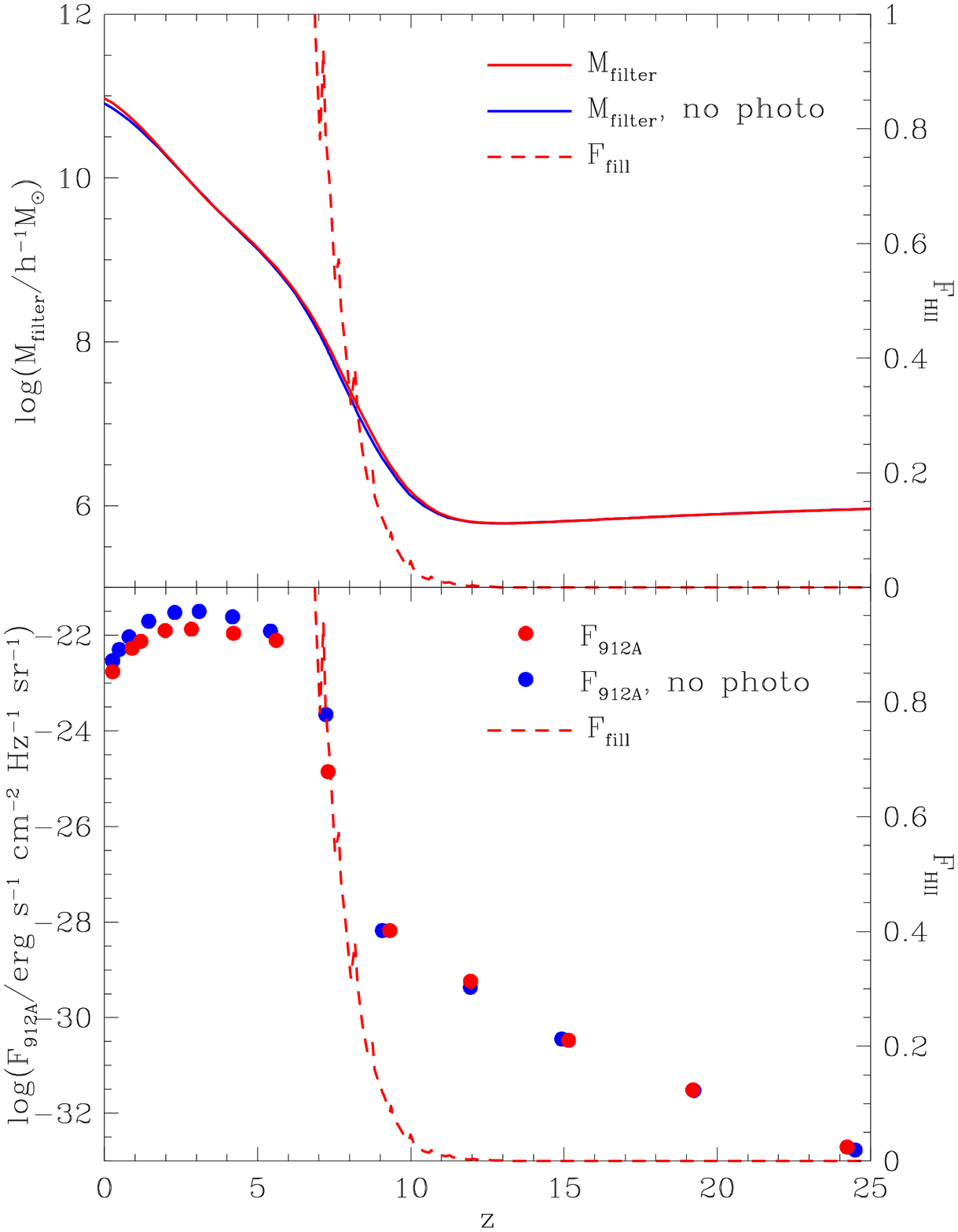,width=90mm}
\caption{Indicators of the strength of photoionization feedback
mechanisms as a function of redshift for the Standard model in the
WMAP RSI cosmology. Results are shown for this single model and
cosmology only as they are typical of all of the models we
consider. The solid red (blue) line in the upper panel shows the
filtering mass (scale on left-hand axis) when photoionization feedback
is (is not) included. In the lower panel red (blue) circles show the
ionizing background intensity (scale on left-hand axis) when
photoionization feedback is (is not) included. The dashed red line in
both panels indicates the \HII\ filling factor (scale on right-hand
axis) for a model including photoionization feedback.}
\label{fig:feedmechs}
\end{figure}

It can be seen that the inclusion of photoionization and
photodissociation feedback has only a small effect on the redshift of
reionization. This is not surprising, as the mechanisms through which
photoionization feedback works (photoheating of gas in haloes by the
ionizing background and an increase in the filtering mass) become
important only \emph{after} reionization. Figure~\ref{fig:feedmechs}
shows the evolution of the \HII\ filling factor\footnote{The 
``spikes'' seen in the \HII\ filling factor are an artifact of the
semi-analytic model. As in Paper~I, we compute the reionization
process by drawing a set of halo masses from the halo mass function at
some redshift $z$ and constructing merger trees for each halo back to
very high redshifts. The galaxy formation calculation then proceeds in
each tree. Our merger tree algorithm is not completely accurate in the
sense that, when used to produce merger trees that span large redshift
ranges, it does not exactly reproduce the expected distribution of
progenitor halo masses. This problem becomes more severe as the
redshift range spanned increases. Consequently, we repeat our
calculations for several starting redshifts $z_i$ (ensuring that at
each $z_i$ we have the correct distribution of halo masses). The
reionization histories shown are constructed by splicing together
results from each calculation, using quantities between $z_i$ and
$z_{i+1}$ from each calculation. At each $z_i$ we find discontinuities
in the predicted properties due to this splicing process.}, filtering
mass and ionizing background for the standard model in the WMAP RSI
cosmology with and without the effects of photoionization feedback
included. The filtering mass begins to rise shortly before
reionization occurs, when photoheating first dominates over cooling
of gas due to the expansion of the Universe, but does not achieve any
significant value until after the Universe is already
reionized. Similarly, the ionizing background cannot build up to any
significant level (and thereby efficiently photoheat gas in haloes)
until the Universe has already been ionized and thereby become
transparent to ionizing radiation. \scite{benson02} found that it was
the filtering mass which caused the greatest suppression of galaxy
formation, seemingly in contradiction with the above
result. \scite{benson02} were, however, considering the effects of
photoionization feedback on the galaxy population at $z=0$, by which
time the filtering mass had had time to fully respond to the change in
the IGM temperature during reionization. When considering the effects
of photoionization feedback on the epoch of reionization itself, as we
do here, the filtering mass is much less important as it does not have
time to fully respond to the heating of the IGM until well after
reionization is over. The ionizing background, however, rises very
rapidly as reionization occurs, allowing photoheating to have some
(albeit small) effect on the redshift of reionization.

\subsubsection{Effects of $H_2$ Dissociation?}

\begin{figure}
\psfig{file=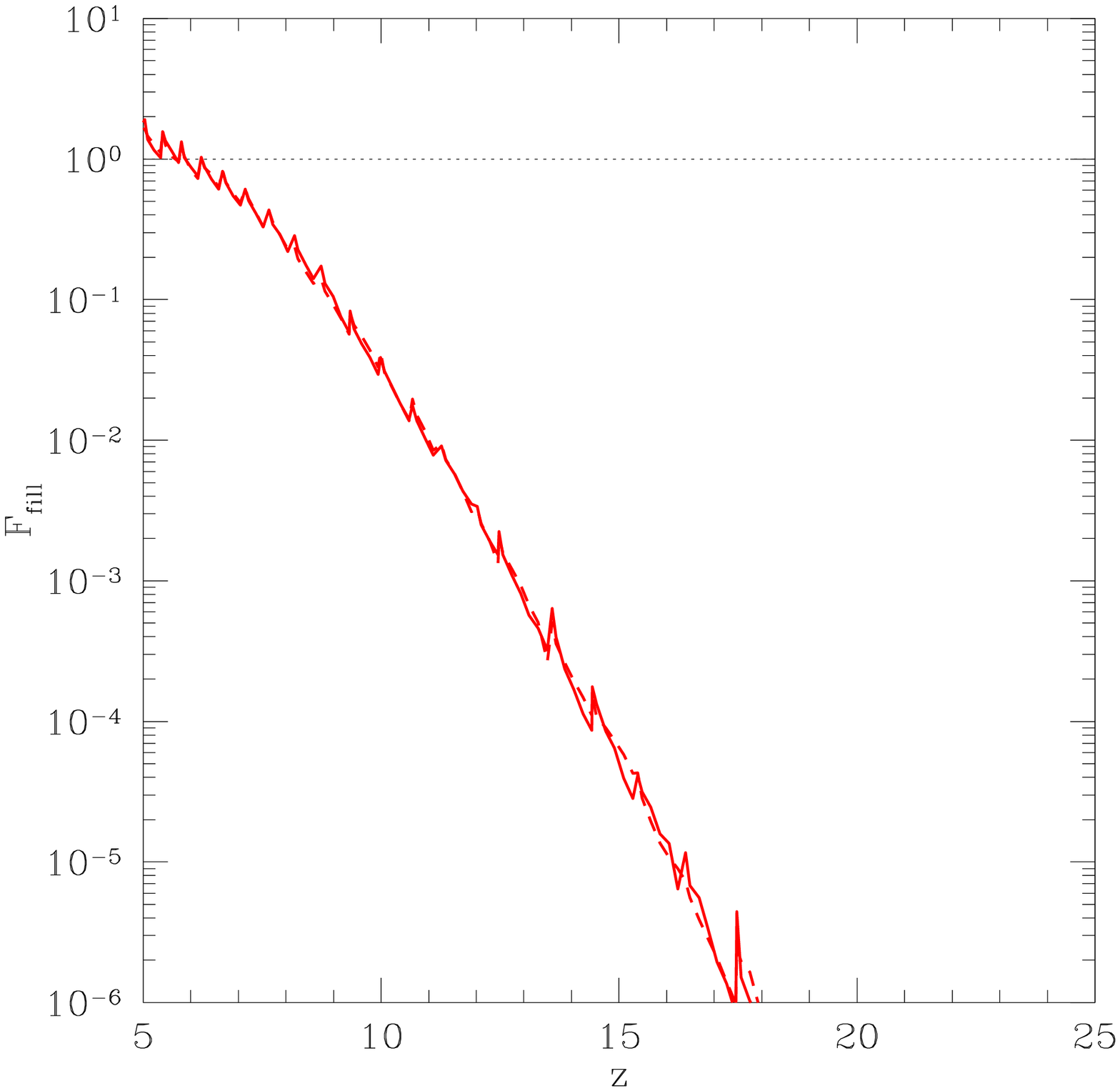,width=80mm}
\caption{The \HII\ filling factor as a function of redshift for model
H$_2$ with (solid line) and without (dashed line) dissociation of
$H_2$ molecules included in our calculations. Results are shown for
the WMAP RSI cosmology only, but are typical of all other cosmologies
and models.}
\label{fig:nodissoc}
\end{figure}

In general, we find that stars formed via $H_2$ cooling contribute
little to the reionization of the Universe, if they form with a normal
IMF. Therefore, it is not surprising that the inclusion or otherwise
of $H_2$ dissociation in our model makes no significant difference to
the end results on the ionized filling factor, as shown in
Fig.~\ref{fig:nodissoc}.

\subsection{Enhanced Emission from Early Generations of Stars}
\label{sec:enhance}

\begin{figure*}
\begin{tabular}{cc}
\psfig{file=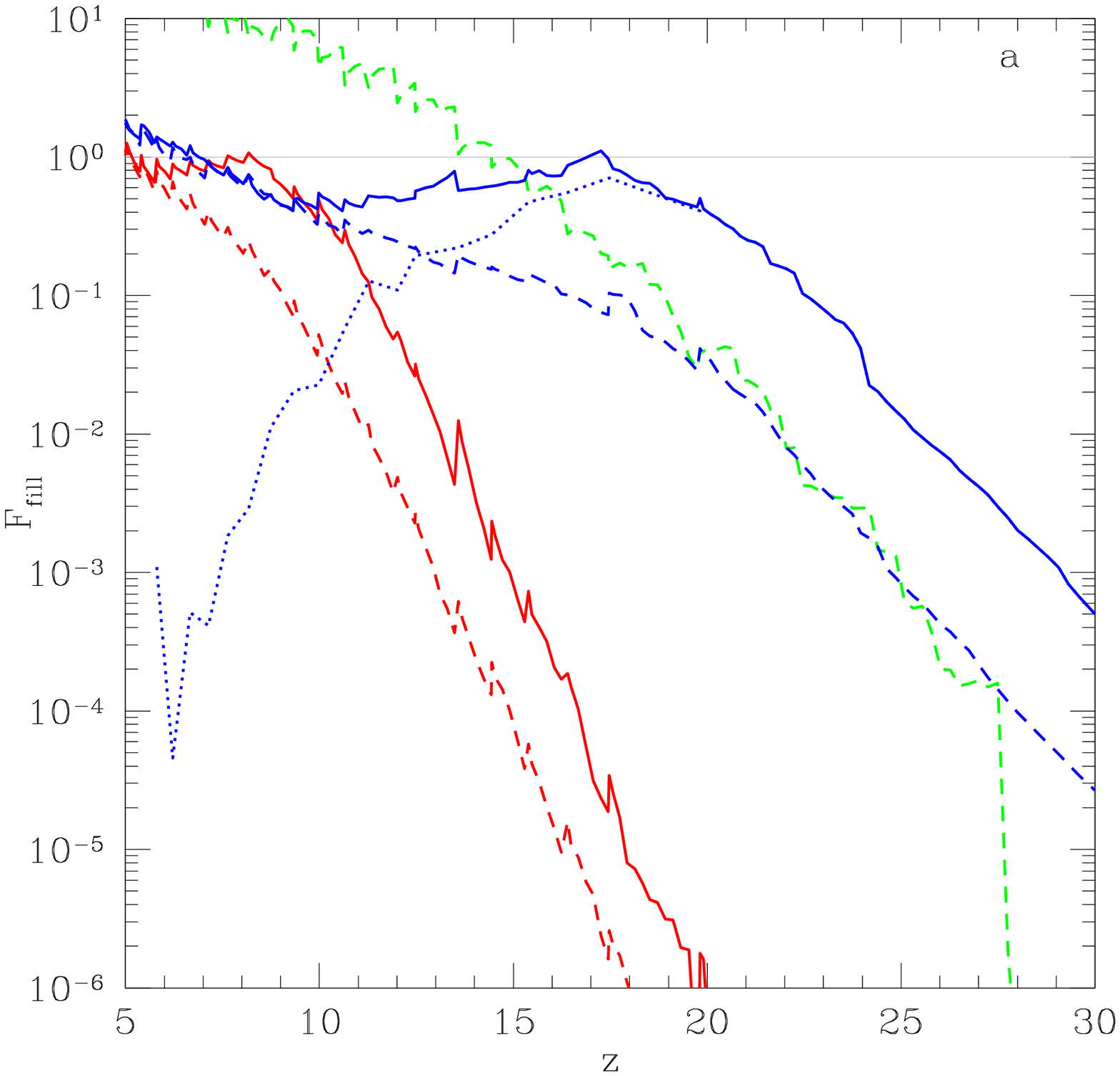,width=80mm} & \psfig{file=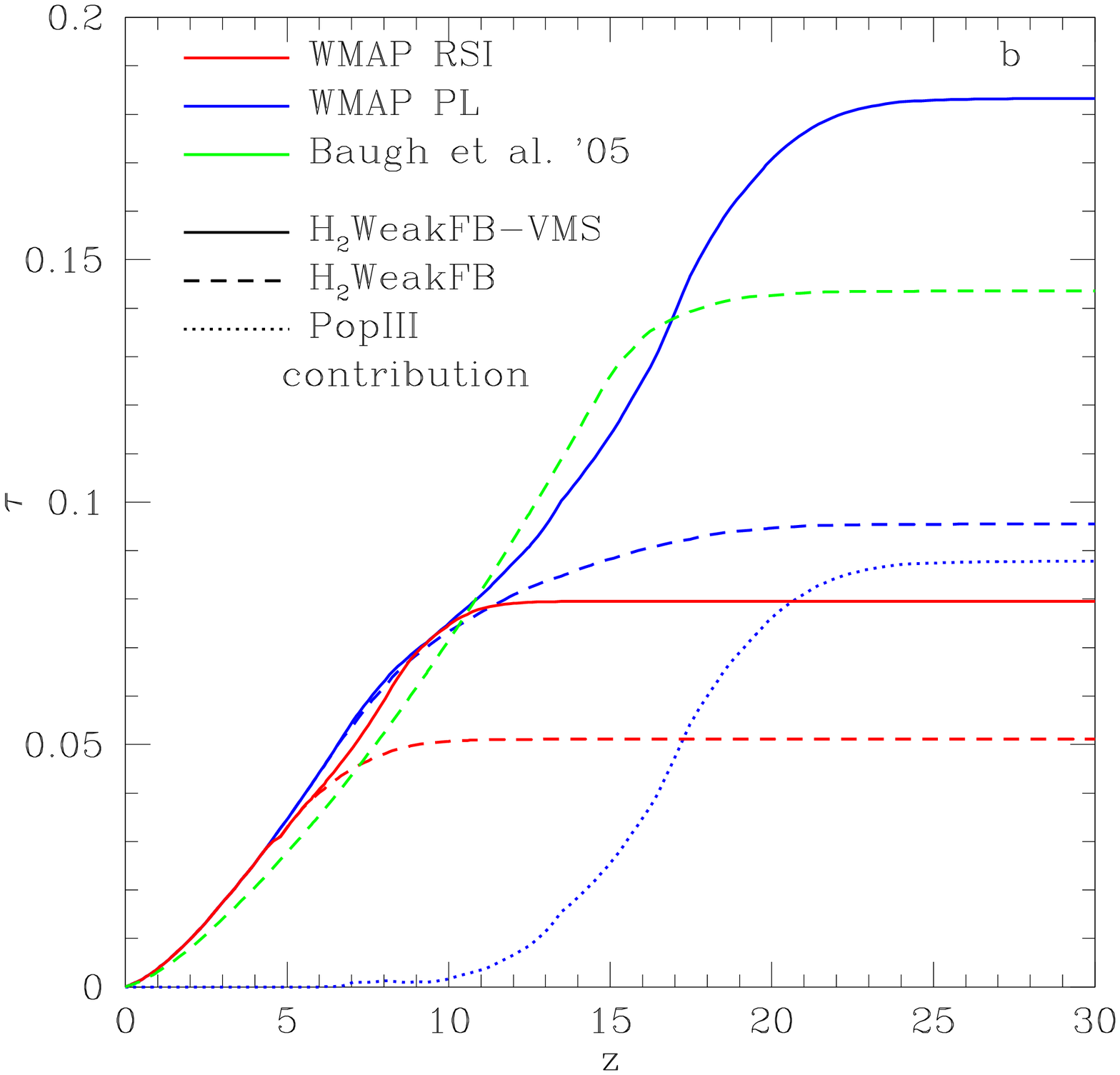,width=80mm}
\end{tabular}
\caption{\emph{Left-hand panel:} The \HII\ filling factor as a function of redshift for models H$_2$WeakFB-VMS (solid lines) and
H$_2$WeakFB (dashed lines) in our three cosmologies (see key in right-hand panel). The dotted line shows, for the WMAP PL
cosmology only, the contribution to the \HII\ filling factor from stars forming in haloes with virial temperatures below
$10^4$K. Escape fractions have been chosen so as to produce double reionizations in each model, apart from the Baugh et al. (2005)
model (for which we do not compute the H$_2$WeakFB-VMS model) in which case we simply choose $f_{\rm esc}=1$. The escape fractions
required are: $f_{\rm esc}=0.25$, 0.15 and 1.0 for the WMAP RSI, WMAP PL and Baugh et al. (2005) cosmologies
respectively. \emph{Right-hand panel:} The electron scattering optical depth as a function of redshift for the same models.}
\label{fig:VMS}
\end{figure*}

In model ``H$_2$WeakFB-VMS'' we enhance the UV emission from stellar
populations which form via \H2\ cooling (i.e. those which form in dark
matter haloes with virial temperatures below $10^4$K) by a factor of
20 relative to a standard IMF, to account for the possibility that
these stars might be super-massive ``pop~III'' stars.  The left-hand
panel of Figure~\ref{fig:VMS} shows the resulting \HII\ filling
factors vs. redshift for this model and for model H$_2$WeakFB (which
is identical apart from having no enhancement of UV emission due to
supermassive stars) for the WMAP cosmologies. The results are plotted
for an escape fraction chosen to give a double reionization: 25\%
(WMAP RSI) and 15\% (WMAP PL). This results
in all models producing full reionization at $z\approx 6$. Note
however, that model H$_2$WeakFB-VMS displays an extended period of
partial reionization back to higher redshifts (from $z\approx 8$ for
the WMAP RSI cosmology to $z\approx 17$ for the WMAP PL cosmology).
As a consequence, model H$_2$WeakFB-VMS has an increased electron
scattering optical depth as shown in the right-hand panel of
Figure~\ref{fig:VMS}.  For example in the WMAP PL cosmology we find
$\tau = 0.15$, compared to $0.08$ for model H$_2$WeakFB. Values of
$\tau$ are lower for the other cosmologies, but still come
significantly closer to the WMAP measurement of $\tau=0.17\pm
0.04$. As such, model H$_2$WeakFB-VMS is in much better agreement with
the data obtained from the WMAP satellite \cite{kogut03}. Enhanced UV
emission from stellar populations formed via \H2\ cooling is therefore
able to reconcile the apparent disparity between a large optical depth
and evidence for reionization at $z\approx 6$ in the WMAP PL
cosmology, by inducing a period of partial reionization.

The dotted line in Fig~\ref{fig:VMS}(a) shows the contribution to the
\HII\ filling factor from stars forming via \H2\ cooling in the
H$_2$WeakFB-VMS model for the WMAP PL cosmology. Before $z\approx 20$
these stars dominated the reionization of the Universe, as the
steepness of the dark matter halo mass function results in there being
very few haloes hot enough to cool via atomic processes at these
redshifts. At lower redshifts the abundance of haloes with virial
temperature above $10^4$K increases, while galaxy formation in lower
temperature haloes becomes inefficient due to increased cooling times
and the rising filtering mass\footnote{The filtering mass in the
H$_2$WeakFB-VMS model in the WMAP PL cosmology rises above the mass
corresponding to a $10^4$K halo at $z=15$.}. Therefore, haloes hotter
than $10^4$K rapidly become the dominant contributors to the \HII\
filling factor.

Obtaining a period of extended partial reionization, or a double
reionization, requires an interplay of several key factors. Firstly,
the cosmological model must produce enough haloes with virial
temperatures below $10^4$K to permit a significant \HII\ filling
factor to be obtained from the early generation of stars (assumed to
be supermassive). The enhancement factor (taken to be 20 in this work)
is also important: if it is too high then reionization may happen once
only, if it is too low the early generation of stars will not be able
to produce a significant \HII\ filling factor. The decline in the
contribution from early generation stars as time progresses must also
be sufficiently rapid (otherwise the filling factor will continue to
rise, leading to a single reionization event)---requiring either that
the filtering mass grows rapidly or that cooling times for haloes
cooler than $10^4$K increase rapidly shortly after the first
reionization. Despite this requirement that many factors combine to
produce a double reionization, we find, for quite reasonable values of
the UV enhancement factor, that double reionization is a natural
outcome of our H$_2$WeakFB-VMS model in all cosmologies considered. An
extensive survey of parameter space would be required to determine how
common such double reionizations are, but this is beyond the scope of
this work.

While our modelling of this enhanced UV emission from early stars is
very simplified (e.g. we assume that supernova feedback is the same
for Very Massive Stars and for normal stellar populations), and based
on uncertain assumptions (e.g. that all haloes which cool by H$_2$
produce Very Massive Stars), it is still relatively sophisticated
compared to previous studies of double reionization
\cite{cen03,wylo03} to the extent that ours is the only model which
follows the full hierarchical merging history of each forming galaxy
and, furthermore, computes the fraction of baryons able to cool and
the fraction able to turn into stars using physically motivated
prescriptions rather than simply making these fractions parameters of
the model.

\section{Observable quantities}
\label{sec:2res}

\subsection{The Epoch of Reionization}
\label{sec:epoch}

Table~\ref{tb:epochs} lists the redshifts of reionization in all of
the models considered (all results shown include the effects of
photoionization and photodissociation feedback), along with the
corresponding electron-scattering optical depths. Observations of the
cosmic microwave background suggest a large value for $z_{\rm
reion}$. \scite{kogut03} used WMAP measurements of the
temperature-polarization (TE) cross-power spectrum to estimate the
optical depth $\tau$. They found a best-fit value $\tau=0.17$, with a
95\% confidence range $0.09\leq \tau \leq 0.28$. For the WMAP
cosmology, this corresponds to a best-fit value of the reionization
redshift $z_{\rm reion} = 17$, and a 95\% confidence range $11< z_{\rm
reion} < 24$, if one assumes instantaneous reionization, i.e. if the
ionized fraction increased from 0 to 1 at $z=z_{\rm
reion}$. \scite{spergel03} found very similar results from their
multi-parameter fit to the whole WMAP dataset combined with data on
galaxy clustering.

Since our models predict a gradual rather than an instantaneous
reionization, we compare our models with the WMAP data using the value
of $\tau$ rather than the value of $z_{\rm reion}$.  From
Table~\ref{tb:epochs} we see that all of our models based on the WMAP
PL cosmology and assuming $f_{\rm esc}=1$ are consistent with the
measured $\tau$ (at 95\% confidence), apart from the H$_2$WeakFB-VMS
model, which predicts $\tau=0.30$, marginally too high. These models
predict $z_{\rm reion}$ in the range 11--22. However, when $f_{\rm
esc}$ is reduced to 15\%, the H$_2$WeakFB-VMS model in this cosmology
predicts $\tau=0.15$ (and $z_{\rm reion}=17$), in good agreement with
the WMAP measurement.  In contrast, all but one of our models based on
the WMAP RSI cosmology, and assuming $f_{\rm esc}=1$, are excluded (at
95\% confidence) by the WMAP measurement of $\tau$, due to the
predicted optical depths being too low, the one exception being the
H$_2$WeakFB-VMS model, which (with $\tau=0.10$) is marginally
compatible. The models in the WMAP RSI cosmology predict $z_{\rm
reion}$ in the range 6--11. Thus, while the WMAP RSI cosmology cannot
be completely excluded based on the WMAP measurement of $\tau$, it
does appear very unlikely. If the escape fraction $f_{\rm esc}$ is in
fact significantly less than unity, then the predicted reionization
epochs will be shifted to considerably lower redshifts in all of the
models. For the Baugh et al. '05 cosmology, we find that the Standard
and \H2\ models are marginally consistent with the WMAP constraint,
while models with weak feedback are able to match that constraint
within current errors.

\begin{table*}
\caption{The epochs of reionization for all models considered in this
work. Columns 1 through 3 specify input parameters of the models:
column 1 specifies the cosmological model while column 2 specifies the
galaxy formation model used and column 3 specifies the value of
$f_{\rm esc}$ used. The remaining columns list output
quantities. Column 4 lists the redshift of reionization---defined as
the redshift at which the filling factor in the ionized sphere
reionization calculation reaches unity. The first value gives $z_{\rm
reion}$ as found directly from our calculations, while the second
value accounts for the finite upper mass limit of the haloes which we
simulate. The true $z_{\rm reion}$ should lie between these two
values. The next column gives the electron scattering optical depth
(assuming that helium is singly ionized at the same time as hydrogen
is ionized, and that helium is doubly ionized below $z=3$). The two
values again refer to the calculation without and with the correction
for the finite upper mass limit of simulated haloes. The next two
columns list the redshifts at which the neutral hydrogen fraction in
the IGM drops to $10^{-2}$ and $10^{-5}$ respectively, as determined
using the IGM evolution model of \protect\scite{benson02}. The final
column lists the redshift at which the mean Gunn-Peterson optical
depth reaches a value of 3}
\label{tb:epochs}
\begin{center}
\begin{tabular}{lccr@{--}lr@{--}lccc}
\hline
\textbf{Cosmology} & \textbf{Model} & {\boldmath $f_{\rm esc}$} & \multicolumn{2}{c}{\boldmath ${z_{\rm reion}}$} & \multicolumn{2}{c}{\boldmath ${\tau}$} & {\boldmath ${z_{\rm f_{HI}=10^{-2}}}$} & {\boldmath ${z_{\rm f_{HI}=10^{-5}}}$} & {\boldmath $z_{\tau_{\rm GP}=3}$} \\
\hline
WMAP RSI & Standard & 1.00 & 6.9 & 6.9 & 0.061 & 0.061 & 7.36 & 7.25 & 7.10 \\
WMAP RSI & MetFree & 1.00 & 7.3 & 7.4 & 0.064 & 0.068 & 7.60 & 7.50 & 7.38 \\
WMAP RSI & TopHeavy & 1.00 & 8.7 & 8.7 & 0.079 & 0.080 & 9.02 & 8.94 & 8.79 \\
WMAP RSI & WeakFB & 1.00 & 8.1 & 8.1 & 0.073 & 0.079 & 8.50 & 8.45 & 8.33 \\
WMAP RSI & H$_2$ & 1.00 & 6.1 & 6.6 & 0.055 & 0.063 & 6.91 & 6.69 & 6.51 \\
WMAP RSI & H$_2$WeakFB & 1.00 & 7.8 & 8.1 & 0.074 & 0.075 & 8.44 & 8.38 & 8.22 \\
WMAP RSI & H$_2$WeakFB & 0.25 & 4.4 & 4.4 & 0.048 & 0.055 & 7.60 & 7.34 & 7.17 \\
WMAP RSI & H$_2$WeakFBVMS & 1.00 & 10.6 & 10.6 & 0.102 & 0.102 & 8.46 & 8.39 & 8.21 \\
WMAP RSI & H$_2$WeakFBVMS & 0.25 & 4.4 & 4.4 & 0.076 & 0.080 & 7.55 & 7.36 & 7.13 \\
\hline
WMAP PL & Standard & 1.00 & 10.8 & 10.8 & 0.141 & 0.141 & 11.38 & 10.74 & 10.54 \\
WMAP PL & MetFree & 1.00 & 11.4 & 11.4 & 0.157 & 0.157 & 11.92 & 11.20 & 10.93 \\
WMAP PL & TopHeavy & 1.00 & 15.3 & 15.3 & 0.213 & 0.213 & 14.54 & 14.46 & 14.33 \\
WMAP PL & WeakFB & 1.00 & 14.1 & 14.1 & 0.191 & 0.191 & 13.73 & 12.77 & 12.42 \\
WMAP PL & H$_2$ & 1.00 & 9.2 & 10.8 & 0.131 & 0.141 & 7.87 & 7.79 & 7.73 \\
WMAP PL & H$_2$WeakFB & 1.00 & 14.1 & 14.1 & 0.206 & 0.206 & 13.87 & 12.56 & 12.31 \\
WMAP PL & H$_2$WeakFB & 0.15 & 6.1 & 8.3 & 0.095 & 0.107 & 7.99 & 7.85 & 7.84 \\
WMAP PL & H$_2$WeakFBVMS & 1.00 & 22.2 & 22.2 & 0.299 & 0.299 & 13.77 & 12.71 & 12.40 \\
WMAP PL & H$_2$WeakFBVMS & 0.15 & 6.9 & 8.8 & 0.183 & 0.196 & 7.98 & 7.85 & 7.84 \\
\hline
Baugh et al. (2005) & Standard & 1.00 & 11.9 & 11.9 & 0.108 & 0.108 & 7.97 & 7.44 & 7.39 \\
Baugh et al. (2005) & H$_2$ & 1.00 & 12.5 & 12.5 & 0.118 & 0.118 & 7.55 & 7.23 & 6.87 \\
Baugh et al. (2005) & H$_2$WeakFB & 1.00 & 14.4 & 14.4 & 0.144 & 0.144 & 8.62 & 8.19 & 7.76 \\

\hline
\end{tabular}
\end{center}
\end{table*}

Observations of the spectra of the highest redshift quasars at $z\sim
6$ show evidence of Gunn-Peterson absorption by neutral hydrogen in
the IGM \cite{becker01,djorg01,white03,fan03,fan06}, which has been
interpreted as evidence that the Universe reionized at redshifts only
slightly above 6, since simple theoretical models predict that the
transition from the IGM being almost completely neutral to almost
completely ionized should have happened over a rather narrow redshift
range. A low $z_{\rm reion}\sim 7$ is in apparent contradiction with
the results from WMAP. This interpretation of the observed quasar
spectra has however been questioned by some authors
(e.g. \pcite{songaila04}). Even if current measurements of the
Gunn-Peterson optical depth $\tau_{\rm GP}$ at $z\sim 6$ are correct,
they can be explained by very small neutral hydrogen fractions in the
IGM, e.g. \scite{white03} measure $\tau_{\rm GP}\sim 5-20$ for two
quasars at $z\approx 6.3$, which can be produced with neutral
fractions of only $f_{\rm HI} \sim 1.5-6 \times 10^{-5}$. Therefore,
in principle the IGM might have ceased to be mainly neutral at much
higher redshifts, and still produce detectable Gunn-Peterson
absorption at $z\sim 6$. Note that the only way in which we are able
to produce a ``double-reionization'' is by assuming that stars forming
via \H2\ cooling (PopIII stars) are very massive objects producing
around 20 times more ionizing photons per unit mass than for a
standard IMF. Nevertheless, under such an assumption,
double-reionization is a natural consequence of our model.

\begin{figure*}
\begin{tabular}{cc}
\psfig{file=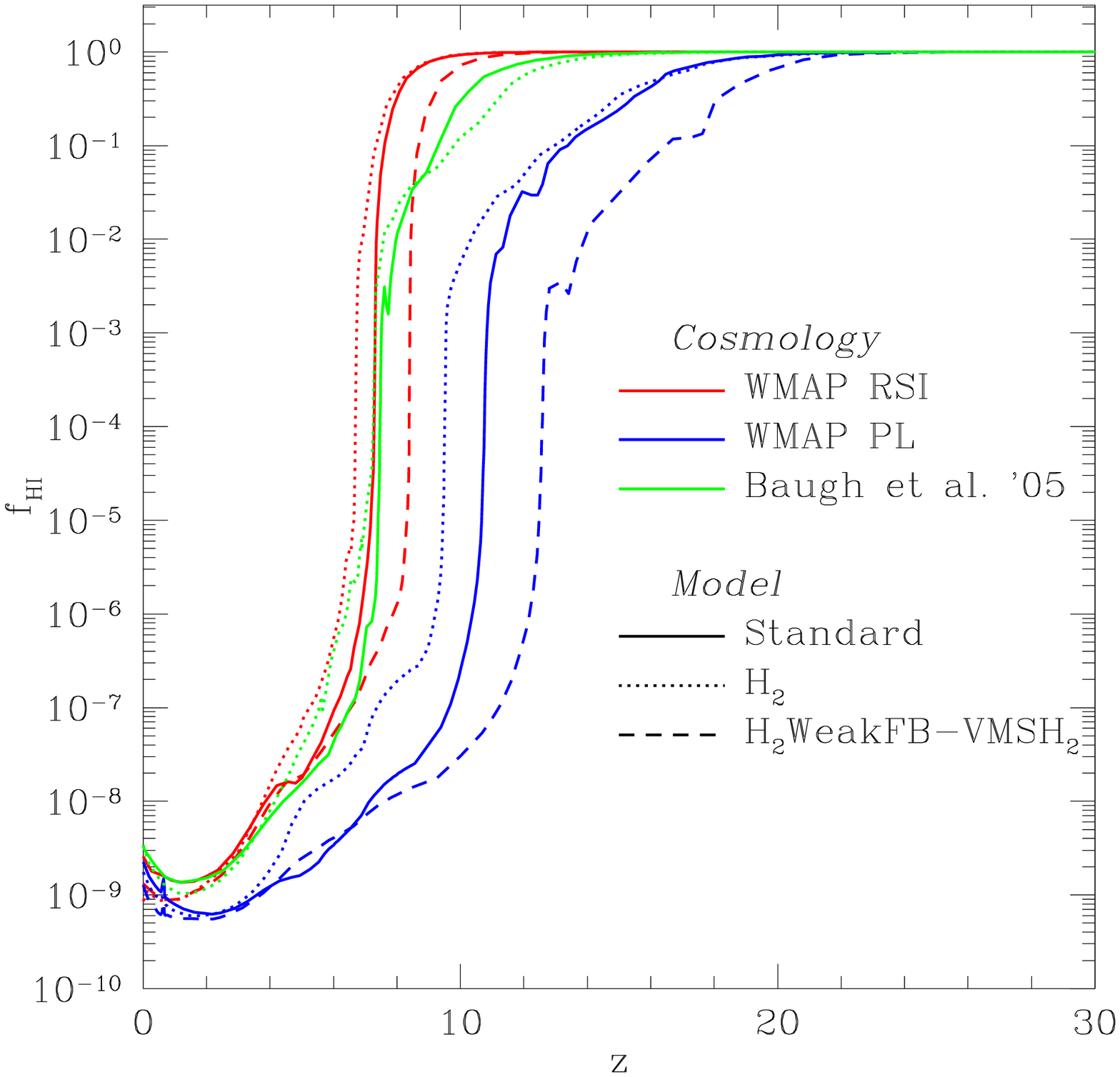,width=80mm} & \psfig{file=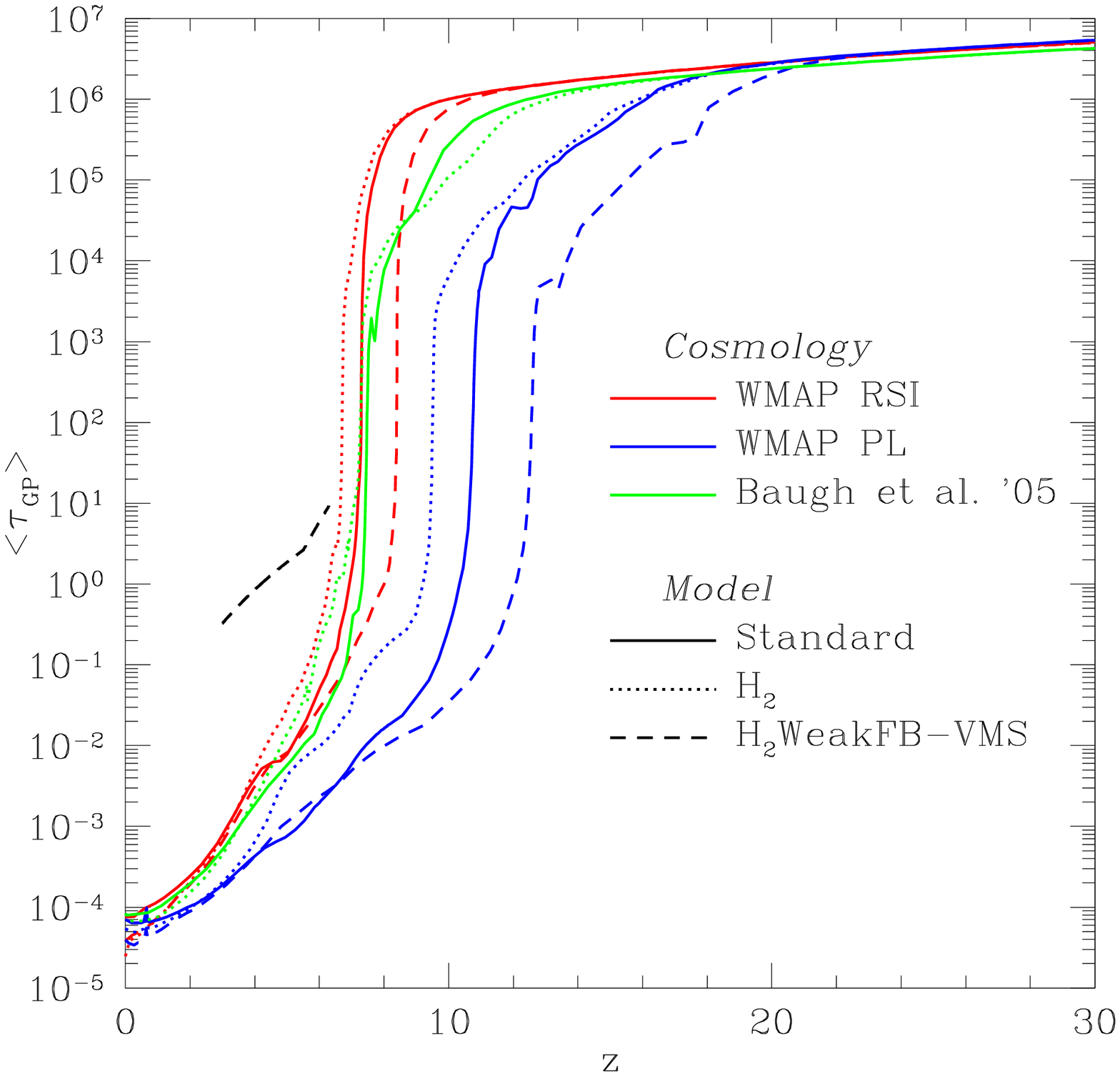,width=80mm}
\end{tabular}
\caption{The mean mass-weighted \HI\ fraction (left-hand panel) and
Gunn-Peterson optical depth (right-hand panel) as functions of
redshift for various models in our three cosmologies (see key in
figure for colour and line type coding). The model results plotted
here all assume $f_{\rm esc}=1$. The dashed black line in the
right-hand panel indicates the fit to the mean optical depth reported
by \protect\scite{fan06}.}
\label{fig:tauGP}
\end{figure*}

To allow a rough comparison of our models with observations of Gunn-Peterson absorption, we have calculated the average neutral
hydrogen fraction in the IGM, $f_{\rm HI}$ (defined as the ratio of volume averaged values of the abundances of neutral hydrogen
$n_{\rm HI}$ and total hydrogen $n_{\rm H}$, which is the quantity which is most directly comparable to measurements derived from
the Gunn-Peterson effect\footnote{That is, $f_{\rm HI}=\int n_{\rm HI} \d V / \int n_{\rm H} \d V$. This is equivalent to a
\emph{mass weighted} average of the local neutral fraction $x_{\rm HI}$.}), and used this to predict the mean Gunn-Peterson
optical depth $\langle \tau_{\rm GP} \rangle$, both as functions of redshift. Results for these quantities for several of our
models (using $f_{\rm esc}=1$) are plotted in Fig.~\ref{fig:tauGP}. We also give in Table~\ref{tb:epochs} for each model the
(lowest) redshift at which the mean neutral fraction drops to $f_{\rm HI}=10^{-2}$ and $10^{-5}$ respectively, and the redshift
for which $\langle \tau_{\rm GP} \rangle = 3$.

The results for $f_{\rm HI}$ and $\langle \tau_{\rm GP} \rangle$ for
the different models favour the models in the WMAP RSI cosmology if
$f_{\rm esc}=1$, or alternatively $f_{\rm esc}$ appreciably smaller
than unity if the WMAP PL cosmology is assumed. Note, however, that
none of our models produce a neutral hydrogen fraction at $z=6$ which
is sufficiently high to explain the observed Gunn-Peterson optical
depths (e.g. \pcite{white03,fan06}) when $f_{\rm esc}=1$. (The \H2\
model in the WMAP RSI cosmology comes closest, achieving $\tau_{\rm
GP}=3$ by $z=6.5$, but by $z=6$ the optical depth has already dropped
to $0.35$.)  The neutral fraction falls very rapidly during
reionization as can be seen from the fifth and sixth columns of
Table~\ref{tb:epochs}, which list the redshifts at which neutral
fractions of $10^{-2}$ and $10^{-5}$ are reached. Neutral fractions
comparable to those required by the observational data could be
obtained if $f_{\rm esc}$ were reduced, at the expense of a slightly
lower reionization redshift.

We find that the \scite{baugh05} model achieves reionization at
$z\approx 12$ for an escape fraction of 100\%, while adding \H2\
cooling to this model raises the reionization redshift to $z\approx
12.5$. The corresponding values of the electron scattering optical
depth are $\tau=0.108$ and $0.118$ respectively. This model is
therefore consistent with the WMAP measurement of the optical depth,
given the rather large errors on that datum. However, the neutral
fraction in this model is too low at $z\approx 6$ to explain
observations of the Gunn-Peterson trough (the optical depth at $z=6$
for the \scite{baugh05} model lies in the range $0.01$ to $0.1$ for
the various models considered here). While this could be rectified by
lowering the escape fraction in order to reduce the reionization
redshift to be closer to $z\approx 6$, this would also reduce the
optical depth.

\subsection{CMB Anisotropies}

\begin{figure*}
\begin{tabular}{cc}
\makebox[75mm]{WMAP PL} & \makebox[75mm]{WMAP RSI} \\
\multicolumn{2}{c}{\psfig{file=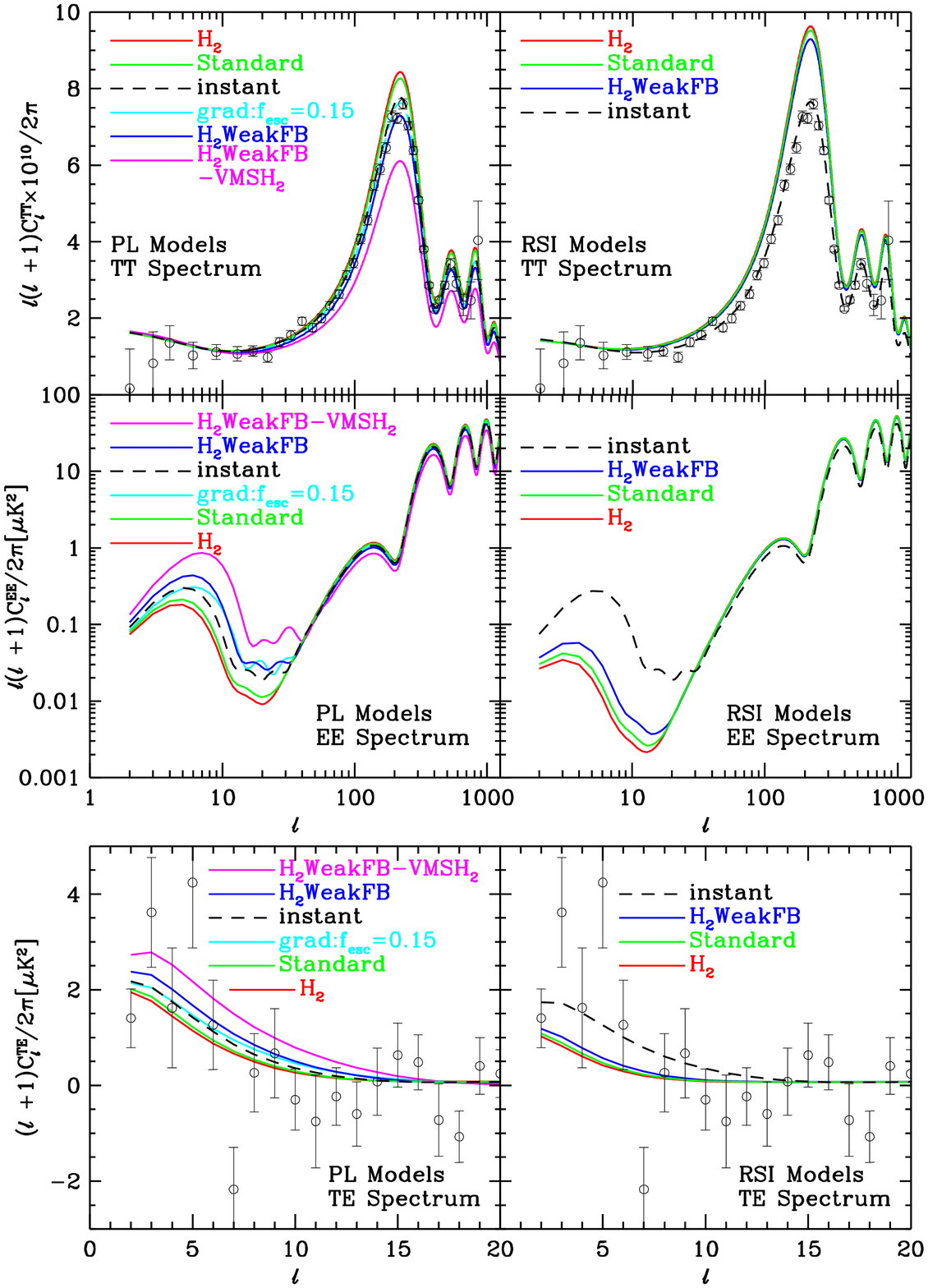,width=150mm}}
\end{tabular}
\caption{CMB anisotropy power spectra for the WMAP PL (left-hand
panels) and WMAP RSI (right-hand panels) cosmologies. The power spectra
of temperature fluctuations (TT) are shown in the upper panels, while
the middle and lower panels show polarization power spectra (EE in the
middle panels and TE in the lower panels). The circles with error bars
show observational data from the WMAP satellite
\protect\cite{hinshaw03,kogut03}. The dashed black lines (labelled
instant) show results for a simple toy model in which reionization is
instantaneous, and occurs at a redshift selected to best fit the WMAP
data. The coloured lines show the predictions for various models from
this work, assuming $f_{\rm esc}=100\%$ unless indicated
otherwise. The cyan lines (labelled grad) show results for the
H$_2$WeakFB-VMS model with an escape fraction of 15\%, which exhibits
gradual reionization.}
\label{fig:CMB}
\end{figure*}

We have computed the temperature and polarization power spectra
predicted by our model reionization histories. Figure~\ref{fig:CMB}
shows results for the WMAP PL and WMAP RSI cosmologies\footnote{The
Baugh et al. '05 cosmology was not designed to accurately match
available CMB data and so is not considered here.}.  To calculate CMB
temperature spectra, we employed the Boltzmann code of
\scite{sugiyama95}, which has been used to calibrate the accuracy of
publicly available codes, e.g., CMBFast \cite{seljak03}.  The models
shown in the figure have $\sigma_8=0.9$ for the WMAP PL cosmology and
$\sigma_8=0.83$ for the WMAP RSI cosmology. Furthermore, all models
are shown for $f_{\rm esc}=1$, except for the model labelled
``grad:$f_{\rm esc}=0.15$'' for which we plot the $\rm H_2WeakFB-VMS$
model with $f_{\rm esc}=0.15$ (which exhibits a double
reionization). For comparison, we also show the CMB anisotropies
predicted for a toy model of instant reionization, in which the IGM
instantaneously becomes fully reionized at $z_{\rm reion}=17$, chosen
to give the optical depth $\tau=0.17$ that best fits the WMAP data.

For the WMAP PL cosmology, it is clear that the $ \rm H_2WeakFB-VMS$
model with $f_{\rm esc}=1.0$ is ruled out from the TT-spectrum.  The
reason for the failure of this model is the combination of the large
optical depth (which suppresses CMB temperature anisotropies on small
scales) and $\sigma_8=0.9$. One could try to make this model agree
with the TT data by increasing $\sigma_8$ by $\sim 20\%$.  However,
this increase leads to galaxy formation at higher redshifts and hence
a larger optical depth. Thus, while increasing $\sigma_8$ helps to
reconcile this model with the high-$\ell$ part of the TT spectrum, it
produces a very large polarization signal at small $\ell$, which is
inconsistent with that measured by WMAP.

The $\rm H_2$ and Standard models in the WMAP PL cosmology also do not
agree well with the WMAP TT-spectrum, due to their optical depths
being too low.

It is very interesting that the gradual reionization model (the $\rm
H_2WeakFB-VMS$ model with $f_{\rm esc}=0.15$; light blue line in
Fig.~\ref{fig:CMB}) shows a unique behaviour in the EE-spectrum
compared with the $\rm H_2WeakFB$ and instant reionization models,
while the optical depths of these three models are similar.  In
particular, this model shows a feature (a bump) at $\ell \approx
20$. The location of the $\ell \approx 8$ peak and the width of this
peak are slightly different from other models of comparable optical
depth (i.e.  the width is slightly greater and the peak location is
shifted to the right).  These differences are due to the fact that the
reionization process occurs over a longer period in this model. On the
other hand, there are only small differences in TT and TE spectra. We
expect the EE-spectrum to be the best probe of the details of
reionization such as the duration and the epoch~\cite{Holder}.

None of the models in the WMAP RSI cosmology fits the WMAP CMB data,
since we employ the $\sigma_8$ normalization here
(i.e. $\sigma_8=0.83$ in all of the RSI models plotted), and the
optical depths of all of the models are very small.

The calculations here do not include the effects of patchiness in the
reionization process. Patchy reionization can produce secondary
anisotropies in the CMB on small angular scales, $\ell \gsim 3000$
(Paper~I). These anisotropies must be studied in detail using
numerical simulations which provide the spatial distribution and
peculiar motions of the ionized gas. We defer this work to a future
paper.

\section{Discussion}
\label{sec:disc}

We have described the most detailed and physically realistic
semi-analytic calculation of the reionization of the Universe that has
been carried out to date. These calculations rely on a physical model
of galaxy formation which includes cooling through Compton and
molecular hydrogen channels, a physically motivated model of star
formation and feedback, and the feedback exerted on galaxy formation
by the process of reionization itself.

We find that, in the WMAP PL cosmological model, the Universe can be
reionized by $z\approx 14$--$15$ either if early generations of stars
are preferentially very massive or if feedback in the high-redshift
Universe is much weaker than it is today. A direct measurement of the
reionization redshift is not yet possible. Instead, we can compute,
from our reionization histories, optical depths for electron
scattering which can be compared directly with the measurement of this
quantity from WMAP. Reionization at these redshifts produces optical
depths in the range $\tau=0.19$--$0.21$. This is consistent with the
WMAP result of $\tau=0.17 \pm 0.04$. In the WMAP RSI cosmology
however, reionization cannot occur before $z\approx 9$, due to the
much smaller number of massive haloes which are able to form at high
redshifts in this model. As such, the WMAP RSI cosmological model is
barely consistent with the WMAP result for $\tau$, achieving at most
$\tau \approx 0.1$. This suggests one of three possibilities:
\begin{enumerate}
\item the WMAP estimate of $\tau$ is too high (either due to
systematic effects or a simple statistical fluctuation);
\item the RSI cosmology is incorrect (perhaps due to problems in
determining the normalization of the Lyman-$\alpha$ forest power
spectrum; see, for example, \pcite{viel04})
\item our understanding of galaxy formation, or the CDM hierarchical
model itself, are incorrect at high redshifts.
\end{enumerate}

Option (iii) is almost certainly true to some degree---there are
clearly large gaps remaining in our understanding of galaxy
formation. We include within option (iii) the possibility that some
neglected galaxy formation physics, for example a contribution to the
ionizing emissivity from super-massive black holes forming in high
redshift galaxies \cite{rico04}, affects our results. However, in the
WMAP RSI cosmology, the basic problem is that there are simply too few
haloes existing at high redshifts to allow enough galaxies to form and
so cause reionization. Only if 100\% of the baryons which have the
potential to cool in high redshift haloes are allowed to turn
instantly into stars can a sufficiently large $\tau$ be attained in
the WMAP RSI cosmology \cite{hh03,rs03}. Within our models, even with
rather extreme assumptions about the strength of feedback, the IMF of
early generations of stars, and the escape fraction of ionizing
photons, we are unable to achieve reionization prior to $z=9$ for the
WMAP RSI cosmology. Thus, the WMAP RSI cosmology seems hard to
reconcile with our understanding of galaxy formation in the CDM
hierarchical model.

The recent three year data release from the WMAP satellite confirms (i) above. Spergel et al. (2006) report $\tau = 0.093\pm
0.029$. This optical depth is well within the reach of our models with $f_{\rm esc}=1$, although physically motivated escape
fractions (see Appendix~\ref{app:fesc}) would result in models still struggling to reach such large optical depths.

Using the model of \scite{baugh05}, which was designed to match the
properties of low and high-redshift galaxies, we find that
reionization can occur by $z\approx 12$ for $f_{\rm esc}=1$, producing
an optical depth of $\tau=0.11$ which is marginally consistent with
the WMAP data. This optical depth can be increased to $\tau=0.12$ by
including \H2\ cooling (which does not affect the ability of this
model to match local galaxy data), and can be increased further if the
strength of feedback is reduced (which \emph{would} affect the ability
of the model to match local galaxy data).

Recently, the Boomerang experiment has provided an improved
measurement of the cosmological parameters \cite{boom03}. We have
computed reionization models using these revised cosmological
parameters (specifically, we use the ``CMBall+B03+LSS'' parameter set
from \scite{boom03}, as listed in Table~\ref{tb:cosmos}). We find that
at high redshifts ($\gsim 20$) the Boomerang cosmology produces a
lower \HII\ filling factor than the WMAP PL cosmology (for all of our
different galaxy formation model parameter sets). However, this
difference becomes less towards lower redshifts, and we find that the
epoch of reionization in the Boomerang cosmology is usually very close
to that for the corresponding model in the WMAP PL cosmology. The one
exception is for the H$_2$WeakFB-VMS model, in which reionization
occurs significantly later in the Boomerang cosmology (since
reionization happens early in the H$_2$WeakFB-VMS model, the
differences between the Boomerang and WMAP cosmologies are more
pronounced). Nonetheless, double reionization can still occur in our
model, even in the Boomerang cosmology. We conclude, therefore, that
the conclusions we present for the WMAP cosmology can also be applied
to the revised cosmological parameter set suggested by the Boomerang
data.

\begin{figure}
\psfig{file=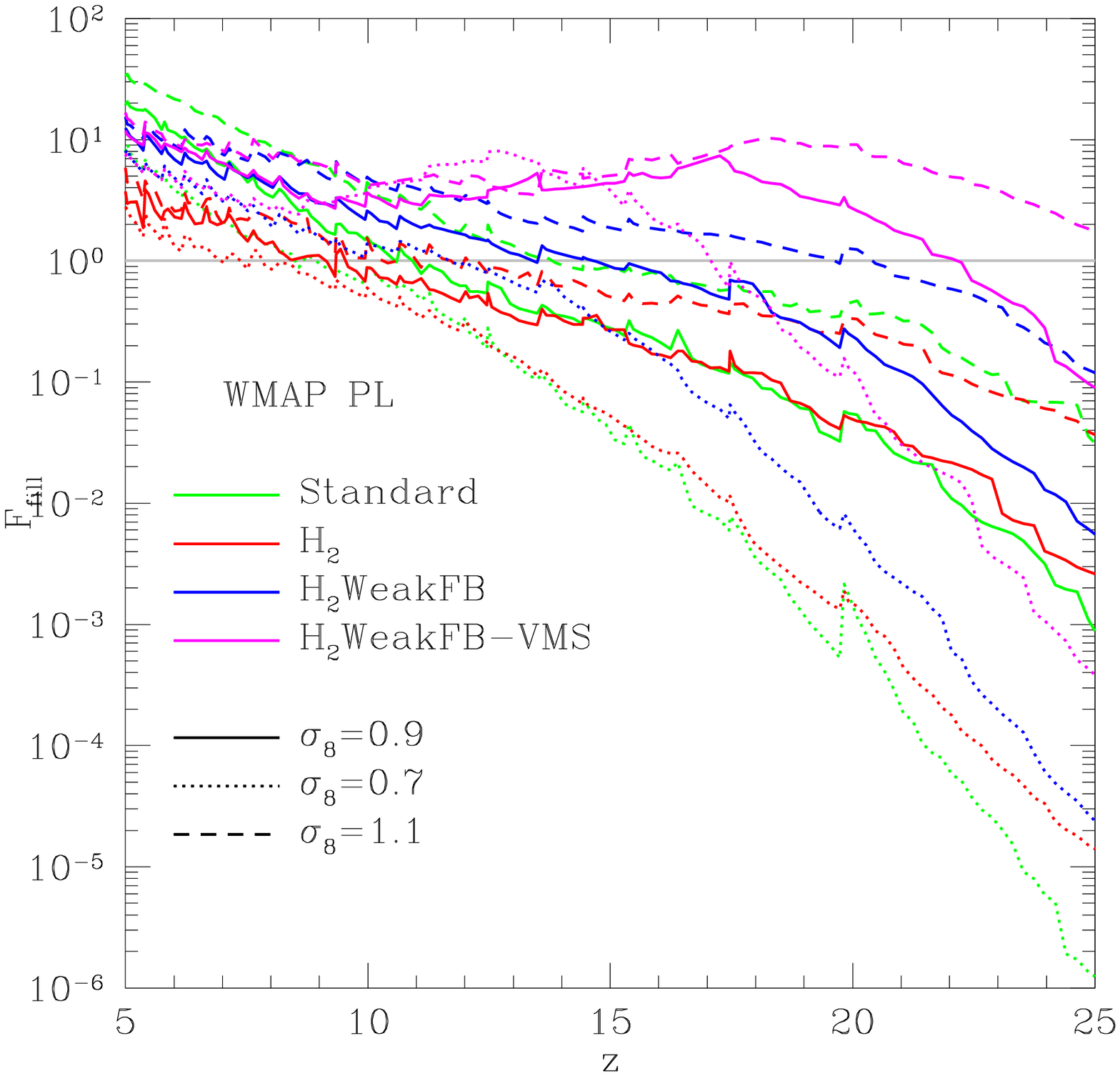,width=80mm}
\caption{The \HII\ filling factor as a function of redshift in various
models (distinguished by line colour---see key in figure) for
different values of $\sigma_8$ (distinguished by line type---see key
in figure). We show results for the WMAP PL cosmology only---results
for other cosmologies are qualitatively similar. All results are shown
for an escape fraction of 100\%.}
\label{fig:sigma8}
\end{figure}

Throughout this work we have used the values of $\sigma_8$ which best fit the WMAP data ($\sigma_8=0.90$ and $0.83$ for the PL and
RSI models respectively, \pcite{spergel03}). However, the value of $\sigma_8$ remains somewhat uncertain, with plausible values
lying between about $0.7$ to $1.1$. In particular, \scite{spergel06} report $\sigma_8 = 0.76\pm 0.05$. We have re-run our WMAP PL
models using these two extreme values of $\sigma_8$. Results for the ionized filling factor are shown in
Fig.~\ref{fig:sigma8}. Not surprisingly, lowering $\sigma_8$ reduces the redshift of reionization, while increasing $\sigma_8$
increases it. The effect is greater for models in which reionization happens at higher redshifts, since here those haloes which
are able to form galaxies and emit ionizing photons correspond to very rare fluctuations in the initial Gaussian density field. As
such, their abundance is very sensitive to the normalization of that density field. Typically, models which reionize at $z\approx
10$ have their epoch of reionization reduced/increased by $\Delta z\approx 1$ for a reduction/increase of $0.2$ in $\sigma_8$,
while models with reionization at $z\approx 20$ experience a $\Delta z$ of 4--5 for the same change in $\sigma_8$ (these values of
$\Delta z$ do depend also on the parameters of the galaxy formation calculation). Thus, $\sigma_8$ can have a large effect on
predictions of the epoch of reionization. 

These results show that, even for the low $\sigma_8$ reported by \scite{spergel06} sufficiently early reionization can be achieved
(i.e. to match the optical depth also reported by \pcite{spergel06}) providing $f_{\rm esc}=100\%$.

Most of the results we present here are based on the assumption that
the fraction of ionizing photons which escape from galaxies into the
IGM is $f_{\rm esc}=1$. In the local Universe, and at $z\sim 3$,
observational estimates give much smaller values. If $f_{\rm esc}$
were low at higher redshifts also, then our estimates of $z_{\rm
reion}$ would be substantially too high. A value of $f_{\rm esc}=0.1$
would reduce the reionization redshift in even our best models to
$z\lsim 10$. In Appendix~\ref{app:fesc} we present reionization
histories computed using a simple physically-motivated model for the
escape fractions. Clearly, a better theoretical understanding of
$f_{\rm esc}$ is crucial if we are to use measures of the reionization
epoch to put interesting constraints on the process of galaxy
formation.

We have found that extended periods of partial reionization and double
reionization can occur in models in which early generations of stars
formed via \H2\ cooling are very massive (with enhanced ionizing
luminosities relative to a standard IMF). Such models are able to
delay full reionization until $z\approx 6$--7, while achieving an
electron scattering optical depth large enough to be consistent with
the results from WMAP. While a more extensive study of such models
would be required to determine whether or not they can match all of
the current observational constraints, they appear to be a promising
route to reconciling the apparently discrepant estimates of the
reionization redshift from the CMB polarization measured by WMAP and
from the Gunn-Peterson absorption detected in the spectra of $z\sim 6$
quasars. It should be noted, however, that these models rely on the
assumption of very weak feedback from supernovae. Within CDM
cosmogonies, it has long been argued that feedback from supernovae
must be very effective in order to reduce the number of low-luminosity
galaxies at low redshifts which are otherwise produced (see
\pcite{benson03} and references therein). Thus, if double reionization
(or an extended period of partial reionization) is required to meet
observational constraints, then it appears that the strength of
supernova feedback must vary with redshift, or else some other
feedback mechanism must act at low redshifts to reduce the number of
faint galaxies formed. This underlies the urgent need for the
inclusion of physical recipes for feedback into semi-analytic models
of galaxy formation, rather than the phenomenological feedback rules
currently employed.

In conclusion, the WMAP RSI cosmology is very difficult to reconcile with our current understanding of galaxy formation (although
a wide parameter space of models remains to be searched) if the optical depth were as large as suggested by the first year WMAP
data. The WMAP PL model, in contrast, is able to achieve a sufficiently high optical depth with reasonable assumptions about the
nature of galaxy formation at high redshifts. The three year WMAP data imply an optical depth significantly lower than that found
in the first year data (i.e. $\tau = 0.093\pm 0.029$; \pcite{spergel06}). Reionization can then be achieved sufficiently early in
both the WMAP PL and WMAP RSI cosmologies, providing $f_{\rm esc}$ is close to 1 and $\sigma_8$ is not too low. We will present a
more detailed analysis of the implications of the 3-year WMAP data in a follow-up paper.

In principle, the value of the optical depth to reionization is one of the few strong constraints on galaxy formation at high
redshifts. The uncertainty mentioned in option (i) has already been reduced by the WMAP three year data.Thus, the ability to
reionize the Universe sufficiently early will now become one of the key tests that any model of galaxy formation must pass before
it can be considered to be viable.

\section*{Acknowledgements}

AJB acknowledges support from a Royal Society University Research
Fellowship. NS is supported by a Grant-in-Aid for Scientific Research
from JSPS (No. 17540276). This work was also supported by the PPARC
rolling grant for extragalactic astronomy and cosmology at Durham. We
wish to thank Carlton Baugh, Shaun Cole and Carlos Frenk for
permitting us to use the {\sc galform} model of galaxy formation in
this work. We also acknowledge helpful suggestions from an anonymous
referee.

\appendix

\section{Revised Cooling Model}
\label{sec:newcool}

In this Appendix, we discuss how we incorporate cooling by molecular
hydrogen and Compton cooling into our model and how we apply the
effects of the filtering mass.

\subsection{Cooling by Molecular Hydrogen}
\label{sec:H2cool}

We incorporate cooling due to molecular hydrogen in our model
following the algorithms of \scite{yoshi03}. Below, we give a brief
account of our calculation of the \H2\ contribution to the cooling
rate. The reader is referred to \scite{yoshi03} and references therein
for a full description.

We begin by estimating the fraction, $f_{\rm H_2}$, of molecular
hydrogen which would be present if there were no background of
\H2-dissociating radiation from stars. As shown by \scite{tegmark97},
at the redshifts of interest here, \H2 is formed mostly via the
formation of ${\rm H}^-$ through the reaction ${\rm H} + {\rm e}^-
\rightarrow {\rm H}^-$, with rate coefficient
\begin{equation}
k_{\rm H^-} = 1.83\times 10^{-18} (T/{\rm K})^{0.88} \hbox{ cm}^3 \hbox{s}^{-1}
\label{eqn:rate1}
\end{equation}
\cite{hutchins76}. The rate of formation of ${\rm H}^-$ depends on the
abundance of free electrons, which depends on recombinations ${\rm
H}^+ + {\rm e}^- \rightarrow {\rm H} + \gamma$ with rate coefficient
\begin{equation}
k_1 = 1.88\times 10^{-10} (T/{\rm K})^{-0.64} \hbox{ cm}^3\hbox{s}^{-1}
\label{eqn:rate2}
\end{equation}
\cite{hutchins76}. (We ignore photoionization and collisional
ionization of hydrogen in this part of the calculation, since cooling
by \H2 is only important in low-temperature haloes at high redshifts.)
\scite{tegmark97} solved the rate equations for electron recombination
and \H2 formation to obtain a characteristic \H2 abundance $f_{\rm
H_2,c}$ for gas with hydrogen number density $n_{\rm H}$ and
temperature $T$ at redshift $z$:
\begin{eqnarray}
f_{H_2,c}  &=& 3.5\times 10^{-4} T_3^{1.52} \left[1 + (7.4\times 10^8 /n_{\rm H,1}) \right. \nonumber \\
&& \left. \times (1+z)^{2.13} \exp(-3173/[1+z]) \right]^{-1}, 
\end{eqnarray}
where $n_{\rm H,1}$ is the hydrogen number density in units of
cm$^{-3}$ and $T_3$ is temperature in units of 1000K. (The above
result depends on the values of various rate coefficients, and is
consistent with the rate coefficients given in eqns.~\ref{eqn:rate1},
\ref{eqn:rate2} and \ref{eqn:rate3}.) The factor in square brackets
corresponds to photodissociation of ${\rm H}^-$ by CMB photons, but is
important only for $z\gsim 100$. For our own estimate of the \H2
abundance in the absence of a photodissociating background from stars,
we then use
\begin{eqnarray}
f_{\rm H_2,0} & = & \hbox{max} \left[ f_{\rm H_2,prim}, f_{\rm H_2,c} \right], 
\end{eqnarray}
where $f_{\rm H_2,prim}$ is the primordial \H2 fraction, which we take
to be $3.5\times 10^{-7}$ \cite{ann96} (although our results are
highly insensitive to this choice of value).

Next, we account for the effects of a background of \H2-dissociating
radiation from stars. If we assume equilibrium between \H2 production
via ${\rm H}^-$ and \H2 destruction by photodissociation, we obtain an
\H2 abundance
\begin{equation}
f_{\rm H_2,eq} = k_{\rm H^-} x_{\rm e} n_{\rm H} / k_{\rm diss},
\label{eq:fH2}
\end{equation}

We calculate the electron fraction, $x_{\rm e}$, appearing in eqn.~(\ref{eq:fH2}) from
\begin{equation}
x_{\rm e} = {x_{\rm e,0} \over 1 + x_{\rm e,0}t_{1/2}/\tau_{\rm rec}},
\label{eq:efrac}
\end{equation}
based on eqn.(6) from \cite{tegmark97}, where
\begin{equation}
\tau_{\rm rec} = 1/(k_1 n_{\rm H}).
\end{equation}
This estimate of $x_{\rm e}$ assumes that recombinations have been
proceeding in gas at its present density and temperature for a time
$t_{1/2}$ equal to half the age of the Universe at redshift $z$,
starting from the primordial post-recombination electron fraction
$x_{\rm e,0}$. We calculate $x_{\rm e,0}$ using {\sc recfast}
\cite{recfast}. Eqn.~(\ref{eq:efrac}) accounts crudely for the age of
the halo, and therefore the time available for recombinations. The
electron fraction is never allowed to fall below the value
corresponding to collisional ionization equilibrium at the specified
temperature. Note that, post-reionization, the electron fraction
should in fact tend towards the equilibrium value for photoionization
equilibrium (as opposed to collisional ionization equilibrium). Our
current model does not account for this fact, but this is unimportant
here as the determination of electron fractions in the
post-reionization Universe does not affect our determination of the
epoch of reionization.

For the photodissociation rate in eqn.~(\ref{eq:fH2}), we assume
\begin{equation}
k_{\rm diss} = 1.1\times 10^{-13} J_{21} F_{\rm shield} \hbox{ s}^{-1},
\label{eqn:rate3}
\end{equation}
\cite{macha01}, where $J_{21}$ is average flux in units of
$10^{-21}$ergs/cm$^2$/s/Hz in the Lyman-Werner bands (with photon
energy around 12.87~eV), and is computed from the radiation background
tracked by our model of IGM evolution \cite{benson02}. We account for
the effects of self-shielding using the algorithm of
\scite{yoshi03}. Specifically, we estimate the shielding factor
\begin{equation}
F_{\rm shield} = \hbox{min}\left[ 1 , \left( {N_{\rm H_2}\over 10^{14}\hbox{cm}^{-2} } 
\right)^{-3/4} \right],
\end{equation}
which depends on the \H2 column density $N_{\rm H_2}$. In order to
avoid the computational cost and complexity of solving iteratively for
the radial abundance profile of \H2 within each halo, we estimate this
shielding factor using the abundance $f_{\rm H_2,0}$ computed in the
absence of a photodissociating background, using
\begin{equation}
N_{\rm H_2} = C f_{\rm H_2,0} \, {\overline n}_{\rm H,vir} \, R_{\rm vir}
\end{equation}
with $C=0.2$, where $R_{\rm vir}$ is the virial radius of the dark
matter halo and ${\overline n}_{\rm H,vir}$ is the mean hydrogen
density within $R_{\rm vir}$. (For a more detailed justification of
this approach, see \scite{yoshi03}.)

We then calculate the final \H2 abundance used in our cooling calculation from
\begin{equation}
f_{\rm H_2} = \hbox{min} \left[ f_{\rm H_2,0}, f_{\rm H_2,eq} \right] \exp\left(-{T \over 51,920{\rm K}}\right).
\end{equation}
We have introduced an exponential cut-off in the molecular hydrogen
fraction at temperatures corresponding to the dissociation energy of
\H2, to account for collisional dissociation of
\H2. (\scite{tegmark97} ignored this because they were interested only
in the relatively low temperature regime. The exact form and position
of the cut-off are unimportant, as the cooling function is dominated
by atomic processes at the temperatures for which collisional
dissociation is important.)

Finally, we compute the cooling timescale due to molecular hydrogen
using\footnote{Using the more recent estimate of the \H2\ cooling
function from \scite{galpal98} makes only small differences to our
results.} \cite{tegmark97}
\begin{eqnarray}
\tau_{\rm H_2} & = & 48,200\hbox{ yr} \left( 1 + {10 T_3^{7/2} \over 60 + T_3^4} \right)^{-1} \nonumber \\ & & \times \exp(512{\rm
 K}/T)(f_{\rm H_2}n_{\rm H,1})^{-1}.
\end{eqnarray}
The net radiative cooling timescale for gas, $\tau_{\rm cool}$ is then
given by
\begin{equation}
\tau_{\rm cool}^{-1} = \tau_{\rm H_2}^{-1} + \tau_{\rm other}^{-1},
\end{equation}
where $\tau_{\rm other}$ is the cooling timescale due to all other cooling mechanisms (e.g. atomic processes, Compton cooling).

\scite{yoshi03} recommend the inclusion of a ``dynamical heating''
term which accounts for the accretion of new material onto a
halo. This accretion is assumed to heat the gas already in the halo
(as the kinetic energy of the infalling material is converted into
thermal energy), reducing the rate at which it can
cool. \scite{yoshi03} find that the inclusion of this term in their
analytic model of gas cooling in haloes produces good agreement with
the results of their gas-dynamical calculations. We have experimented
with including such a term in our calculations. We note, however, that
our semi-analytic model already effectively includes such dynamical
heating. Whenever a halo's mass doubles we consider the halo to have
reformed \cite{cole00}. The temperature of the gas in the halo is then
recomputed, and the cooling clock is reset to zero. It is simple to
show that this results in an effective heating rate for the gas which
is comparable to that supplied by the dynamical heating term of
\scite{yoshi03}. For example, following the approach of
\scite{yoshi03} and taking the virial temperature of a halo to be
given by
\begin{equation}
T = \gamma M^{2/3}(1+z),
\label{eq:Tvir}
\end{equation}
where $\gamma$ is a normalization constant, we find by differentiation that
\begin{equation}
\dot{T} = {2 \over 3} \gamma {(1+z) \over M^{1/3}} \left( \dot{M} + {3 \over 2} M {\dot{z} \over 1+z}\right).
\end{equation}
(Note that \scite{yoshi03} include only the first of the two terms on
the right-hand side of this equation. However, we expect them to be of
similar magnitude.)  In the semi-analytic calculation, the virial
temperature is also given by eqn.~(\ref{eq:Tvir}) and is updated after
each halo reformation event. Consequently, when averaged over a halo
lifetime, the amount of dynamical heating in the semi-analytic model
must approximately equal that proposed by \scite{yoshi03}.

Not surprisingly therefore, we find that the explicit inclusion or
otherwise of a dynamical heating term (of the form proposed by
\pcite{yoshi03}) makes no significant difference to the results of our
calculations.

While in primordial plasma \H2\ must form in the gas phase via the
H$^-$ channel described above, in enriched gas there is the
possibility of \H2\ formation on the surfaces of dust grains. While we
do not include this formation mechanism in our cooling model we have
checked that it makes no significant difference to the evolution of
the filling factor. To do this, we used the methodology of
\scite{cazaux} to compute the ratio of \H2\ formation rates on dust
grains and in the gas phase. We then scaled the abundance of \H2\ in
our cooling model appropriately and computed a selection of our models
using this revised cooling model. We found no significant change in
the evolution of the \H2\ filling factors.

\subsection{Compton Cooling}
\label{sec:Compt}

The Compton cooling timescale does not explicitly depend on the gas
density, but does depend on the electron abundance, and therefore on
the ionization state of the gas. The electron abundance, in turn,
\emph{does} depend on the gas density, due to the density dependence
of the electron recombination timescale.

Prior to halo formation, the electron abundance in the IGM gas is
computed from our model of the ionization state of the IGM. Prior to
IGM reionization, the electron abundance is low, corresponding to that
left over from recombination. Post-reionization, the electron fraction
becomes large as the Universe is almost entirely ionized. During and
after halo formation, the electron fraction will move towards the
appropriate equilibrium value as described in \S\ref{sec:H2cool}. The
Compton cooling timescale is then computed using the recombined
electron abundance from eqn.(\ref{eq:efrac}), or the equilibrium
electron abundance, whichever is the larger.

\subsection{Filtering Mass}
\label{app:filt}

The filtering mass is similar to the Jeans mass, but is the relevant
quantity for describing the effect of pressure on the gravitational
collapse of gas in an expanding Universe with a time-varying temperature. It
is defined by
\begin{equation}
M_{\rm F} = (4\pi/3)\rho_0(2\pi a/k_{\rm F})^3,
\end{equation}
where $\rho_0$ is the mean density of the Universe, $a$ is the expansion factor and
\begin{equation}
{1 \over k_{\rm F}^2(t)} = {1 \over D(t)} \int_0^t {\rm d}t^\prime a^2(t^\prime) 
{\ddot{D}(t^\prime)+2H(t^\prime)\dot{D}(t^\prime)\over k_{\rm J}^2(t^\prime)} 
\int_{t^\prime}^t {{\rm d}t^{\prime\prime} \over a^2(t^{\prime\prime})},
\end{equation}
where $D(t)$ is the linear theory growth factor, $H(t)$ is the Hubble
variable and $k_{\rm J}(t)$ is the Jeans wavenumber, defined in the
usual way. The collapse of gas into halos with masses below $M_{\rm
  F}$ is greatly suppressed relative to the case of zero pressure.

\section{Escape Fractions}
\label{app:fesc}

\begin{figure*}
\begin{center}
\begin{tabular}{cc}
\psfig{file=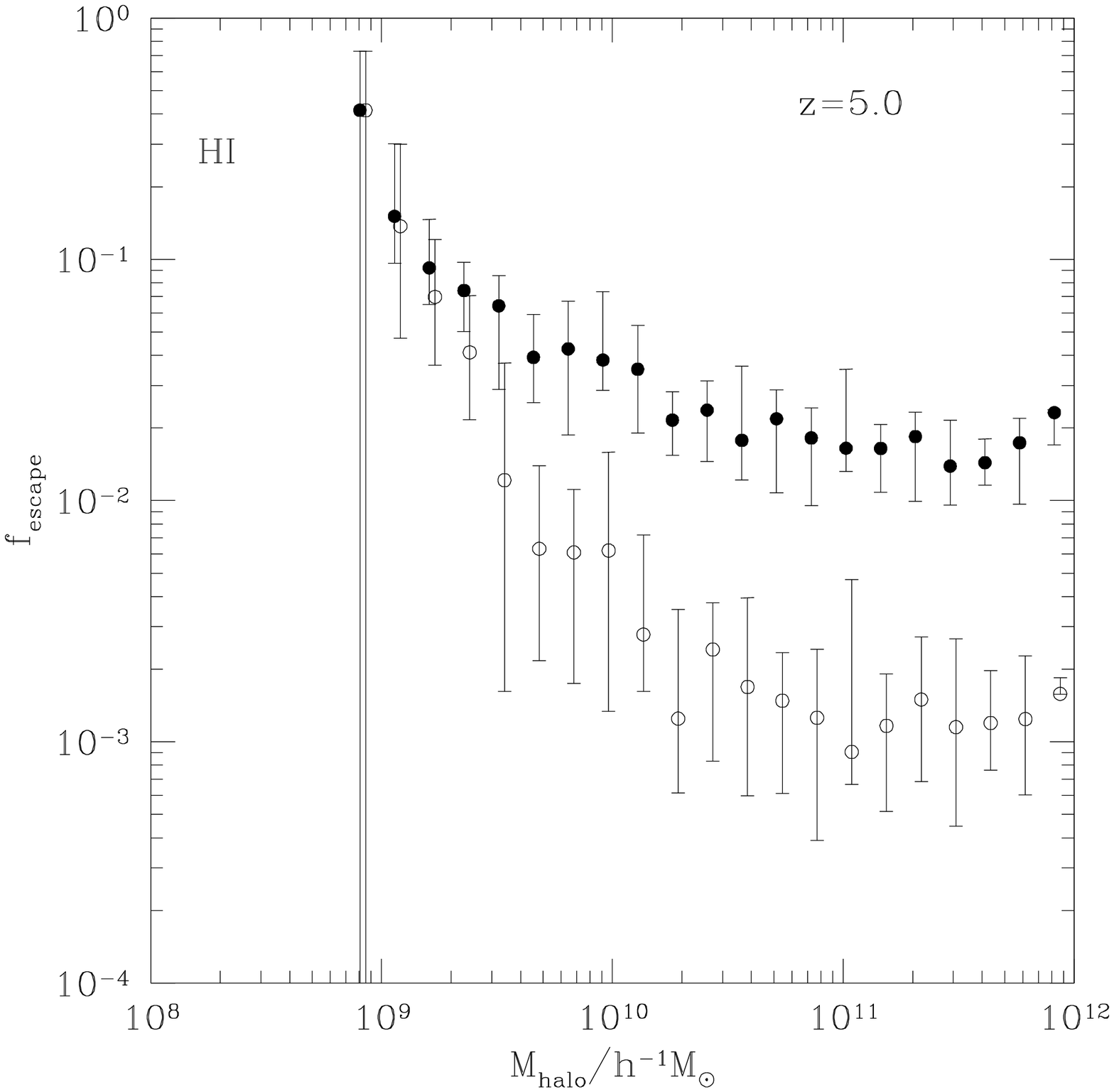,width=80mm} 
& \psfig{file=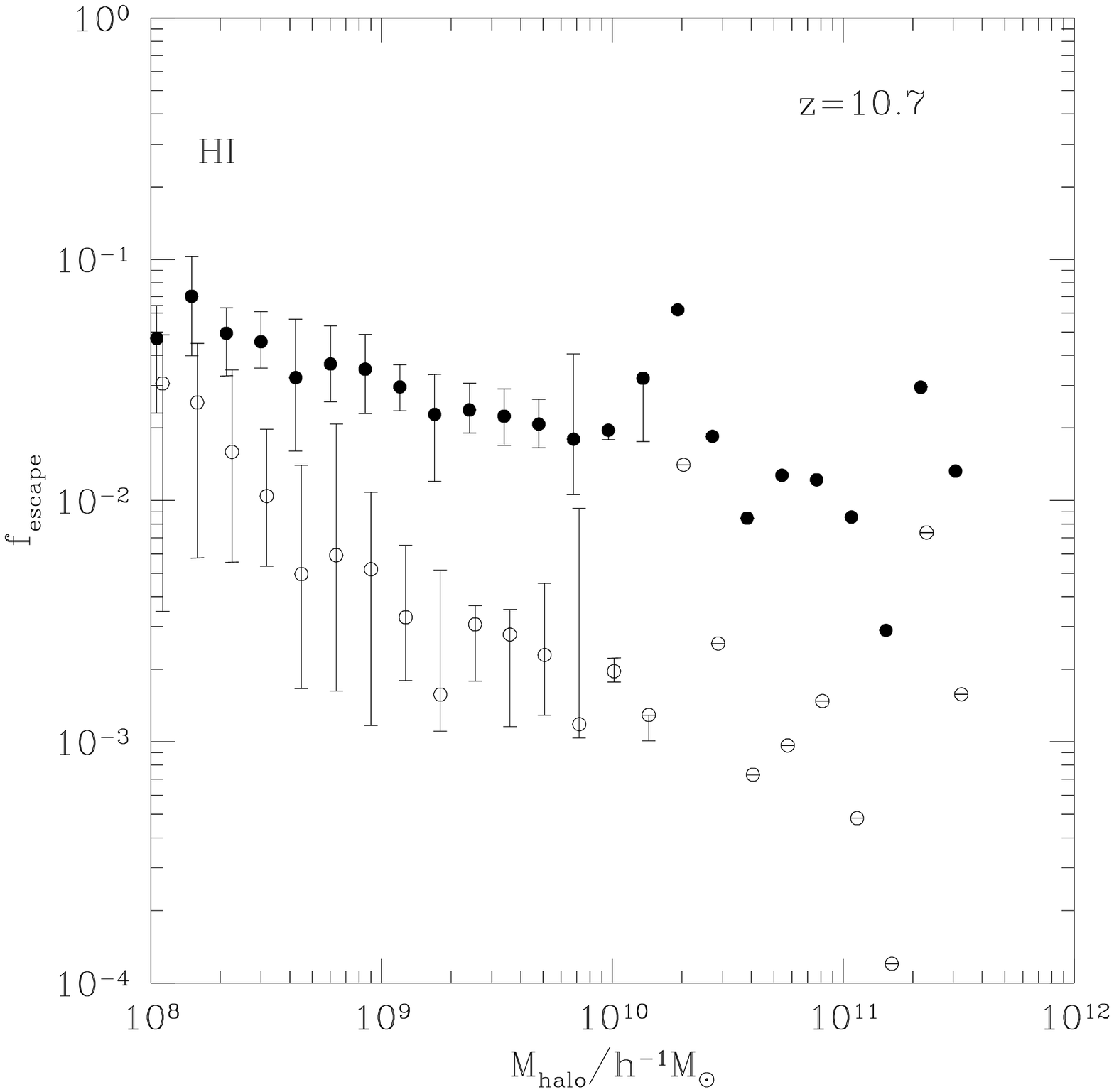,width=80mm}
\end{tabular}
\end{center}
\caption{Ionizing luminosity-weighted mean \HI\ escape fractions for
haloes as a function of halo mass at $z=5$ (left-hand panel) and
$z=10.6$ (right-hand panel) for the Standard model in the WMAP RSI
cosmology. The points show the median escape fraction for all haloes
of this mass, while the error bars show the upper and lower quartiles
of the distribution. Filled points include only the effects of
absorption by neutral gas, while open points also include the effects
of dust extinction.}
\label{fig:fesc}
\end{figure*}

Here, we apply simple, physical models to predict the escape fraction
for each galaxy in our model.  We follow Paper~I and use the model of
\scite{ds94} (and the development thereof carried out in Paper~I) to
compute the amount of ionizing radiation absorbed by the gas in each
galaxy. The resulting escape fractions are shown in
Table~\ref{tb:fesc} and in Fig.~\ref{fig:fesc}.

\begin{table}
\caption{Ionizing luminosity-weighted escape fractions for galaxies at
two different redshifts ($z=5$ and 11) are given for all cosmologies
and models. Redshift is specified in the first column, with cosmology
and model specified in columns 2 and 3. Column 4 specifies the escape
fractions for \HI\ ionizing radiation when dust is ignored, while
column 5 specifies the same escape fractions when dust extinction is
included.}
\label{tb:fesc}
\begin{center}
\begin{tabular}{cllcc}
\hline
 & & & \multicolumn{2}{c}{\boldmath{$f_{\rm esc}$}}\\
\boldmath{$z$} & {\bf Cosmology} & {\bf Model} & {\bf No dust} & \bf {With dust}  \\
\hline
11 & WMAP RSI & Standard & 0.06 & 0.019 \\
 5 & WMAP RSI & Standard & 0.03 & 0.005 \\
11 & WMAP RSI & MetFree & 0.05 & 0.012 \\
 5 & WMAP RSI & MetFree & 0.03 & 0.005 \\
11 & WMAP RSI & TopHeavy & 0.05 & 0.006 \\
 5 & WMAP RSI & TopHeavy & 0.02 & 0.002 \\
11 & WMAP RSI & WeakFB & 0.02 & 0.002 \\
 5 & WMAP RSI & WeakFB & 0.02 & 0.002 \\
11 & WMAP RSI & H$_2$ & 0.06 & 0.016 \\
 5 & WMAP RSI & H$_2$ & 0.05 & 0.011 \\
11 & WMAP RSI & H$_2$WeakFB & 0.06 & 0.007 \\
 5 & WMAP RSI & H$_2$WeakFB & 0.02 & 0.002 \\
11 & WMAP RSI & H$_2$WeakFB-VMS & 0.04 & 0.001 \\
 5 & WMAP RSI & H$_2$WeakFB-VMS & 0.03 & 0.004 \\
11 & WMAP PL & Standard & 0.03 & 0.005 \\
 5 & WMAP PL & Standard & 0.03 & 0.004 \\
11 & WMAP PL & MetFree & 0.03 & 0.004 \\
 5 & WMAP PL & MetFree & 0.04 & 0.008 \\
11 & WMAP PL & TopHeavy & 0.03 & 0.001 \\
 5 & WMAP PL & TopHeavy & 0.02 & 0.001 \\
11 & WMAP PL & WeakFB & 0.02 & 0.001 \\
 5 & WMAP PL & WeakFB & 0.03 & 0.001 \\
11 & WMAP PL & H$_2$ & 0.05 & 0.014 \\
 5 & WMAP PL & H$_2$ & 0.07 & 0.022 \\
11 & WMAP PL & H$_2$WeakFB & 0.02 & 0.001 \\
 5 & WMAP PL & H$_2$WeakFB & 0.03 & 0.002 \\
11 & WMAP PL & H$_2$WeakFB-VMS & 0.02 & 0.001 \\
 5 & WMAP PL & H$_2$WeakFB-VMS & 0.03 & 0.002 \\
11 & Baugh et al. 05 & Standard & 0.05 & 0.036 \\
 5 & Baugh et al. 05 & Standard & 0.08 & 0.076 \\
11 & Baugh et al. 05 & H$_2$ & 0.08 & 0.066 \\
 5 & Baugh et al. 05 & H$_2$ & 0.13 & 0.130 \\
11 & Baugh et al. 05 & H$_2$WeakFB & 0.06 & 0.062 \\
 5 & Baugh et al. 05 & H$_2$WeakFB & 0.12 & 0.117 \\
11 & Baugh et al. 05 & H$_2$WeakFB-VMS & 0.07 & 0.069 \\
 5 & Baugh et al. 05 & H$_2$WeakFB-VMS & 0.12 & 0.120 \\

\hline
\end{tabular}
\end{center}
\end{table}

The mean escape fraction for \HI\ ionizing photons for galaxies due to
absorption by neutral gas are found to be a few percent. These numbers
agree reasonably well with the limited observational determinations of
escape fractions, e.g. \scite{leith95} find that less than 3\% of
ionizing photons escape from low redshift starburst galaxies, and
similar results have been obtained by \scite{tumlinson99}, while
\scite{steidel01} estimate escape fractions $\sim 10\%$ for
Lyman-break galaxies at $z=3.4$. More recently, \scite{bergvall06}
found an escape fraction of 4--10\% for a local starbursting dwarf
galaxy.

It should be noted that these escape fraction calculations are
extremely simplified---they model the interstellar medium (ISM) as a
smooth distribution of gas (a clumpy distribution could enhance the
escape fraction as noted by \pcite{wood00}) and ignore dynamical
effects (such as expanding superbubbles as considered by \scite{dsf00}
which significantly lower the escaping fraction). More detailed
calculations, employing hydrodynamical simulations, suggest that high
redshift dwarf galaxies might have much higher escape fractions
\cite{fujita03}. A better theoretical understanding of escape
fractions is a crucial missing component of accurate calculations of
reionization.

We estimate the effects of dust extinction on the ionizing radiation
escape fractions by following the methods of Paper~I. Briefly, we
extrapolate the Milky Way extinction curve to the wavelengths
corresponding to the ionization threshold of each species, \HI, \HeI\
and \HeII, and compute an extinction averaged over galaxy inclinations
using the calculations of \scite{ferrara99}. The results of including
dust extinction on escape fractions are shown in Table~\ref{tb:fesc}
and in Fig.~\ref{fig:fesc}. It is seen that dust can cause severe
(frequently an order of magnitude) attenuation in the escaping
fraction, at least in this highly simplified calculation of a smooth
ISM. Clearly, a more detailed analysis of the effects of dust on the
escaping fraction is urgently needed if we are to gain a better
understanding of how the IGM was reionized.

\begin{figure*}
\psfig{file=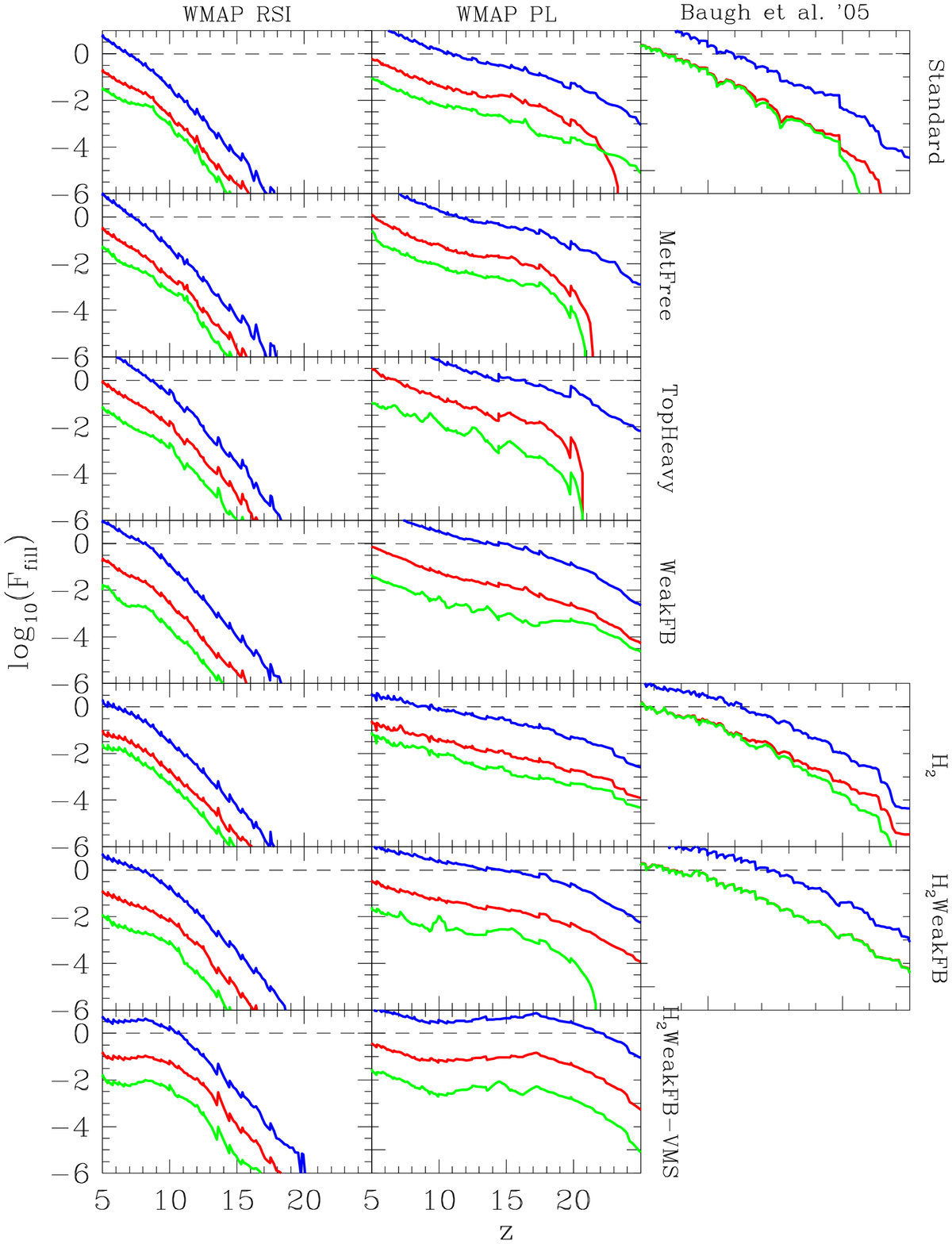,width=180mm}
\vspace{-20mm}
\caption{The \HII\ filling factor for all cosmologies and models for a
fixed escape fraction of $f_{\rm esc}=1$ (blue lines) and $f_{\rm
esc}$ computed using the model described in
Appendix\protect\ref{app:fesc} both for absorption by neutral gas only
(red lines), and for absorption by gas and dust (green lines).}
\label{fig:vfesc}
\end{figure*}

The effects on the ionized filling factor of using different
assumptions for the escape fraction are shown in Fig.~\ref{fig:vfesc},
for different models and different redshifts.  It is interesting to
note that, in the WMAP cosmologies, only the TopHeavy model in the
WMAP PL cosmology is able to achieve reionization by $z=6$, and then
only if the effects of dust on the escape fraction are ignored. In the
Baugh et al. '05 the majority of ionizing photons are produced during
bursts of star formation. As such, dust (assumed to be located in
galaxy disks in our simple model) has little effect on the escape
fraction. It should be noted that our model for the escape fraction is
likely to be very unreliable in cases where most of the ionizing
photons are produced in bursts rather than in quiescent disks.

\section{Tests of the Model}
\label{app:tests}

In this Appendix we present results of various tests of our model. We
begin by comparing to the results of Paper~I after which we check that
our methodology leads to model calculations that are suitably
converged.

\subsection{Comparison with Paper I Results}
\label{app:p1comp}

We begin by comparing our current results with those of Paper~I. Our
galaxy formation model, {\sc galform}, has undergone significant
development over that used in Paper~I. As such, we expect some
differences in the results obtained. To assess the degree of change in
our results we have performed the same calculations as in
Paper~I. Specifically, we adopted the same model parameters as in that
paper and computed the \HII\ filling factor as a function of
redshift. The models of Paper~I use cosmological parameters
$(\Omega_0,\Lambda_0)=0.3, 0.7$ and $\sigma_8=0.9$, while using
similar galaxy formation prescriptions and parameters as used in this
work. The Paper~I models are similar to our Standard model, with major
differences being a weaker feedback, $V_{\rm hot}=150$km/s as opposed
to 200km/s as used here, and a simpler model for galaxy merging than
used here (Paper~I uses the model of \scite{cole00} whereas the
current work uses that of \scite{benson04} which gives somewhat longer
merger timescales). Furthermore, in Paper~I a fraction of all star
formation was assumed to result in the formation of brown dwarfs, such
that the effective ionizing luminosity per unit mass of stars formed
was reduced by a factor of 0.65.

Calculations were performed for both the low and high baryon fraction
models used in Paper~I ($\Omega_{\rm b}=0.02$ and $0.04$
respectively).

\begin{figure*}
\begin{center}
\begin{tabular}{cc}
\psfig{file=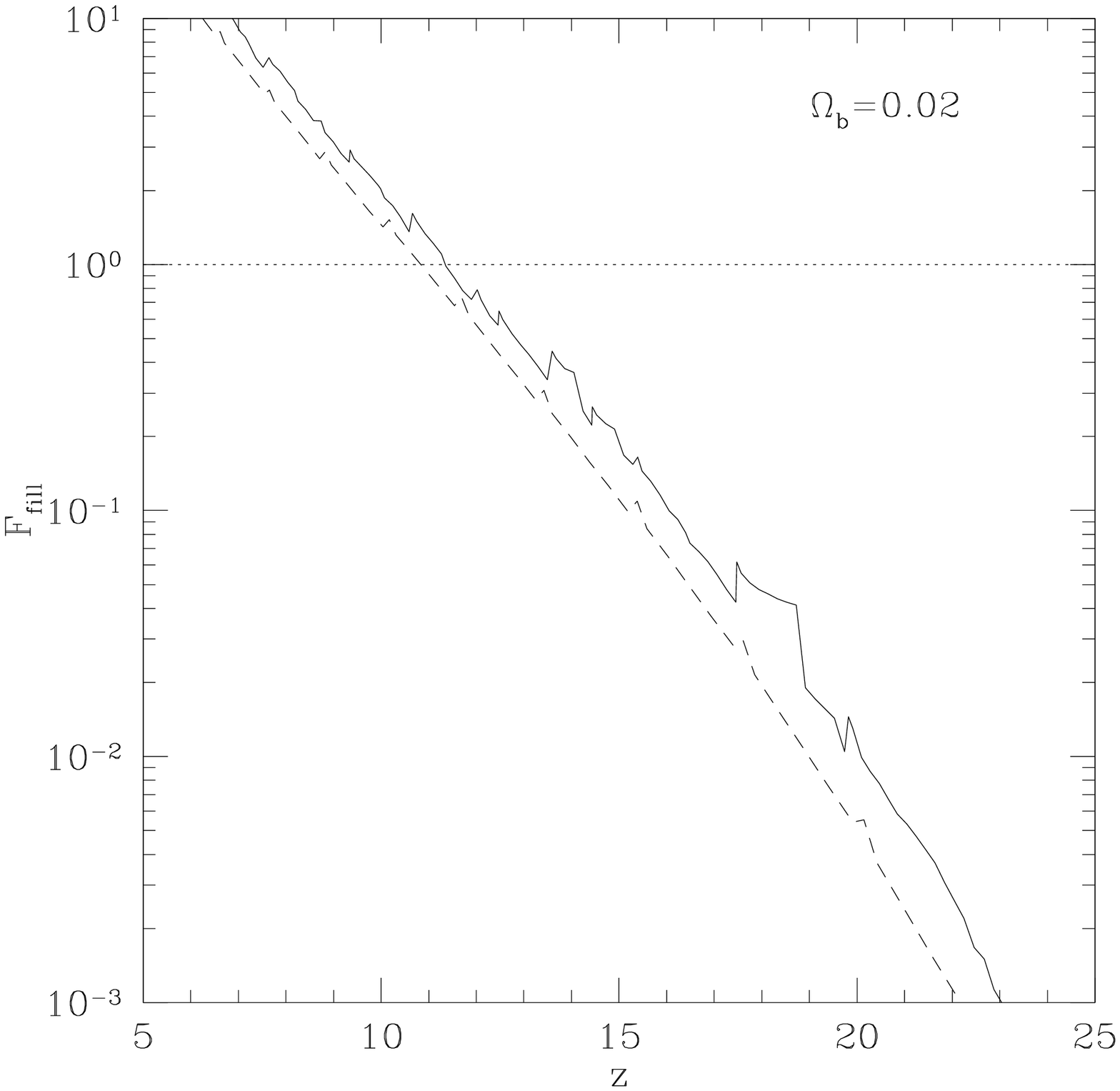,width=80mm} & \psfig{file=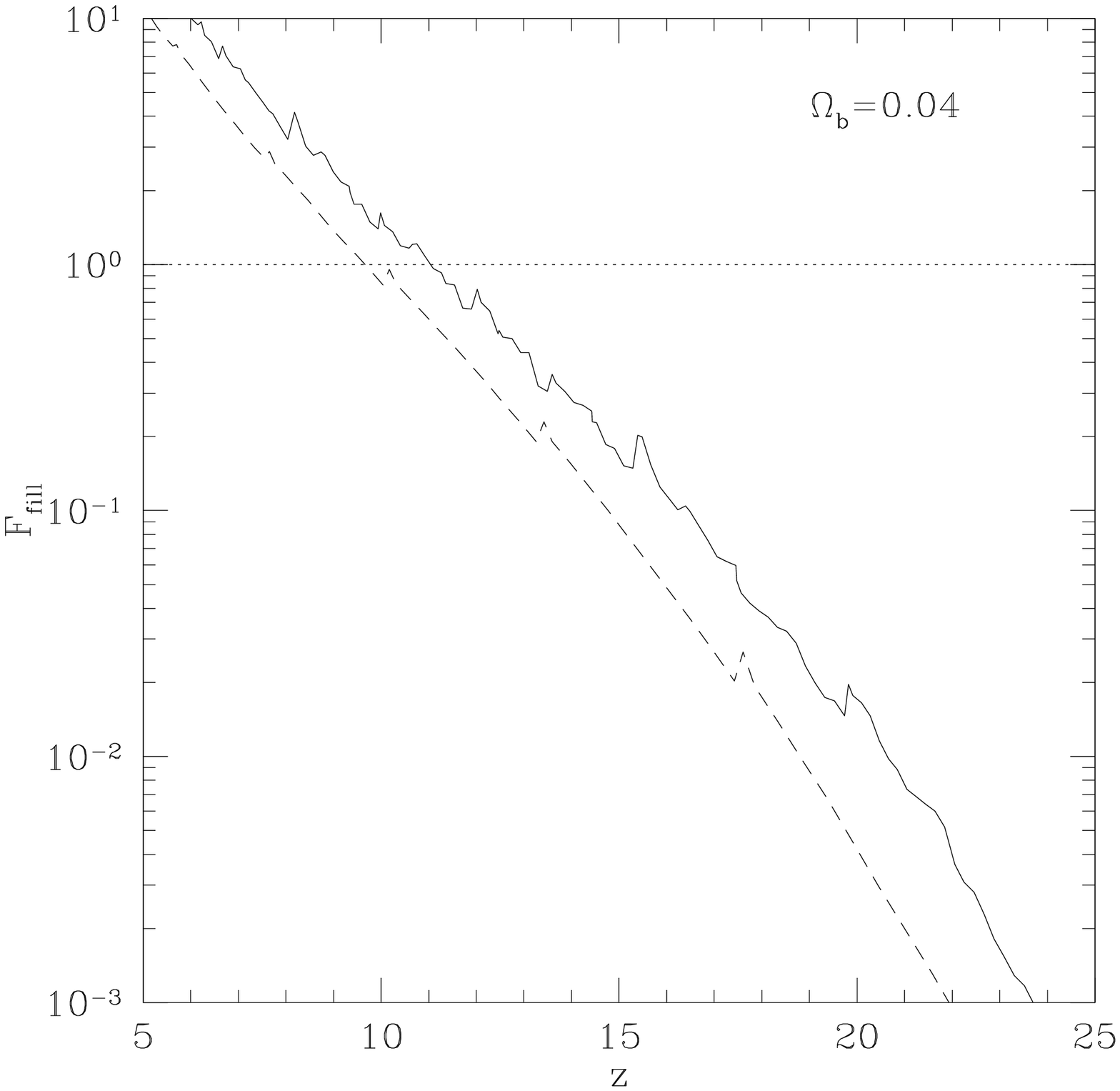,width=80mm} 
\end{tabular}
\end{center}
\caption{Comparison of \HII\ filling factors using the parameters of
Paper~I. Dashed lines show the results from Paper~I (for an escape
fraction of 100\% and no dust absorption). Solid black lines show the
results from this work using ionized spheres. \emph{Left panel:} Model
with $\Omega_{\rm b}=0.02$. \emph{Right panel:} Model with
$\Omega_{\rm b}=0.04$.}
\label{fig:old}
\end{figure*}

As can be seen in Fig.~\ref{fig:old}, differences do exist between the
calculations of this work and those of Paper~I. The revised modelling
tends to produce slightly earlier reionization (by $\Delta z \approx
1$). These changes are due to several improvements and corrections
made to the semi-analytic model and our calculation of ionized sphere
growth made since Paper~I.

\subsection{Tests of Model Convergence}

We have examined two possible sources of numerical error in our
computational method. We outline both below.

\subsubsection{Convergence of Ionizing Emissivity}

As noted in \S\ref{sec:model} we use an iterative procedure to produce
a self-consistent model of galaxy formation and IGM evolution. To be
self-consistent, the ionizing emissivity as a function of redshift
produced by the galaxy formation model should be identical to that
used to evolve the IGM in the previous iteration. We have checked, for
all models used in this work, that such agreement has in fact been met
to within the accuracy of the model calculation (which is limited by
the number of merger trees we are able to simulate).

\subsubsection{Upper Limit to Halo Masses}
\label{sec:ulim}

\begin{figure}
\psfig{file=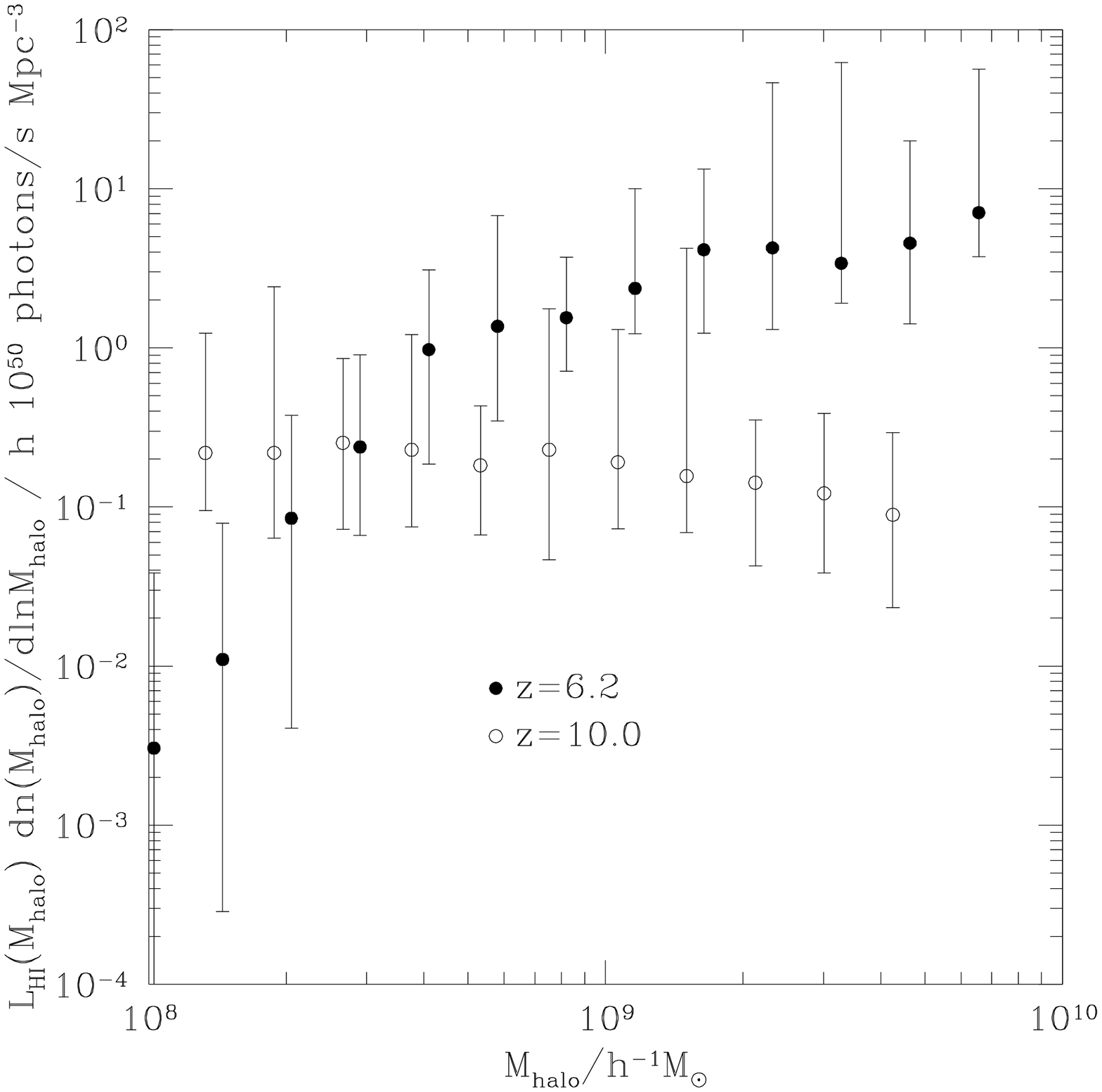,width=80mm}
\caption{The \HI\ ionizing luminosity of haloes of given mass weighted
by the halo mass function, ${\rm d}n/{\rm d}\ln M_{\rm halo}$. The
points show the median value at fixed mass while the errorbars enclose
80\% of the distribution. Results are shown for model H$_2$ in the
WMAP RSI cosmology. Filled points correspond to $z=6.2$ while open
points correspond to $z=10$. The highest halo masses shown correspond
to the most massive haloes included in our galaxy formation
calculations at the specified redshift.}
\label{fig:lumin}
\end{figure}

With finite computing resources, we can only simulate haloes with
masses up to some maximum value, if we always resolve the formation
histories of haloes down to the same minimum progenitor mass. Our
approach is to ensure that the resolution of the merger trees used in
the semi-analytic model is sufficient to resolve all haloes in which
gas could cool (i.e. haloes with virial velocities above 13km/s for
models with no \H2\ cooling or 1.3km/s for models with \H2\
cooling). In our standard runs, the most massive haloes we consider
are a factor $1.3\times 10^5$ more massive than the lowest mass
haloes. The contribution to the ionizing emissivity from galaxies in
more massive haloes is therefore lost from our
calculation. Figure~\ref{fig:lumin} shows the median \HI\ ionizing
luminosity from haloes as a function of mass weighted by the halo mass
function, ${\rm d}n/{\rm d}\ln M_{\rm halo}$. This therefore gives an
indication of the relative contribution to the total ionizing
luminosity from haloes of different masses. At high redshifts, the
contribution to the total luminosity declines with increasing halo
mass (as the halo mass function is steeply declining), so we may
expect the contribution from higher mass haloes to be negligible. At
lower redshifts the contribution is an increasing function of halo
mass. However, at the highest masses shown the relation is quite flat
and is expected to turn over at still higher masses once the
exponential break in the mass function is reached.

We can estimate the effects of the non-inclusion of higher mass haloes
as follows. We find that, for all models, the dependence of the mean
ionizing luminosity per halo on halo mass can be well represented by a
power-law.\footnote{For even larger masses this relation flattens
(see, for example, Figure~8 of \protect\scite{benson00} which shows
that, for high mass haloes, galaxy formation becomes gradually less
efficient as halo mass increases), but this will merely make our
estimate of the effects of ignoring high mass haloes a conservative
one.} We can use this relation, combined with the known mass function
of dark matter haloes to determine the filling factor we would have
obtained had we been able to model haloes of arbitrarily large
mass. We have checked the accuracy of this correction by repeating
some of our calculations including haloes up to a factor of 10 more
massive than in our standard calculations\footnote{While this is
possible, the computational time involved makes it impractical to do
this for all of our models.}. We find that our correction does indeed
overestimate the filling factor (as described above), but is typically
accurate to better than 10\%. The correction is smaller for higher
redshifts (due to the ever more rapidly declining halo mass function
found at those redshifts), and is therefore less important prior to
reionization (which is the period of interest in this work).

\begin{figure}
\psfig{file=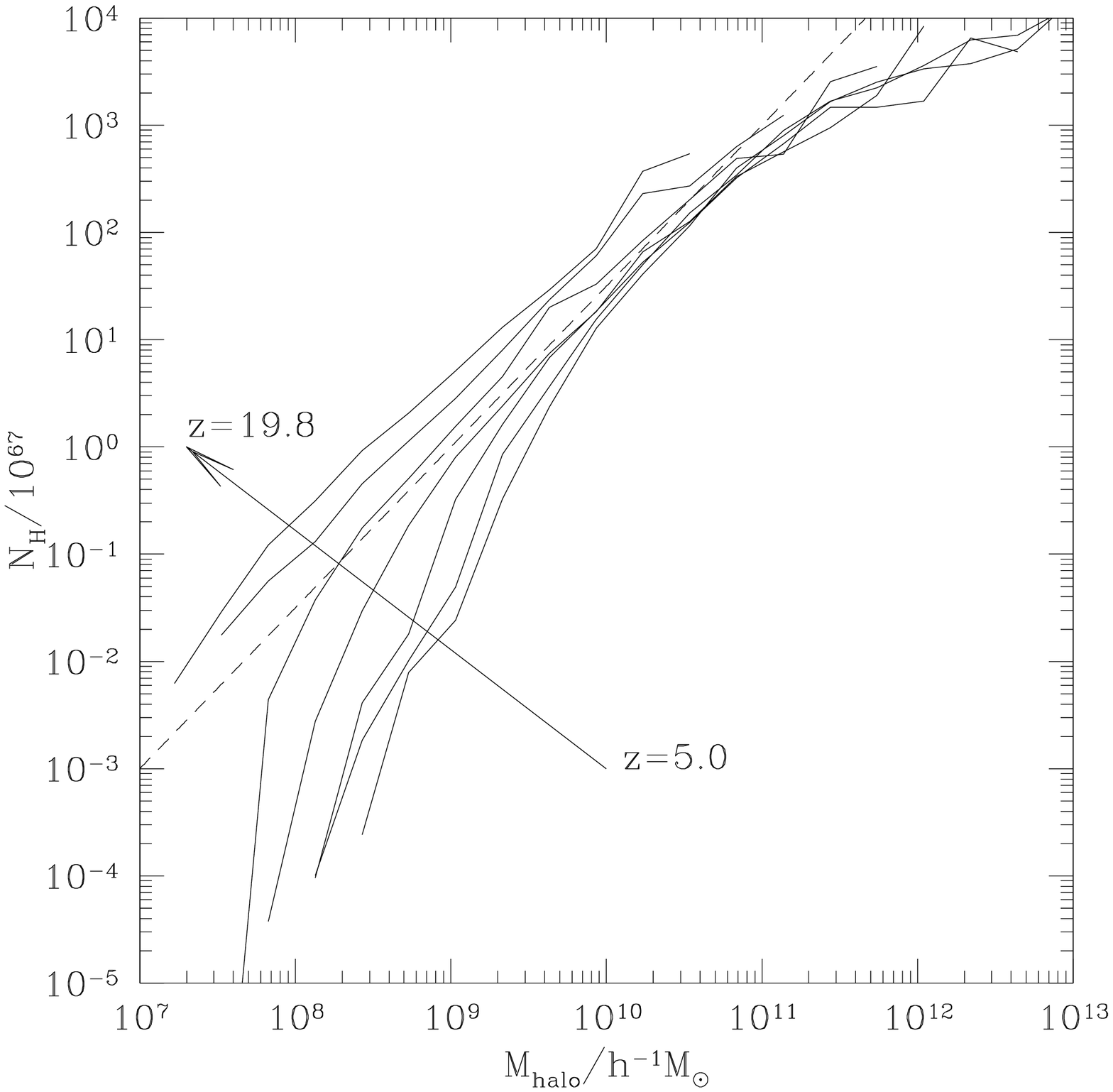,width=80mm}
\caption{The median number of hydrogen atoms ionized by sources in a
dark matter halo as a function of halo mass. Results are shown for the
Standard model in the WMAP PL cosmology. Solid lines indicate the
median number for redshifts $5.0$, $6.2$, $8.2$, $10.7$, $12.5$,
$15.4$ and $19.8$ (increasing from right to left as indicated by the
arrow). The dashed line shows the relation $N_{\rm H} \propto M^{1.5}$
for comparison.}
\label{fig:volvmass}
\end{figure}

The relation between halo mass and ionizing luminosity which arises
from our semi-analytic modelling of galaxy formation is non-trivial as
can be seen in Fig.~\ref{fig:volvmass}. While it is well represented
by a power-law for the highest mass haloes which we consider ($N_{\rm
H}\propto M_{\rm halo}^{0.5-0.7}$ for haloes in the mass range
$10^{12}$ to $10^{13}h^{-1}M_\odot$) it steepens for lower mass
haloes, reaching $N_{\rm H}\propto M_{\rm halo}^{1.5-3.0}$ depending
on redshift. The form of this relation arises from the various
physical ingredients (e.g. cooling times, supernovae feedback etc.)
which go into our calculations. Other authors
(e.g. \pcite{hh03,wylo03,furlanetto04}) have used highly simplified
assumptions (such as that the number of hydrogen atoms ionized is
simply proportional to the halo mass; \pcite{furlanetto04}) which may
be expected to give significantly different results.

\begin{figure}
\psfig{file=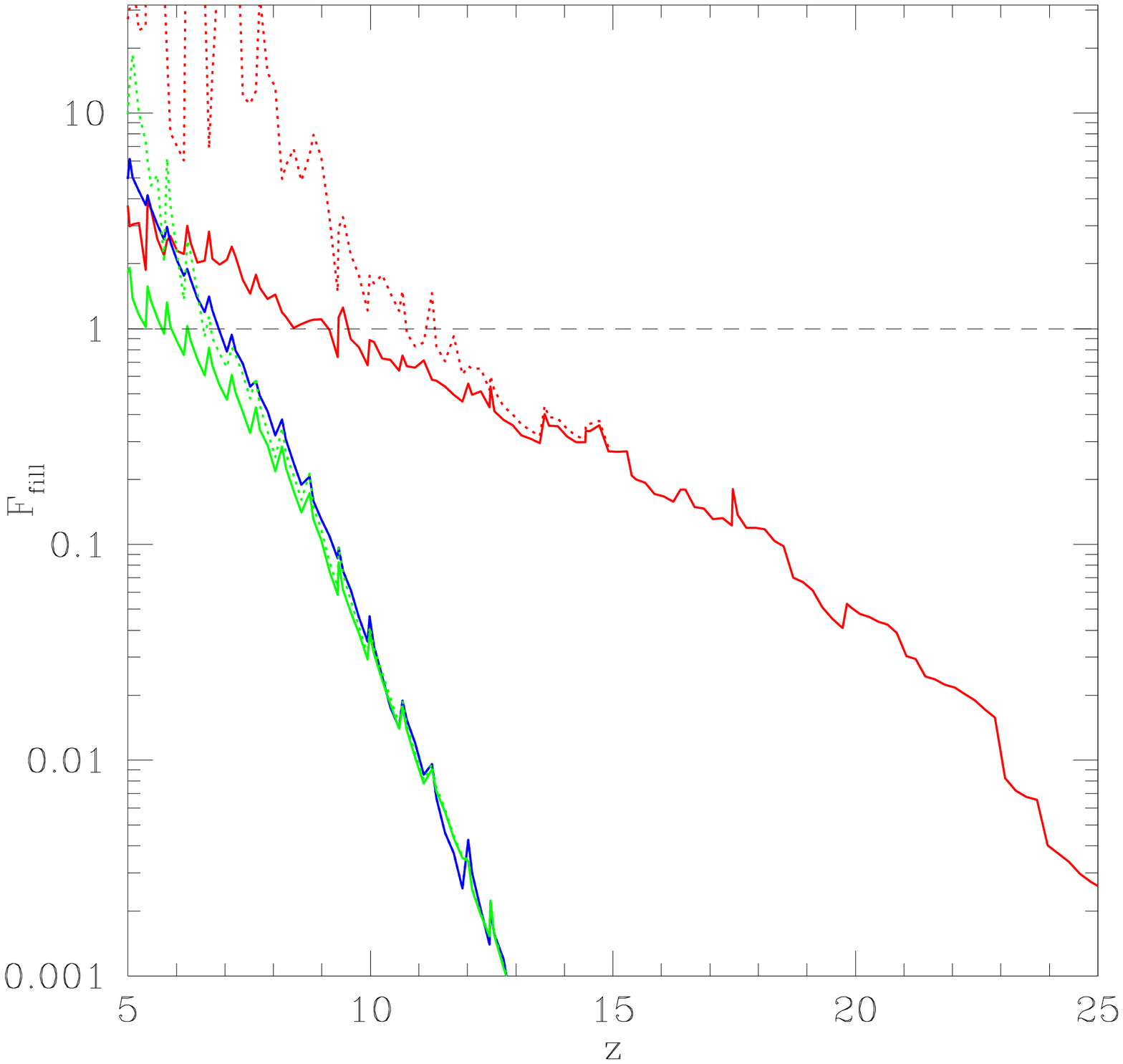,width=80mm}
\caption{The \HII\ filling factor as a function of redshift for the
``\H2'' model in the WMAP PL cosmology is shown by red lines. The
solid line shows the result direct from our model calculation while
the dotted line shows the effects of correcting for the lack of the
highest mass haloes in our calculation. The horizontal dashed line
indicates a filling factor of unity. Note that this model has the
largest fractional correction to the \HII\ filling factor out of those
listed in Table~\protect\ref{tb:GFmods}. For comparison, we show
results from the Standard model in the WMAP RSI cosmology (blue lines)
which is least affected by the lack of high mass haloes and from the
H$_2$ model in the WMAP RSI cosmology (green lines) in which the
effect of missing the highest mass haloes is more typical.}
\label{fig:jmmaxeffect}
\end{figure}

We find that the loss of the highest mass haloes affects our
determination of the epoch of reionization significantly only for
models including \H2\ cooling in the WMAP PL
cosmology. Figure~\ref{fig:jmmaxeffect} illustrates the effects of
missing the highest mass haloes in such a model (red lines). For
models without \H2\ cooling our calculations span a sufficient range
of halo masses to capture nearly all of the ionizing emissivity, while
for the WMAP RSI cosmology the space density of high mass haloes is so
small that even models including \H2\ cooling probe a sufficient range
of halo masses. The typical correction to the \HII\ filling factor in
our models is only 3--5\% for $z>10$ in models including \H2\
cooling. For $z\approx 6$ the correction reaches a factor of 2--3 for
models including \H2\ cooling. In models which ignore \H2\ cooling the
correction is less than 1\% for $z>6$. Figure~\ref{fig:jmmaxeffect}
also shows results for the model which is minimally affected by the
lack of high mass halos (Standard model in the WMAP RSI cosmology;
blue lines) and for a model in which the effect of missing high mass
halos is more typical (model ``H$_2$ in the WMAP RSI cosmology; green
lines).

When quoting results for the epoch of reionization we   list
the value computed directly from our calculations along with the value
obtained after the above correction has been applied. The true epoch
of reionization should then lie between these two values.

\section{Condensed Fractions}
\label{app:fcond}

\begin{figure*}
\psfig{file=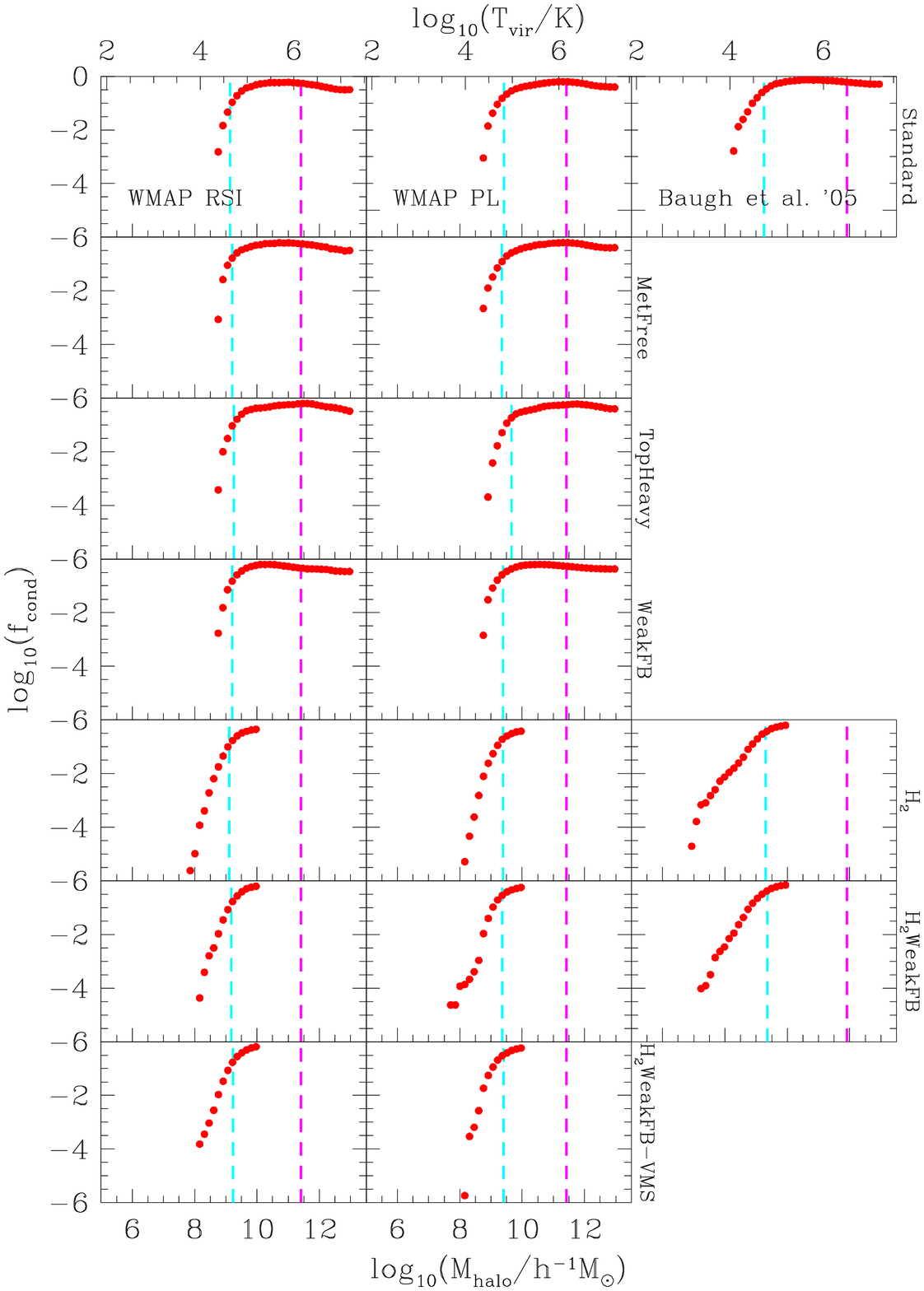,width=180mm}
\vspace{-17mm}
\caption{The condensed fraction, $f_{\rm cond}$, defined as the mass
of material in a halo which has condensed into the galaxy phase,
divided by $\Omega_{\rm b}M_{\rm halo}/\Omega_0$ is shown by red
circles. Results are shown for $z=5$ and for all three cosmological
models (one per column as indicated by the labels in the upper row of
panels), and for all models computed (one model per row as indicated
at the right-hand edge of each row). The upper x-axis indicates the
virial temperature corresponding to halo mass. Vertical dashed magenta
lines indicate the virial temperature corresponding to the feedback
parameter $V_{\rm hot}$ in each model, while vertical cyan lines
indicate the filtering mass in each model.}
\label{fig:fcond5}
\end{figure*}

\begin{figure*}
\psfig{file=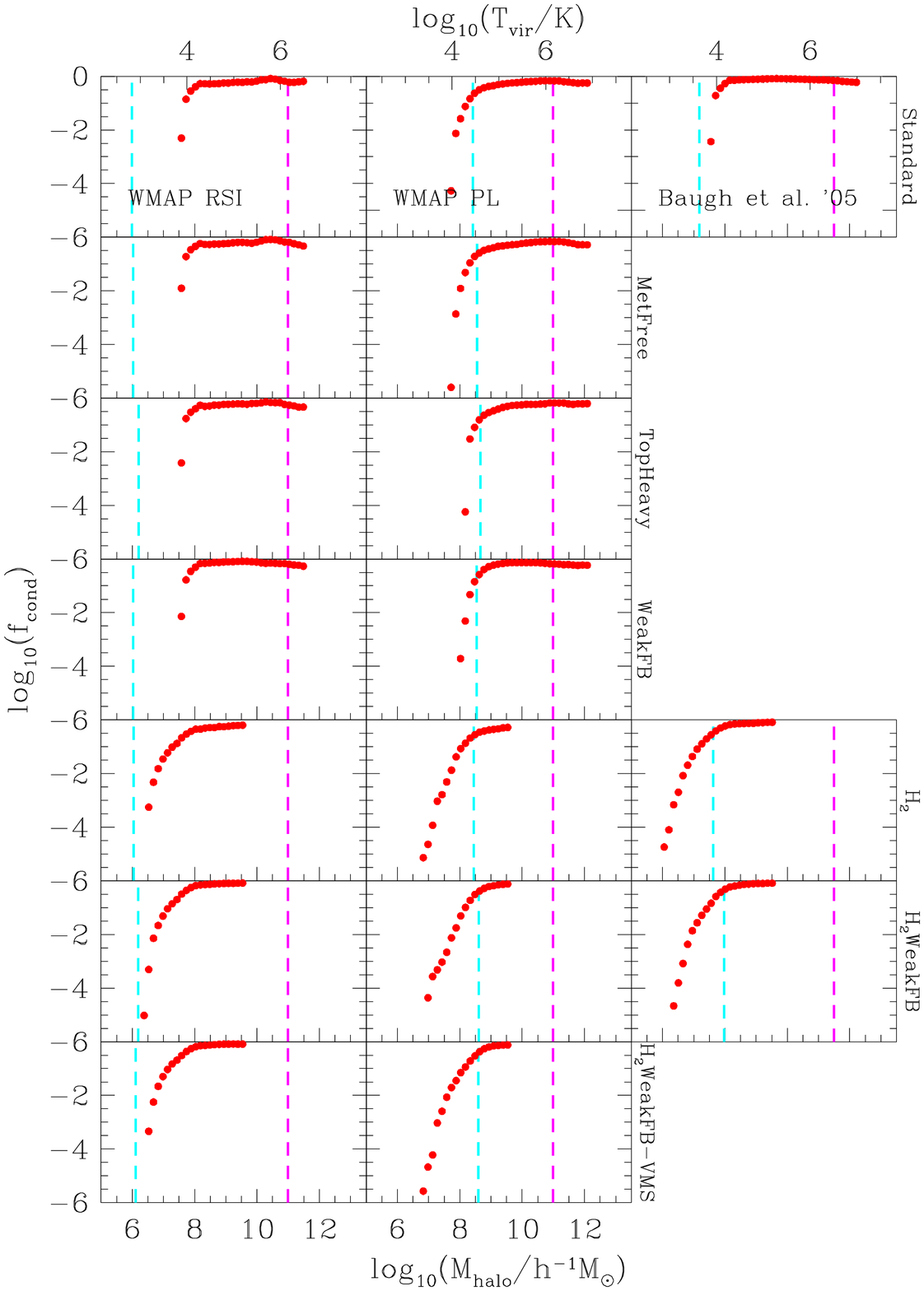,width=180mm}
\vspace{-17mm}
\caption{As Fig.~\protect\ref{fig:fcond5} but for $z=10$.}
\label{fig:fcond10}
\end{figure*}

Figures~\ref{fig:fcond5} and \ref{fig:fcond10} show condensed fraction
(i.e. the mass of material condensed into the galaxy phase divided by
$\Omega_{\rm b}M_{\rm halo}/\Omega_0$) as a function of halo mass
$M_{\rm halo}$ for redshifts $z=5$ and 10 respectively. Columns
indicate the cosmology (see labels in upper row) while rows indicate
the model (see labels at right-hand edge of each row). In each panel,
the vertical dashed magenta line indicates the virial temperature
corresponding to the feedback parameter $V_{\rm hot}$ while the
vertical dashed cyan line indicates the filtering mass. Red points
indicate the condensed fraction (note that the vertical scale is
logarithmic).

The effects of including \H2\ cooling can be clearly seen as the
presence of a tail of low $f_{\rm cond}$ for low mass halos. The
effects of weak feedback are also apparent, resulting in a small
increase in $f_{\rm cond}$ in low mass halos.

\begin{thebibliography}{}
\bibitem[Abel, Bryan \& Norman <2000>]{abel00}Abel~T., Bryan~G.~L., Norman~M.~L., 2000, ApJ, 540, 39
\bibitem[Abel, Bryan \& Norman <2002>]{abel02}Abel~T., Bryan~G.~L., Norman~M.~L., 2002, Science, 295, 93
\bibitem[Anninos \& Norman <1996>]{ann96}Anninos~P., Norman~M.~L., 1996, ApJ, 460, 556
\bibitem[Baugh et al. <1998>]{baugh98}Baugh~C.~M., Cole~S., Frenk~C.~S., Lacey~C.~G., 1998, ApJ, 498, 504
\bibitem[Baugh et al. <2005>]{baugh05}Baugh~C.~M., Lacey~C.~G.,
  Frenk~C.~S., Granato~G.~L., Silva~L., Bressan~A., Benson~A.~J.,
  Cole~S., 2005, MNRAS 356, 1191
\bibitem[Becker et al. <2001>]{becker01}Becker~R.~H. et al., 2001, ApJ, 122, 2850
\bibitem[Benson et al. <2000>]{benson00}Benson~A.~J., Cole~S., Frenk~C.~S., Baugh~C.~M., Lacey~C.~G., 2000, MNRAS, 311, 793
\bibitem[Benson et al. <2001>]{benson01}Benson~A.~J., Nusser~A.,
  Sugiyama~N., Lacey~C.~G., 2001, MNRAS, 320, 153 (Paper I)
\bibitem[Benson et al. <2002a>]{benson02}Benson~A.~J., Lacey~C.~G., Baugh~C.~M., Cole~S., Frenk~C.~S., 2002a, MNRAS, 333, 156
\bibitem[Benson et al. <2002b>]{benson02b}Benson~A.~J., Frenk~C.~S., Lacey~C.~G., Baugh~C.~M., Cole~S., 2002a, MNRAS, 333, 177
\bibitem[Benson et al. <2003a>]{benson03}Benson~A.~J., Bower~R.~G., Frenk~C.~S., Lacey~C.~G.,. Cole~S., 2003a, ApJ, 599, 38
\bibitem[Benson et al. <2003b>]{benson03b}Benson~A.~J., Frenk~C.~S., Baugh~C.~M., Cole~S., Lacey~C.~G., 2003b, MNRAS, 343, 679
\bibitem[Benson et al. <2004>]{benson04}Benson~A.~J., Lacey~C.~G., Frenk~C.~S., Baugh~C.~M., Cole~S., 2004, MNRAS, 351, 1215 
\bibitem[Bergvall et al. <2006>]{bergvall06}Bergvall~N., Zackrisson~E., Andersson~B.-G., Arnberg~D., Masegosa~J., \"Ostlin~G., 2006, astro-ph/0601608
\bibitem[Bond et al. <1991>]{bond91}Bond~J.~R., Cole~S., Efstathiou~G., Kaiser~N., 1991, ApJ, 379, 440
\bibitem[Bower <1991>]{bower91}Bower~R.~G., 1991, MNRAS, 248, 332
\bibitem[Bromm, Kudritzki \& Loeb <2001>]{bromm01}Bromm~V., Kudritzki~R.~P., Loeb~A., 2001, ApJ, 552, 464
\bibitem[Bromm, Coppi \& Larson <2002>]{bromm02}Bromm~V., Coppi~P.~S., Larson~R.~B., 2002, ApJ, 564, 23
\bibitem[Bromm \& Loeb <2004>]{bromm03}Bromm~V., Loeb~A., 2004, Nature, 425, 812
\bibitem[Bruscoli et al. <2000>]{bruscoli00}Bruscoli~M., Ferrara~A., Fabbri~R., Ciardi~B., 2000, MNRAS, 318, 1068
\bibitem[Bruzual \& Charlot <1993>]{bc93}Bruzual~A.~G., Charlot~S., 1993, ApJ, 405, 538
\bibitem[Cazaux \& Spaans <2004>]{cazaux}Cazaux~S., Spaans~M., 2004, ApJ, 611, 40
\bibitem[Cen <2003>]{cen03}Cen~R., 2003, ApJ, 591, L5
\bibitem[Chen \& Kamionkowski <2004>]{chen04}Chen~X., Kamionkowski~M., 2004, Phys. Rev. D, 70
\bibitem[Chiu \& Ostriker <2000>]{chiu00}Chiu~W.~A., Ostriker~J.~P., 2000, ApJ, 534, 507
\bibitem[Ciardi et al. <2000>]{ciardi00}Ciardi~B., Ferrera~A., Governato~F., Jenkins~A., 2000, MNRAS, 314, 611
\bibitem[Ciardi, Ferrara \& White <2003>]{ciardi03}Ciardi~B., Ferrara~A., White~S.~D.~M., 2003, MNRAS, 344, 7
\bibitem[Cole et al. <2000>]{cole00}Cole~S., Lacey~C.~G., Baugh~C.~M., Frenk~C.~S., 2000, MNRAS, 319, 168
\bibitem[Djorgovski et al. <2001>]{djorg01}Djorgovski~S.~G., Castro~S.~M., Stern~D., Mahabal~A.~A., 2001, ApJ, 560, 5
\bibitem[Dove \& Shull <1994>]{ds94}Dove~J.~B., Shull~M.~J., 1994, ApJ, 423, 196
\bibitem[Dove, Shull \& Ferrara <2000>]{dsf00}Dove~J.~B., Shull~M.~J.,
  Ferrara~A., 2000, ApJ, 531, 486
\bibitem[Fan et al. <2003>]{fan03}Fan~X., et al., 2003, AJ, 128, 515
\bibitem[Fan et al. <2006>]{fan06}Fan~X., et al., 2006, astro-ph/0512082 
\bibitem[Ferrara et al. <1999>]{ferrara99}Ferrara~A., Bianchi~S., Cimatti~A., Giovanardi~C., 1999, ApJS, 123, 437
\bibitem[Fujita et al. <2003>]{fujita03}Fujita~A., Martin~C.~L., Mac Low~M.-M., Abel~T., 2003, ApJ, 599, 50
\bibitem[Furlanetto, Zaldarriaga \& Hernquist <2004>]{furlanetto04}Furlanetto~S.~R., Zaldarriaga~M., Hernquist~L., 2004, ApJ, 613, 16
\bibitem[Furlanetto \& Loeb <2005>]{furl05}Furlanetto~S.~R., Loeb~A., 2005, ApJ, 634, 1
\bibitem[Galli \& Palla <1998>]{galpal98}Galli~D., Palla~F., 1998, A\&A, 335, 403
\bibitem[Gnedin \& Ostriker <1997>]{go97}Gnedin~N.~Y., Ostriker~J.~P., 1997, ApJ, 486, 581
\bibitem[Gnedin <2000>]{gnedin00}Gnedin~N.~Y., 2000, ApJ, 542, 535
\bibitem[Granato et al. <2000>]{granato00} Granato, G.L., Lacey, C.G.,
  Silva, L., Bressan, A., Baugh, C.M., Cole, S., Frenk, C.S., 2000,
  ApJ 542, 710
\bibitem[Haiman \& Loeb <1997>]{haiman97}Haiman~Z., Loeb~A., 1997, ApJ, 483, 21
\bibitem[Haiman \& Holder <2003>]{hh03}Haiman~Z., Holder~G., 2003, ApJ, 595, 1
\bibitem[Hansen \& Haiman <2004>]{hansen04}Hansen~S.~H., Haiman~Z., 2004, ApJ, 600, 26
\bibitem[Hinshaw et al. <2003>]{hinshaw03}Hinshaw G. et al., 2003, ApJSupp, 148, 135
\bibitem[Holder et al. <2003>]{Holder}Holder~G.~P., Haiman~Z., Kaplinghat~M., Knox~L., 2003, ApJ, 595, 13
\bibitem[Hutchins <1976>]{hutchins76}Hutchins~J.~B., ApJ, 205, 103
\bibitem[Jenkins et al. <1998>]{jenkins98}Jenkins~A. et al., 1998, ApJ, 499, 20
\bibitem[Kasuya, Kawasaki \& Sugiyama <2004>]{kasuya04}Kasuya~S., Kawasaki~M., Sugiyama~N., 2004, Phys. Rev. D, 69
\bibitem[Kasuya \& Kawasaki <2004>]{kasuya04b}Kasuya~S., Kawasaki~M., 2004, Phys. Rev. D, 70, 103519
\bibitem[Kauffmann et al. <1999>]{kauffmann99}Kauffmann~G.,  Colberg~J.~M., Diaferio~A., White~S.~D.~M., 1999, MNRAS, 303, 188
\bibitem[Kennicutt <1983>]{kennicutt83}
Kennicutt, R.C., 1983, ApJ, 272, 54
\bibitem[Kogut et al. <2003>]{kogut03}Kogut A. et al., 2003, ApJSupp, 148, 161
\bibitem[Le Delliou et al. <2005a>]{LeD05a}
Le Delliou, M., Lacey, C., Baugh, C.M., Guiderdoni, B., Bacon, R.,
Courtois, H., Sousbie, T., Morris, S.L., 2005, MNRAS,  357, L11
\bibitem[Le Delliou et al. <2006>]{LeD05b}
Le Delliou, M., Lacey, C., Baugh, C.M., Morris, S.L., 2006, MNRAS,
365, 712
\bibitem[Leitherer et al. <1995>]{leith95}Leitherer~C., Ferguson~H., Heckman~T.~M., Lowenthal~J.~D., 1995, ApJ, 454, 19
\bibitem[Machacek, Bryan \& Abel <2001>]{macha01}Machacek~M.~E., Bryan~G.~L., Abel~T., 2001, ApJ, 548, 509
\bibitem[MacTavish et al. <2005>]{boom03}MacTavish et al., 2005, submitted to ApJ (astro-ph/0507503)
\bibitem[Nagashima et al. <2005a>]{nagashima05a}
Nagashima, M., Lacey, C.G., Baugh, C.M., Frenk, C.S., Cole, S.,
2005, MNRAS 358, 1247
\bibitem[Nagashima et al. <2005b>]{nagashima05b}
 Nagashima, M., Lacey, C.G., Okamoto, T., Baugh, C.M., Frenk, C.S., Cole, S.,
2005, MNRAS, 363, L31  
\bibitem[Naselsky \& Chaing <2004>]{nasel04}Naselsky~P., Chiang~L.-Y., 2004, MNRAS, 2004, 347, 795
\bibitem[Onken \& Miralda-Escud\'e <2004>]{onken04}Onken~C.~A., Miralda-Escud\'e~J., 2004, ApJ, 610, 1
\bibitem[Peebles <1968>]{peebles68}Peebles~P.~J.~E., 1968, ApJ, 153, 1
\bibitem[Ricotti \& Ostriker <2004>]{rico04}Ricotti~M., Ostriker~J.~P., 2004, MNRAS, 352, 547
\bibitem[Schneider et al. <2002>]{schneider02}Schneider~R., Ferrara~A., Natarajan~P., Omukai~K., 2002, ApJ, 571, 30
\bibitem[Seager, Sasselov \& Scott <2000>]{recfast}Seager~S., Sasselov~D.~D., Scott~D., 2000, ApJS, 128, 407
\bibitem[Seljak et al. <2003>]{seljak03}Seljak~U., Sugiyama~N., White~M., Zaldarriaga~M., 2003, Phys. Rev. D, 68, 083507
\bibitem[Somerville, Bullock \& Livio <2003>]{rs03}Somerville~R.~S.,
  Bullock~J.~S., Livio~M., 2003, ApJ, 593, 616
\bibitem[Songaila <2004>]{songaila04}Songaila, A., 2004, AJ, 127, 2598
\bibitem[Spergel et al. <2003>]{spergel03}Spergel~D.~N. et al., 2003, ApJS, 148, 175
\bibitem[Spergel et al. <2006>]{spergel06}Spergel~D.~N. et al., 2006, submitted
\bibitem[Steidel, Pettini \& Adelberger <2001>]{steidel01}Steidel~C.~C., Pettini~M., Adelberger~K.~L., 2001, ApJ, 546, 665
\bibitem[Sugiyama <1995>]{sugiyama95}Sugiyama~N., 1995, ApJS, 100, 281
\bibitem[Tegmark et al. <1997>]{tegmark97}Tegmark~M., Silk~J., Rees~M.~J., Blanchard~A., Abel~T., Palla~F., 1997, ApJ, 474, 1
\bibitem[Tumlinson et al. <1999>]{tumlinson99}Tumlinson~J., Giroux~M.~L., Shull~M.~J., Stocke~J.~T., 1999, AJ, 118, 2148
\bibitem[Valageas \& Silk <1999>]{vs99}Valageas~P., Silk~J., 1999, A\&A, 347, 1
\bibitem[Viel, Weller \& Haehnelt <2004>]{viel04}Viel~M., Weller~J.,
  Haehnelt~M.~G., 2004, MNRAS, 355, L23
\bibitem[White et al. <2003>]{white03}
White, R.L., Becker, R.H., Fan, X., Strauss, M., 2003, AJ, 126, 1
\bibitem[Wood \& Loeb <2000>]{wood00}Wood~K., Loeb~A., 2000, ApJ, 545, 86 
\bibitem[Wyithe \& Loeb <2003>]{wylo03}Wyithe~J.~S.~B., Loeb~A., 2003, ApJ, 586, 693
\bibitem[Yoshida et al. <2003a>]{yoshi03}Yoshida~N., Abel~T., Hernquist~L., Sugiyama~N., 2003a, ApJ, 592, 645
\bibitem[Yoshida et al. <2003b>]{yoshi03b}Yoshida~N., Sokasian~A., Hernquist~L., Springel~V., 2003b, ApJ, 598, 73
\end{thebibliography}
\end{document}